# Hadronic Regge Trajectories in the Resonance Energy Region

A.E. Inopin

Department of Physics and Technology, V.N. Karazin Kharkov National University, Kharkov 61077, Ukraine & Virtual Teacher Ltd., Vancouver, Canada.

Nonlinearity of hadronic Regge trajectories in the resonance energy region has been proved.

## Contents

# INTRODUCTION

Regge trajectories (RT) in hadron physics have been known for some 40 years. Initially they were introduced by Tullio Regge [1,2], who simply generalized the solution for scattering amplitude by treating the angular momentum *l* as a complex variable. He proved that for a wide class of potentials the only singularities of the scattering amplitude in the complex *l* plane were poles, now called Regge poles. If these poles occur for positive integer values of l they correspond to the bound states or the resonances, and they are also important for determining certain technical aspects of the dispersion properties of the amplitudes. Regge interpreted the simple poles of $a_l(k^2)$ on the complex *l*-plane to be either resonances or bound states. Chew and Frautschi [3] applied the Regge poles theory to investigate the analyticity of $a_l(k^2)$ in the case of strong interactions. They simply postulated that all strongly interacting particles are self-generating (the bootstrap hypothesis) and that they must lie on Regge trajectories (Chew-Frautschi conjecture) [4]. At first, linearity was just a convenient guide in constructing the Chew-Frautschi plots, because data were scarce and there were few a priori rules to direct the mesons and baryons into the same trajectories [5]. Once linearity was found to be a good working hypothesis, justification was given through certain assumptions in the Regge poles theory as follows: for *Rel≥1/2*, the partial–wave components of the scattering amplitude *f* have only simple poles and are functions of $k^2$

$$a_l(k^2) \approx \beta(k^2)/(l - \alpha(k^2)) \qquad (1)$$

where $\beta$ is the residue and $\alpha$ the position (Regge trajectory) of the simple poles. By the end of the 1960s quarks were discovered experimentally and quark-parton model emerged almost immediately [6,7]. In the 1970s quantum chromodynamics (QCD) got a firm ground as a theory of strong interactions. This theory, QCD, has nothing to do with the original framework of Regge and Chew-Frautschi, since it's dealing with different composite objects and dynamical equations.

The aim of the present paper is to review and scrutinize broad class of modern models of hadron spectroscopy, which consider RT. The focus will be the central issue of linear versus nonlinear, and parallel versus nonparallel RT. We will reach consensus on some point between different the theoretical schemes. On the other hand hadronic data itself presents the purest imprint of the hadronic world. Therefore we will scrutinize the last issue of Review of Particle Physics 2000, and reconstruct all possible RT for mesons and baryons. Finally we will reach consensus between the data and theory.

## II. THEORETICAL MODELS

### II.1. ħ-expansion model

In the series of papers [8], the authors, introduced a new procedure for the solution – of the SE for mesons, which is based on the expansion of wave function in the Planck constant ħ. Following the standard scheme of logarithmic perturbation theory, it's easy to go over to the nonlinear Riccati equation

$$\hbar C'(r) + C^2(r) = \frac{\hbar^2 l(l+1)}{r^2} + 2m(V(r) - E),$$ (1)

the solution of which is sought as an asymptotic expansion in Planck's constant :

$$C(r) = \sum_{k=0}^{\infty} C_k(r)\hbar^k,$$ (2)

$$\alpha(E) = \hbar l(E) = \sum_{k=0}^{\infty} \alpha_k(E)\hbar^k.$$

The Cauchy integral of the logarithmic derivative $C(r)$ around closed contour that surrounds only the specified zeros leads to the well–known Zwaan-Dunham quantization conditions [9,10]

$$\frac{1}{2\pi i}\oint C(r)dr = n\hbar, \quad n=0,1,2, \dots$$ (3)

These relations are exact and are used to find the corrections to the WKB approximation. It should be pointed out that the transition from quantum to classical mechanics, carried out formally as passage to the limit $\hbar{\to}0$, is not uniquely defined. The reason for this is that the WF depends on the radial and the orbital quantum number, the behaviour of which must be regulated as $\hbar{\to}0$. The rule for going to the classical limit in the WKB approximation is

$\hbar{\to}0, \ n{\to}\infty$ in such a way that $\hbar n{=}const$ (4).

The proposed method is constructed as complementary to the WKB approximation, and it is realized in accordance with the alternative possibility

$\hbar{\to}0, \ n{=}const$ in such a way that $\hbar n{\to}0$ (5).

From the physical point of view the rule (4) corresponds to classical motion of the particle between the turning points with an energy appreciably above the minimum of the effective potential

$$V_{eff}(r) = V(r) + \frac{\Lambda^2}{2mr^2}$$ (6)

whereas in the case of the classical limit (5) the particle is at the minimum of the effective potential, i.e., it moves in a stable circular orbit with radius $r_0$ determined from the equation

$$mr_0^3 V'(r_0) = \Lambda^2$$ (7).

Here, $\Lambda^2$ is the contribution of the classical orbital angular momentum obtained from the quantum-mechanical expression $\hbar^2 l(l+1)$ in the limit

$\hbar{\to}0, \ l{\to}\infty$ in such a way that $\hbar l{=}const$ (8).

Expansion (2) brings the condition (3) to the form

$$\frac{1}{2\pi i}\oint C_1(r)dr = n; \quad \frac{1}{2\pi i}\oint C_k(r)dr = 0; \qquad k{\neq}1$$ (9).

If we use power-law potential and apply expansion (2) to $q\bar{q}$ SE, we have

$$\alpha(E) = \hbar l(E) = \alpha_0(E) - (1+(2n+1)\sqrt{\nu+2})\frac{\hbar}{2} + \frac{(\nu-2)(\nu+1)}{144\alpha_0(E)}(6n^2+6n+1)\hbar^2 + ... \qquad (10)$$

with $\alpha_0(E) = \sqrt{\nu/2}\left[2E/(\nu+2)\right]^{(\nu+2)/2\nu}$.

These three terms of expansion already provide the accuracy which is enough for our purposes. The leading contribution in (10) can be rewritten as

$$\alpha_0(\mathrm{X}) = \mathrm{A}(\sqrt{x}+b)^\mu \qquad (11),$$

where $\mathrm{A} = \sqrt{\nu/2}(a\mu\nu)^{-\mu}$, $\quad x = \mathrm{M}^2/\mathrm{M}_0^2$, $\quad b = -\dfrac{m_1+m_2+V_0}{\mathrm{M}_0}$, $\qquad \mu = \dfrac{\nu+2}{2\nu}$.

We point out some general features of the RT for the power-law potentials. It can be found that

$$\alpha(\mathrm{M}^2) = \mathrm{A}\left(\frac{\mu}{\mu_0}\right)^\mu - (1+(2n+1)\sqrt{\nu+2})\hbar/2 \text{ as } \mathrm{M}^2 \to \infty .$$

Therefore the parent and daughter RT become parallel in asymptotics.

On performing the differentiation of (11) with respect to $x$ the estimation for the slope of the parent RT is given by

$$\alpha'(\mathrm{M}^2) \approx \frac{\mathrm{A}\mu}{2\mathrm{M}_0^2}\left(\frac{\mathrm{M}}{\mathrm{M}_0}\right)^{\mu-2}\left(1+b\frac{\mathrm{M}_0}{\mathrm{M}}\right)^{\mu-1} \qquad (12).$$

Jet us consider $\rho$-$a$ trajectory with potential $r^{2/3}$, using (12). It is seen that these RT, as a matter of fact, are *nonlinear* (Fig. 1). Moreover, the slope of the parent RT (12) gradually decreases from $\alpha'(\mathrm{M}_{\rho 1}^2) = 1.10 GeV^{-2}$ at the first resonance position, approaching asymptotically the value $\mathrm{A}/\mathrm{M}_0^2 = 0.53 GeV^{-2}$. Thus, the approximate linearity of the RT in the resonance region really has nothing in common with the asymptotical linearity which begins on the mass value $\mathrm{M} \approx 7-8 GeV$, unachievable for light mesons.

It is easy to get from (12) $\alpha'(\mathrm{M}^2 \to \infty)$ for the popular choices of power-law potentials.

$$\begin{cases} \nu = 0.1 & \alpha'(\infty) = \infty \\ \nu = 2/3 & \alpha(\infty) = 0.523 Gev^{-2}{}' \\ \nu = 1 & \alpha'(\infty) = 0 \\ \nu = 2 & \alpha'(\infty) = 0 \end{cases}$$

We should make some remarks:

(a)     From Eq.(10) it follows that all generally used potentials lead to the nonlinear RT (it is valid for a relativistic description, too).

(b)

(c)     In the experimental region, mainly to its comparatively small length, the RT will be rather weakly nonlinear for all permissible potentials with parameters chosen appropriately.

(d)     The asymptotical linearity of the RT resulting from the potential $r^{2/3}$, in the SE framework, has nothing in common with the approximate linearity in the resonance region.

(e)     If the manifestation of an asymptotical behavior influence on the parent RT is expected at $l \approx 6\text{-}10$, that for the daughter RT will be at lower values of $l$, as follows from Eq (10).

(f)     There are neither theoretical non experimental good reasons in favor of any concrete asymptotical behavior of potentials, including also $r^{2/3}$, at present. As has been proved [11], even the potentials which approach a constant at large distances, providing "incomplete quark confinement", describe well the mass spectrum.

Authors [8] investigate charmonium spectroscopy with such screened potential

$$V(r) = b + \frac{a}{\mu}(1 + e^{-\mu r}) \tag{13},$$

and also compute corresponding RT. Since the potential approaches a constant value, the eigenvalues $E$ belong to the interval $0 < E < 1/\lambda$   $(\lambda = \mu/(am)^{1/3})$. In terms of the quarkonium mass this is written as

$$(2m+b) < M < (2m+b+a/\mu) \tag{14}$$

Then it  follows that the considered RT will be limited. The boundary value attained  at $M = 2m + b + a/\mu$ equals

$$l_{\max} \approx \frac{2}{e}\left(\frac{am}{\mu^3}\right)^{1/2} - (n+1)\hbar \tag{15}$$

The numerical estimation of the boundary for the deconfinement region is difficult due to undetermined values of the parameters. Rough estimate for parent RT gives

$$M_{max} \approx 5.2 GeV, \qquad\qquad l_{max} \approx 12. \tag{16}$$

Although there is the possibility of color state creation, it is not realized because of the high threshold value $(M_{max} > 5.2 GeV)$. At energies lower that the threshold value the usual confinement physics correct.

Authors constructed RT for specific potentials widely used in quarkonium physics. For the power-law potential $V(r) = Ar^\nu$ the leading approximation is

$$\alpha_0(E) = \left[\frac{2E}{A(\nu+2)}\right]^{\frac{\nu+2}{2\nu}} \sqrt{\frac{A\nu}{2}} \tag{17}$$

and the parent and daughter RT are given by (10).

For the harmonic oscillator $(v=2)$ and the Coulomb interaction $(v=-1)$, the third term $\alpha_2(E)$ is equal to zero and the first two terms $\alpha_0(E)+\hbar\alpha_1(E)$ reproduce the exact result. Indeed, when $v=2$, $A=\omega^2$

$$\alpha(E) = \frac{E}{2\omega} - \frac{4h+3}{2}\hbar \tag{18}$$

or in the more traditional form

$$E_{n,l} = \hbar\omega(4n+2l+3) \tag{19}$$

If $v=-1$, $A=-ze^2$, then

$$\alpha(E) = \frac{ze^2}{2\sqrt{-E}} - (n+1)\hbar , \tag{20}$$

and $E_{n,l} = -\dfrac{z^2e^4}{4\hbar^2(l+n+1)^2}$ (21)

Note also that expression (10) for $\alpha(E)$ can be inverted at fixed l-values to yield the energy eigenvalues $E_{n,l}$

$$E_{n,l} = \frac{v+2}{2}(2v)^{-\frac{v}{v+2}}\left[(2n+1)\sqrt{v+2} + \sqrt{4\Lambda - \frac{n(n+1)}{3}(v+1)(v-2) - \frac{v^2-v-20}{18}}\right]^{\frac{2v}{v+2}} \tag{22}$$

Applying (10) to Martin's potential $(v=0.1)$, we obtain an energy spectrum consistent with the exact one.

The $\hbar$-expansion method is nonperturbative. Authors proved that for all potentials including funnel-shaped $(V(r)=-a/r+br)$, logarithmic, and power-law, expressions for $\alpha(E)$ work well both for small and large values of a coupling constant. It will be very interesting to see how the $\hbar$-expansion method will work for detailed analysis of all meson/baryon spectra and RT.

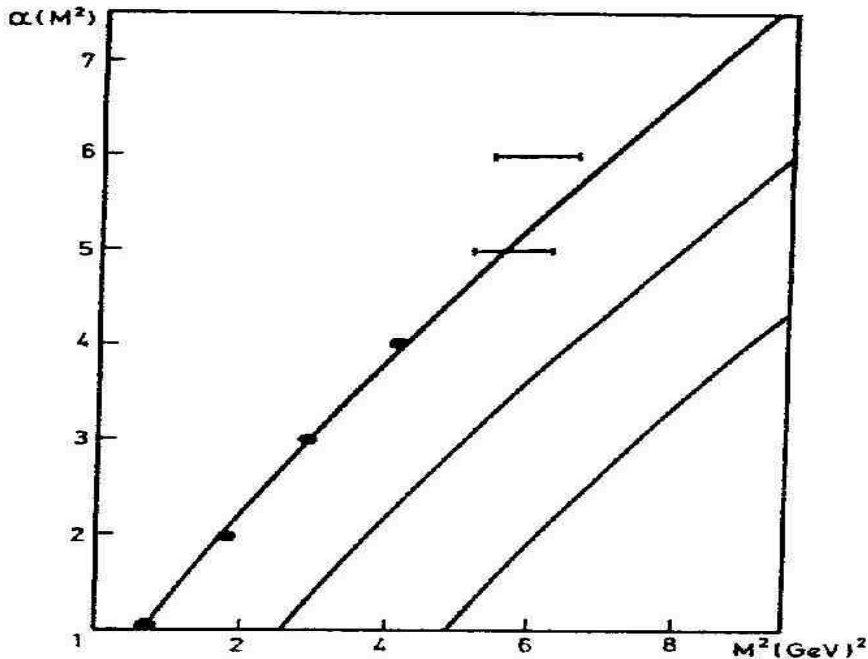

## II.2. K-deformed algebra model

The *k*-deformed algebra could be introduced via the commutation relations between the components of energy-momentum operator $P_\upsilon(\upsilon=0, 1, 2, 3)$, angular momentum operator $M_i$ and Lorentz boosts $L_i$.

$[P_i, P_j]=0,$ $\qquad\qquad\qquad [P_i, P_0]=0$

$[M_i, P_j]=i\varepsilon_{ij}kP_k,$ $\qquad\qquad [M_i, P_0]=0$

$[L_i, P_0]=iP_i,$ $\qquad\qquad\qquad [L_i, P_j]=i\varepsilon^{-1}\delta_{ij}sinh(\epsilon P_0),$

$[M_i, M_j]=i\varepsilon_{ijk}M_k,$ $\qquad\qquad [M_i, L_j]=i\varepsilon_{ijk}L_k,$

$[L_i, L_j]=-i\varepsilon_{ijk}[M_k\,cosh(\epsilon p_0)\text{-}1/4\epsilon P_kP_lM_l],$ $\qquad\qquad\qquad\qquad$ (1)

where $i, j, k = 1, 2, 3$. Eqs. (1) are written in terms of the inverse of the deformation parameter, $\epsilon\equiv k^{-1}$, a quantity with dimension of length. In the limit $\epsilon\rightarrow 0$ one recovers the usual undeformed Poincare algebra.

The first Casimir invariant operator, $C_1$, is given by

$$C_1 = \left[\frac{2}{\epsilon}\sinh\left(\frac{\epsilon P_0}{2}\right)\right]^2 - P_iP_i \qquad\qquad\qquad (2)$$

In the rest frame *(Pi=0)* we have

$$C_1 = \left[\frac{2}{\epsilon}\sinh\left(\frac{m\epsilon}{2}\right)\right]^2 \qquad\qquad\qquad (3)$$

Equating (2) and (3) for the Casimir invariant *(E=P₀),* we have

$$\left[\frac{2}{\epsilon}\sinh\left(\frac{\epsilon E}{2}\right)\right]^2 = \vec{P}^2 + \left[\frac{2}{\epsilon}\sinh\left(\frac{m\epsilon}{2}\right)\right]^2 \qquad\qquad\qquad (4)$$

Hence

$$E = \frac{2}{\epsilon}\sinh^{-1}\left[\left(\frac{\epsilon}{2}\right)^2\vec{P}^2 + \sinh\left(\frac{m\epsilon}{2}\right)\right]^{1/2} \qquad\qquad\qquad (5)$$

Authors of this model [12] predicted a *flattening* with angular momentum J of the resonance spectrum of hadrons. That a dynamical symmetry group can be applied to get the spectrum of elementary particles was realized early in the work of Barut [13]. Later Dothan et al [14] and independently Barut and Böhm [15] developed the method of spectrum generating algebra, which was applied to both relativistic and nonrelativistic systems. In nuclear physics this became known as the interacting boson model (IBM) and was generalized to boson-fermion cases. It was realized early that this can be applied to the baryon spectrum and could include mesons too. From [16, 17],

$$E^2 = m^2 + \frac{L}{\alpha'} + \frac{n}{\beta'} + \frac{S}{\gamma'} + \frac{J}{\delta'} \qquad\qquad\qquad (6)$$

where *J*, the angular momentum, and *n*, the quantum number for radial excitation. Formula (6) include spin-spin, spin-orbit and tensor effects through terms involving ***J*** and ***S***. Using (6) we obtain

$$E = \frac{2}{\epsilon} \sinh^{-1} \left[ \left( \frac{\epsilon}{2} \right)^2 \left( \frac{L}{\alpha'} + \frac{n}{\beta'} + \frac{S}{\gamma'} + \frac{J}{\delta'} \right) + \sinh^2 \left( \frac{m\epsilon}{2} \right) \right]^{1/2} \tag{7}$$

such that (6) is reobtained in the undeformed *(ε→0)* limit of this equation. Authors compute angular and radial excitations of pion (Fig. 1, [12]), which show significant departure from linearity. In particular this implies that the density of states for large energy *E* is larger than the conventional linear models.

Detailed fit to charmed and beauty mesons show a *significant curvature* to the Regge type plot of *E²* versus the radial quantum number *n*.

All in all, authors [12] study *π-b, ρ-a, K, K*, φ, ω-f, Υ, Ψ* meson families and *N, Δ, Λ, Σ* baryon families, computing spectra up to *J=17/2*. They found a strong effect of *nonlinearity* for these RT, for rotational and radial excitations, including daughters.

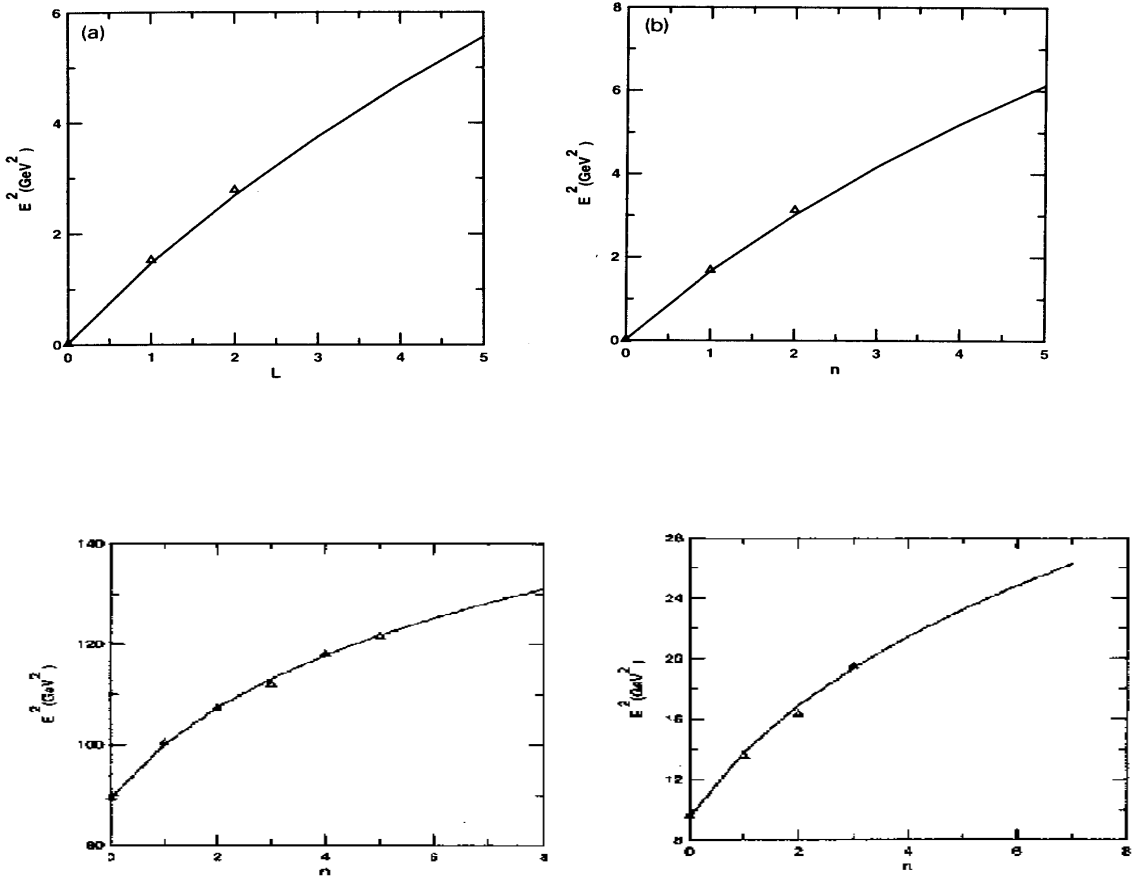

## II.3. STRING MODELS

String models in it's vast variety are one of the most popular models of hadron structure and RT. We will discuss here the variants of string model and results, pertinent to our focus – linear vs. nonlinear RT.

### a) **Olsson model**

This group is among the most fruitful in this field. In series of papers Olsson et al [18-20] used string approach to mesons, employing different equations: generalized spinless Bethe-Salpeter (BS), generalized Klein-Gordon (KG) equation. They considered the possibility of vector and scalar confinement. The exact and quasiclassical solutions has been found. Nonlinearity of the mesonic RT has been established at low $J$. At high $J$ authors deduced mostly linear and parallel RT.

In the paper [18] Olsson proved that variation of quark mass and account for the Coulomb interaction can violate the linearity of RT. It seems inescapable that massless quarks bound by a linear confinement potential generate a family of parallel linear RT.

A generalized spinless Salpeter (GSS) equation with Lorentz-scalar and – vector potentials is given by

$$H\Psi = M\Psi ,$$

$$H = 2\left\{\left[m + \frac{1}{2}S(r)\right]^2 + \vec{p}^2\right\}^{1/2} + V(r) \tag{1}$$

Where for simplicity we consider here equal-mass particles. This equation can also be generalized to the case of unequal masses.

A closely related second-order equation is the generalized Klein-Gordon (GKG) equation

$$\left\{\vec{P}^2 + \left[m + \frac{1}{2}S(r)\right]^2\right\}\Psi = \frac{1}{4}\left[M - V(r)\right]^2 \Psi \tag{2}$$

For a pure scalar potential *(V=0)* the GKG equation is exactly the square of the GSS Eq. (1). When *V≠0* the two wave equation differ but they have in common the same classical and semi classical solution.

From Eq. (2) with $\vec{P}^2 = P_r^2 + J^2 / r^2$ we have

$$(H - V)^2 = S^2 + 4J^2 / r^2 + 4P_r^2 \tag{3}$$

We with to obtain unquantized lowest-energy state for a given angular momentum. As we shall see this solution is a circular orbit and we will call this the classical solution. The condition for a stable circular orbit is that $H(r, J, P_r)$ be a minimum with respect to $r$ and $P_r$. This requires $P_r \equiv 0$ and $H_{1r}(r, J, 0) \equiv \partial H/\partial r = 0$, that is

$$\left[M - V(r)\right]^2 + S^2(r) + 4J^2 / r^2 , \tag{4a}$$

where $M = H(r, J, 0)$, and

$$[M - V(r)]V'(r) + S(r)S'(r) = 4J^2/r^3 \qquad (4b)$$

If the two conditions for circular orbit are imposed, only one of the three quantities $M, J, r$ is independent. This results in relations such as $J = J_C(r)$ [or inversely $r = r_C(J)$], the angular momentum of circular orbits as a function of their radius, and $M = M_C(J)$, the energy of circular orbits as a function of their angular momentum. This last $M_C(J)$ is the *classical yrast* (also the leading RT), that is, the least energy of the system for a given $J$; the *quantum-mechanical yrast* (discussed below) exceeds $M_C(J)$ by the zero-point energy of radial motion.

In WKB approximation for GKG with pure vector potential we have

$$\alpha' M_n^2 = J - J_0 + (2n+1)\left[\frac{J - J_0}{2J}\right]^{1/2} \qquad (5)$$

As we see at small $J$ *nonlinear* RT arise.

For thee pure vector potential *(S=0)* exact solution for GKG is

$$\alpha' M^2 = l + \frac{1}{2} - J_0 + \sqrt{2}\left(n + \frac{1}{2}\right)\left[\frac{l + 1/2 - J_0}{l + 1/2}\right]^{1/2} \qquad (6)$$

In this case the daughter RT are only parallel if $l >> (J_0)$ and we require $J_0 \le l/2$ for real trajectories. For the pure scalar potential exact solution for GKG is

$$\alpha' M_n^2 = l - J_0 + 2n + 3/2 \qquad (7)$$

By comparison (6) and (7) we observe that the daughter spacing is different.

When there is a Lorentz-vector potential the GSS and GKG equations are no longer equivalent wave equations. They have however the same classical and semiclassical solutions so that RT of Eq. (6) hold equally well for the second-order GKG equation.

For the mixed vector and scalar confinement we get

$$\alpha' M_n^2 = J + (2n+1)\left[\frac{3 - D}{2}\right]^{1/2}, \qquad (8)$$

where $D$ is a constant, $1 \le D \le 2$.

There is how increasing evidence that states of different $J$ form into degenerate towers at a given mass. Within the quadratic parametrization the Regge and daughter structure is given by (8). For *tower structure* the coefficient of $J$ and n must be commensurate in the right proportion. The two simple cases where tower structure is most evident are

$$D = 2, \qquad M_n^2 = J + 2n + 1 \qquad (9)$$

This is the scalar confinement case in which states differing by two units of angular momentum are degenerate:

$$D = 5/2, \qquad M_n^2 = J + n + 1/2 \tag{10}$$

For this choice states of angular momentum differences are degenerate. We can however rule out this case since it violates the limit *(D ≤2)* needed to ensure a real scalar confining potential.

We have tried to emphasize here the intimate connection between relativistic kinematics, linear confinement, and straight parallel families of RT. By explicit classical and semiclassical calculation and by exact numerical evaluation we have investigated the energy levels of the GSS equation.

For both scalar and vector confinement, with massless quarks, families of straight parallel RT result. In both the scalar and vector cases we found an enlarged class of potentials which deviate from linear confinement at small radii but nevertheless yield straight RT at large angular momenta.

In another paper [19] Olsson et al quantized flux tube and developed an interesting numerical algorithm for the solution of the equation of motion. Authors used the hydrogen wave functions (WF) basis with big Hilbert space *($N_{max}$≤50)*. They investigated the effect of quark mass on the character of RT and yrast RT in particular.

First they compare RT for *m=0* and *m=0.3 GeV*. There is a very small difference up to *l =20*:

a) Parent and daughter RT are parallel.

*b)* Physical states united into towers of mass degenerate states with odd or even *l*. The degeneracy break down at *l =15*.

c) The nonlinearity of RT is clearly seen.

At *m=1.5 GeV* (two heavy quarks) Olsson compute parent and three daughter RT. Since the solution is a continuous function of *l*, it is straightforward to calculate the dimensionless Regge slope

$$\alpha' = 2\pi a \frac{dl}{dM^2} \tag{11}$$

We observe that for large *l* the slopes seem to approach unity as expected. The daughter RT approach the Nambu slope more slowly as might be anticipated since more energy is contained in radial motion. This is exactly corresponds to our NRQM findings [ ].

Authors consider also heavy–light quark mesons *($m_2$>>$m_1$)*. Yrast solution to the heavy-light equation implies a Regge type slope double the normal Nambu slope:

$$\frac{L}{(M - m_2)^2} = \frac{1}{\pi a} = 2\alpha'_{Nambu} \tag{12}$$

Corresponding dimensionless slope $\alpha' = 2\pi a \dfrac{\partial l}{\partial (M - m_2)^2}$ was calculated for parent and three daughters RT and it's clearly has nonlinear nature. It's remained to be seen how this model would work for real meson and baryon spectra and RT.

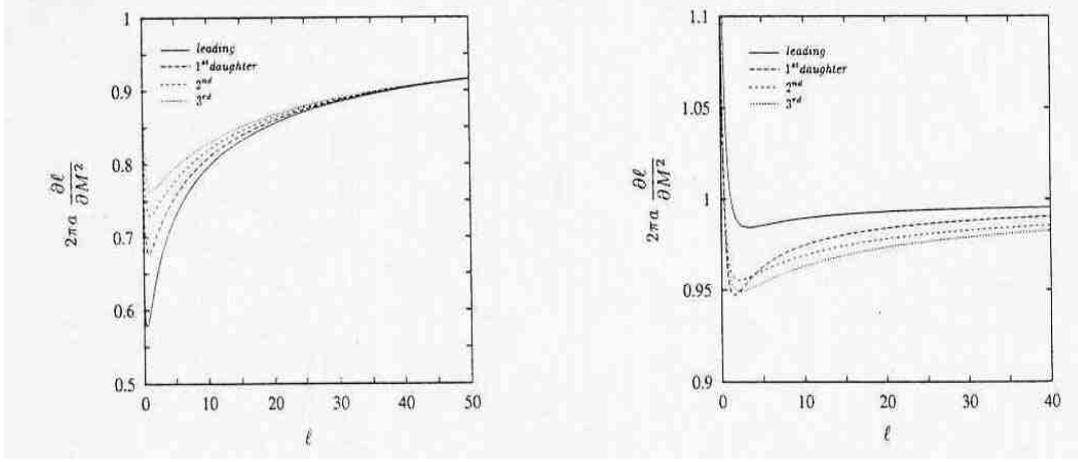

### b) <u>Soloviev's string model (SQM)</u>

During the last few years Soloview has been active on the Regge front [21]. He suggested a relativistic quantum model in which the RT can grow faster than linearly. The model is a rigid string with a Lagrangian given by an exponential function of the string world-sheet curvature. Exactly solvable generalizations of the model are considered. Relativistic quantum models admit any growth rate for the RT. In the string models considered here, this growth is limited by the exponent 3/2.

In the paper [22] Soloviev consider a relativistic quantum model of confined massive spinning quarks and antiquarks which describes the leading RT of mesons. The quarks are described by the Dirac equations and the gluon contribution is approximated by the Nambu-Goto straight-line string. The string tension and the current quark masses are the main parameters of the model. Additional parameters are phenomenological constants which approximate nonstring short-range contributions. A comparison of the measured meson masses with the model predictions allows one to determine the current quark masses (in MeV) to be $m_s=227\pm5$, $m_c=1440\pm10$, and $m_b=4715\pm20$. The chiral $SU_3$ model makes it possible to estimate from the $u$- and $d$-quark masses to be $m_u=6.2\pm0.2 MeV$ and $m_d=11.1\pm0.4 MeV$.

In the newest paper [23] author applied the above model to all mesons, from pion to $\Upsilon$, lying on the leading RT. The model describes the meson mass spectrum, and a comparison with measured meson masses allows one to determine the parameters of the model: current quark masses, universal string tension, and phenomenological constants describing the nonstring short-range interaction. The meson RT are in general nonlinear; only the trajectories for light-quark mesons with the nonzero lowest spins are practically linear. The model predicts masses of many new higher-spin mesons. A new $K^*(l')$ meson is predicted with a mass of *1910 MeV*. In some cases the masses of new low-spin mesons are predicted by extrapolation of the phenomenological short-range parameters in the quark masses. In this way the model predicts the mass of $\eta_b(1S)$ $(0^{-+})$ to be *9500±30 MeV*, and the mass of $B_C(0^-)$ to be *6400±30 MeV*.

The relativistic WF of the composite mesons allow one to calculate the energy and spin of mesons. The average quark-spin projections in polarized ρ meson are twice as small as the nonrelativistic quark model predictions. The spin structure of $K^*$ reveals an 80% violation of the flavor *SU(3)*. The $\Upsilon$ diameter in this string quark model (SQM) is *0.02 fm*, 1/5 that of the pion.

The *X(1920)* meson, found in GAMS and VES experiments at IHEP, Protvino, agrees quite well with SQM predictions and may be a $2^{++}$ trajectory partner of *$a_0(980)$* (Fig 6, [23]).

Soloviev also consider glueballs in the SQM [24]. It is shown that the eigenstates of the quantized simplest closed (elliptic) Nambu-Goto string, called glueballs, have quantum numbers $I^G j^{PC}=0^+ j^{++}$. Lightest glueballs have spins *j=0,1,* and *2* and the same mass *1500±20 MeV*. They correspond to
*$f_0$ (1500), $f_1$ (1510)* and *$f_2$ (1565)* − mesons. Next glueballs have *j=0, 1, 2, 3, 4* and the same mass *2610±20 MeV*. The slope of the glueball RT's is twice as small as for $q\bar{q}$ -mesons. The intercept of the leading glueball trajectory – the pomeron RT is *1.07±0.03*.

Let us consider the motion of the string in the frame where it is at rest as a whole, $\vec{P}=0$. The laboratory time is

$$t = X^0 - X_0^0 = dt \,, \qquad\qquad d = \frac{m}{2\pi a} \qquad\qquad (1)$$

The string behavior essentially depends on the pseudospins $\vec{L}_1$ and $\vec{L}_2$, $\left|\vec{L}_1\right|=\left|\vec{L}_2\right|=L$ . We have

$$m^2 = 8\pi a L \,, \qquad\qquad \vec{J} = \vec{L}_1 + \vec{L}_2 \qquad\qquad (2)$$

In case $\vec{L}_1 = -\vec{L}_2$, the string spin is zero, the string lies in the plane orthogonal to $\vec{L}_i$ and represents a circumference with an oscillating radius, from maximal value d to zero and back.

In the general case, when the angle between $\vec{L}_1$ and $\vec{L}_2$ is in the limits *0<2α<π,* the value of the string spin is

$$J = \frac{m^2}{4\pi d}\cos\alpha \qquad\qquad (3)$$

The string represents an ellips with half-axises

$$A = dN \,, \qquad B = d\sin\alpha\cos(\tau + \tau_0) \qquad\qquad (4)$$

(the large half-axis *A* is orthogonal to the spin and the small one is parallel to the spin) and rotates around the spin with the angular velocity

$$\left|d\vec{f} / dt\right| = d^{-1}\cos\alpha N^{-2} \qquad\qquad (5)$$

The half-axises are maximal: $A=d$, $B=d\sin\alpha$, and the instant angular velocity is minimal: $d^{-1}\cos\alpha$, when string is in the plane orthogonal to the plane of $\vec{L}_i$. Rotating, the ellips shrinks and accelerates. When it reaches the plane of $\vec{L}_i$, it shrinks into a straight-line with half-length $d\cos\alpha$. It's angular velocity at this moment is maximal and is equal to inverse of its half-length. Then the ellips expands and slows down and so on.

In the other extreme case, $\vec{L}_1=\vec{L}_2$, the string spin is maximal

$$J=\frac{m^2}{4\pi a} \tag{6}$$

(and twice as small as for an open straight-line string of the same mass), the string is compressed into a straight line with half-length d and rotates with a constant angular velocity $d^{-1}$.

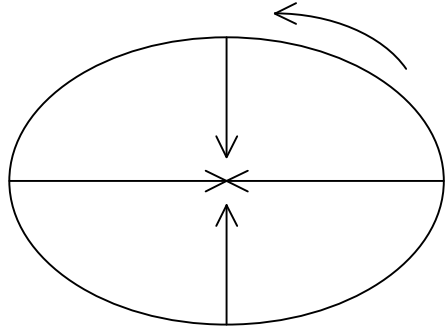

To quantize the classical solution, one has to know the domain of its stability. It is not difficult to show that the above solution is stable for all values of the initial conditions except for $L=0$. For $L=0$ the string reduces to a pointline system with zero mass, however for arbitrary small $L\neq0$ we have system with small mass but with all degrees of freedom of the elliptic string. Therefore quantization of our system has no meaning for values of pseudospins close to zero.

After quantization we see that the glueballs are space-parity even for all $j$ and $l$. Since they are electrically neutral and charge-parity even, their quantum numbers are

$$I^G j^{PC}=0^+ j^{++} \tag{7}$$

The glueball mass depends on $l$ only. Introducing $K=0$ for the leading RT, $K=1$ for the first daughter RT, $K=2$ for the second one and so on, we can write down for the spin

$$j=2l-k \tag{8}$$

Then the glueball RT'S are equal to

$$\sqrt{(j+k)(j+k+2)}=d_0+\frac{1}{4\pi a}m^2 \tag{9}$$

Glueballs with even spins lie on the leading RT, on the second daughter RT and so on. Glueballs with odd spins lie on the first, third daughter RT and so on. All these RT $j(m^2, K)$ as functions of $m^2$ have at large j the slope

$$j^{'}(\infty, K) = \frac{1}{4\pi a} = 0.452 GeV^{-2} \qquad (10)$$

which is twice as small as for the $q\bar{q}$ states and are nonlinear at small $j$.

The glueball quantum numbers (7) and mass degeneracy of the states with different spins $j$ at the same pseudospin $l$ are remarkable properties of the glueballs in the model. The lightest states with $l = 1$ and quantum numbers $0^+0^{++}$, $0^+1^{++}$ and $0^+2^{++}$ can be identified with the mesons

$f_0 (1500)$,  $0^+0^{++}$,  $m=1500\pm10$

$f_1 (1510)$,  $0^+1^{++}$,  $m=1518\pm5$ $\qquad (11)$

$f_2 (1565)$,  $0^+2^{++}$,  $m=1542\pm22$

These mesons are not the quark-antiquark ones. Let us conservatively estimate the lightest glueball mass with $l = 1$ to be

$$m_1 = 1500 \pm 20 \qquad (12)$$

This allows one to obtain the constant $a_0$ from (9)

$$a_0 = 1.81 \pm 0.04 \qquad (13)$$

Now, when we know the model parameters we can fix the glueball RT (9) and to predict masses and quantum numbers of heavier glueballs. For $l = 2$ we have spin-parities and mass

$0^{++}$, $1^{++}$, $2^{++}$, $3^{++}$, $4^{++}$; $m_2=2610\pm20$ $\qquad (14)$

For the next glueballs the model predicts

$0^{++}$, $1^{++}$, ..., $5^{++}$, $6^{++}$; $m_3=3360\pm20$ $\qquad (15)$

From Eq. (9) we get the glueball leading RT – the Pomeron trajectory

$$j(m^2,0) \equiv j(m^2) = \sqrt{\left(d_0 + \frac{1}{4\pi a}m^2\right)^2 + 1} \; -1 \qquad (16)$$

with the intercept

$j(0)=1.07\pm0.03,$ $\qquad (17)$

which corresponds to the high-energy data on the hadron scattering.

Let us remark in conclusion that glueball decays can be considered within this approach if interacting meson fields introduced instead of meson wave functions (second quantization).

Thus, the SQM provides a single relativistic approach for description of quark-antiquark and glueball meson states including prediction of the Pomeron RT, with wide area of critical comparison with experiment.

## c) **Burakovsky - Goldman string model**

Burakovsky present a new generalized string model for RT'S $J=J(E^2)$. Author demonstrate that this model is not to produce linear RT, in contrast to the standard Nambu-Goto string, but generally nonlinear trajectories, which in many cases be given in analytic form. As an example, we show how the model generates square-root, logarithmic and hyperbolic trajectories that have been discussed in the literature.

The string model with constant tension is known to predict linearly rising RT: $J=E^2/(2\pi\gamma)$. The string trajectories are exactly linear in the case of the mass less ends, and asymptotically linear in the case of the massive ands having some curvature in the region $E \gtrsim m_1 + m_2$. However, the realistic RT extracted from data are nonlinear. Indeed the straight line which crosses the $\rho$ and $\rho_3$ squared corresponds to an intercept $\alpha_\rho(0)=0.48$, whereas the physical intercept is located at $0.55$, as extracted by Donnachie from the analysis of $pp$ and $\bar{p}p$ scattering data in a simple pole exchange model [26]. The nucleon RT extracted from the $\pi^+p$ backward scattering data is [27]

$$\alpha_N(t) = -0.4 + 0.9t + \frac{1}{2}0.25t^2 ,$$

and contains positive curvature.

Recent UA8 analysis of the inclusive differential cross sections for the single-diffractive reactions $p\bar{p} \to px, \ p\bar{p} \to x\bar{p}$ at $\sqrt{s}=630GeV$ reveals a similar curvature of the Pomeron trajectory [28]:

$$\alpha_p(t) = 1.10 + 0.25t + \frac{1}{2}\left(0.16 \pm 0.02\right)t^2$$

An essentially nonlinear $\alpha_2$ trajectory was extracted in [29] for the process $\pi^-p \to \eta n$.

This model is the generalization of the standard string model. Such generalization is done by the modification of the standard string tension into the effective string tension which is a function of $|x|$, as follows:

$$S_{gen} = -\int_{t_1}^{t_2} dt \int_0^\pi d\sigma \gamma\left(|x|\right)\sqrt{X'^2\left(1-\dot{X}^2\right)+\left(\dot{X}X'\right)^2} - \sum_{i=1,2} m_i \int_{t_1}^{t_2} dt\sqrt{1-\dot{X}_i^2} \tag{1}$$

In the nonrelativistic limit the effective string tension is the derivative of the interaction potential between the massive ends of the string. Therefore, different choices of the effective string tension would be related to different potentials, which makes it possible to deal, among the others, with color-screened potentials, i.e., potentials that approach constant values at large separations; e.g., $V(\rho) \equiv \gamma / \mu\left(1-\exp\left(-\mu\rho\right)\right)$ which is used to fit the lattice QCD data.

The energy and orbital momentum of the generalized massless string, $m_1=m_2=0,$ are given by

$$E = 2\int_0^R \frac{d\rho\gamma(\rho)}{\sqrt{1-\omega^2\rho^2}}, \quad J = 2\int_0^R \frac{d\rho\gamma(\rho)\omega\rho^2}{\sqrt{1-\omega^2\rho^2}} \tag{2}$$

where $R=1/\omega$ is half of the string length for a given $\omega$. By eliminating $\omega$ from (2) one can obtain $J$ as function of $E^2$, the Regge trajectory. In the most important cases of analytic nonlinear RT the corresponding potentials happen to be recovered analytically. Bellow we present such potentials for three examples of analytic nonlinear RT that have been discussed in the literature.

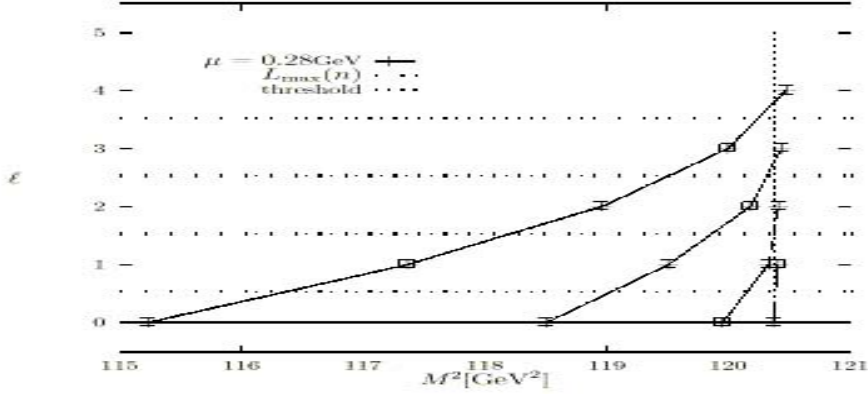

The square-root RT, $J \sim E_{th} - \sqrt{E_{th}^2 - E^2}$, where $E_{th}$ is the trajectory energy threshold is the simplest choice of a trajectory for dual amplitude with Mandelstam analyticity. The corresponding potential is ($\gamma, \mu = const, V(\rho) \to \gamma/2\mu$ as $\rho \to \infty$, and hence $E \to E_{th} = \gamma(\mu)$)

$$V(\rho) = \frac{\gamma}{\pi\mu} \arctan(\pi\mu\rho), \tag{3}$$

for which

$$\gamma(\rho) = \frac{dV(\rho)}{d\rho} = \frac{\gamma}{1+(\pi\mu\rho)^2} \tag{4}$$

and leads, via (2), to

$$E = \frac{\pi\gamma}{\sqrt{\omega^2+\pi^2\mu^2}}, \qquad J = \frac{\gamma}{\pi\mu^2}\left(1 - \frac{\omega}{\sqrt{\omega^2+\pi^2\mu^2}}\right) \tag{5}$$

Eliminating $\omega$ from the relations gives

$$J = \frac{1}{\pi\mu}\left(\gamma/\mu - \sqrt{(\gamma/\mu)^2 - E^2}\right), \tag{6}$$

i.e., the square-root RT. For $E << \gamma/\mu$, it reduces to an approximate linear RT, $J \simeq E^2/(2\pi\gamma)$.

The logarithmic RT, $J \propto - log(1-E^2/E_{th}^2)$, is the ingredient of dual amplitude with logarithmic trajectories. The corresponding potential is (again $\gamma, \mu = const, V(\rho) \to \gamma/2\mu$ as $\rho \to \infty$, and hence $E \to E_{th} = \gamma/\mu)$).

$$V(\rho) = \frac{\gamma}{2\pi\mu} \left( 2\arctan(2\pi\mu\rho) - \frac{\log[1 + (2\pi\mu\rho)^2]}{2\pi\mu\rho} \right), \tag{7}$$

for which $\gamma(\rho) = \gamma \dfrac{\log[1 + (2\pi\mu\rho)^2]}{(2\pi\mu\rho)^2}$ , $\tag{8}$

and leads to

$$E = \frac{\gamma}{2\pi\mu^2} \left( \sqrt{\omega^2 + 4\pi^2\mu^2} - \omega \right), \qquad J = \frac{\gamma}{2\pi\mu^2} \log \frac{\omega + \sqrt{\omega^2 + 4\pi^2\mu^2}}{2\omega} \tag{9}$$

from which eliminating $\omega$ gives

$$J = -\frac{\gamma}{2\pi\mu^2} \log \left( 1 - \frac{E^2}{(\gamma/\mu)^2} \right) \tag{10}$$

For $E << \gamma/\mu$, it again reduces to an approximate linear form, $J \simeq E^2/(2\pi\gamma)$. Note that in the corresponding quantum case the number of states on the trajectory would be infinite. In each of the two examples considered above the potential belongs to the family of the color-screened potentials.

The hyperbolic trajectory $J \propto \cosh(E)-1$ results in the K-deformed Poincare phenomenology [12]. Here the corresponding potential is

$$V(\rho) = \frac{\gamma\rho\Phi\left(-(\pi\mu\rho)^2, 2, 1/2\right)}{4} = \gamma\rho \sum_{k=0}^{\infty} \frac{\left[-(\pi\mu\rho)^2\right]^k}{(2k+1)^2} , \tag{11}$$

where $\Phi(z, s, a) = \sum_{k=0}^{\infty} \dfrac{z^k}{(a+k)^s}$ $\tag{12}$

is the so-called Lerch's transcendent. The corresponding $\gamma(\rho)$,

$$\gamma(\rho) = \gamma \frac{\arctan(\pi\mu\rho)}{\pi\mu\rho}, \tag{13}$$

leads to

$$E = \frac{\gamma}{\mu} \log\left(\pi\mu/\omega + 1 + (\pi\mu/\omega)^2\right), \tag{14}$$

$$J = \frac{\gamma}{\pi\mu^2} \left( \sqrt{1 + (\pi\mu/\omega)^2} - 1 \right), \tag{}$$

from which eliminating $\omega$ gives

$$J = \frac{\gamma}{\pi\mu^2} \left[ \cosh\left(\frac{E}{\gamma/\mu} - 1\right) \right], \tag{15}$$

i.e., the hyperbolic trajectory. For $E << \gamma/\mu$, it reduces to the linear form, as well as in the above two cases: $J \simeq E^2/(2\pi\gamma)$.

The main phenomenological implication of this model is the possibility to obtain RT for an arbitrary (nonrelativistic) potential in general and for a color-screened potential in particular. Since the trajectory for the latter is characterized by an energy threshold, and in some cases by a finite number of states, it may be of relevance to QCD to predict the numbers of states lying on different RT, and the corresponding energy threshold, if a color-screened potential is indeed realized in QCD.

### d) __Sharov string model__

The considered model of baryon consists of three pointlike masses (quarks) bounded pairwise by relativistic strings forming a curvilinear triangle [30]. Classic analytic solutions for this model corresponding to a planar uniform rotation about the system center of mass are found and investigated. These solutions describe a rotating curve composed of segments of a hypocycloid. The curve is a curvilinear triangle or a more complicated configuration with a set of internal massless points moving at the speed of light. Different topological types of these motions are classified in connection with different forms of hypocycloids in zero quark mass limit. An application of these solutions to the description of baryon states on RT is considered.

A free choice of the integer parameters $j_1$, $j_2$, $j_3$, $l$ result in a very large number of different motions of the system distinguishing from each other by their topological structure. A motion or state of the system we will denominate "simple" if the position of the string (section $t=const$) is a curvilinear triangle with smooth sides (Figs. 2-4, [30]). In the opposite case if there are some singular massless points on the sides of the "triangle" we will denominate the state "exotic" (Fig. 5, [30]). These singular points move at the speed of the light.

World surface for the exotic motion has peculiarities $\dot{x}^2 = x'^2 = 0$ on the world lines of singular points (cusps) of the hypocycloid. There are many types of exotic motion differing from each other by the number and positions of these peculiarities. One must differ the peculiar points (cusps) on these curves from the quark positions (Fig. 5, [30]). At the point of quark position two segment of the string are joined at a nonzero angle. For each curve 1-5 in Fig. 5 the first quark is situated at the lowest point, two others are along the string in the counter clockwise direction.

The number of the curve is in the center of rotation. Curve 1 in Fig. 5 represents the simplest exotic state with one peculiar point. The pentagonal line 2 and the starlike curve 3 both contain two singularities and represent two different topological configurations of the string.

In points of self-intersection different parts of the string do not interact. Curves 4-6 in Fig. 5 contain three singular points with various arrangements – in symmetric lines 4 and 6 these cusps alternate with quark positions. The curve 4 is a curvilinear hexagon. Curve 5 has cusp between the second and the third, and two cusps between the third and the first quark.

When considering RT, the following formulas for E and J could be obtained

$$\begin{cases} E = \gamma D \dfrac{a^2 - b^2}{a} + \sum_{i=1}^{3} \dfrac{m_i}{\sqrt{1 - v_i^2}} & \qquad (1) \\[4mm] J = \dfrac{a}{2\omega}\left( E - \sum_{i=1}^{3} m_i \sqrt{1 - v_i^2} \right) & \qquad (2) \end{cases}$$

Expressions (1) and (2) set an implicit nonlinear connection between $E$ and $J$ of the considered system. In the nonrelativistic limit from (1), (2) we get

$$J \simeq \left(\frac{2}{3}\right)^{3/2} \frac{\sqrt{m_1 m_2 m_3}}{\gamma(m_1 + m_2 + m_3)}\left(E - \sum_{i=1}^{3} m_i\right)^{3/2}, \qquad v_i << 1 \qquad (3)$$

if each mass is less that the sum of the others. If one of the masses is larger then the sum of two others, then the nonrelativistic asymptotic case describes a rotation of a rectilinear double string with two masses $m_2$ and $m_3$ at the ends and a mass $m_1$ at the rotational center. In this case we get

$$J \simeq \left(\frac{2}{3}\right)^{3/2} \frac{1}{2\gamma}\sqrt{\frac{m_2 m_3}{m_2 + m_3}}\left(E - \sum_{i=1}^{3} m_i\right)^{3/2}, \ v_i << 1 \qquad (4)$$

In the opposite ultrarelativistic limit $v_i \to 1,\ E \to \infty,\ J \to \infty$ we obtain the RT for a state with an arbitrary type

$$J \approx \alpha' E^2 + \alpha_1 E^{1/2}, \ v_i \to 1, \qquad (5)$$

This is close to the standard linear form $J = \alpha' E^2 + \alpha_0$.

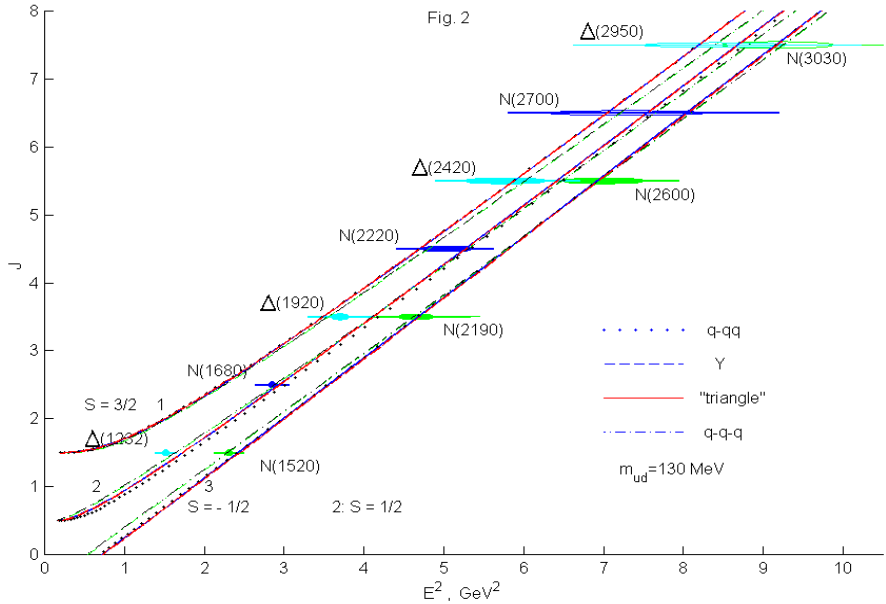

In the paper [31] various string models of baryons treated in terms of the quark-diquark configuration, the linear configuration of quarks, the star *(Y)* configuration, and triangle *(Δ)* configuration have been considered. Classical rotational motions (plane rotations of the system at a constant speed) have been used to describe orbital baryonic excitations lying on RT. The spin-orbit interaction of the quarks has been taken into account as corrections in various versions of the theory. The model-dependent quark masses and other parameters of the model have been estimated on the basis of attempts at describing, within the aforementioned models, the entire body of experimental data on baryon RT. In the present paper we allow the mixing between different string configurations.

The spin-orbit interaction of quarks is taken here into account as a correction to energy. As a result, we arrive at the correction

$$\Delta E_{SL} = \sum_i \left[ 1 - \left( 1 - v_i^2 \right)^{1/2} \right] \left( \Omega S_i \right) \tag{6}$$

For particles considered below, the corrections in (6) are about *100-200 MeV*. Their effect on the dependence *J(E²)* is illustrated in Fig. 2, [31] for various string configurations of baryons.

It is assumed that the diquark mass is equal to the doubled quark mass and that the effective tensions for the various string configurations are given by

$$\gamma_Y = \frac{2}{3}\gamma, \qquad \gamma_\Delta = \frac{3}{8}\gamma, \qquad \gamma_\Delta = \gamma_{q-qq} = \gamma_{q-q-q} = \frac{1.1}{2\pi}\left( GeV^2 \right) \tag{7}$$

We can see that spin corrections to the angular momentum and energy shift the curve *J=J(E²)* to nearly the same extent for the different string configurations.

The model describes reasonably well *N, Δ, Λ, Σ* spectra as well as part of meson spectrum (Fig. 5 [31]). The drawback is that string can't account properly for different parity of resonances.

In the paper [32] for the relativistic string with massive ends (the meson model) and four various string baryon models (*q-qq, q-q-q, Y* and *Δ)* Sharov consider the classical quasirotational motions, which are small disturbances of the planar uniform rotations of these systems. For the string meson model two types of these solutions are obtained. They describe oscillatory motions in the form of stationary waves in the rotational plane and in the orthogonal direction. This approach and the suggested method of determining an arbitrary motion of the system on the basis of initial data let us solve the stability problem for the rotational motions for all mentioned string configurations. It is shown that the classic rotational motions are stable for the string meson model (or its analog *q-qq*) and for the *Δ* baryon configuration, but they are unstable for the string baryon models *Y* and *q-q-q*. For the latter two systems any small asymmetric disturbances grow with increasing time.

The motion of the *q-q-q* configuration becomes more complicated and quasiperiodic but the quarks do not merge. In the case of the *Y* model the evolution of disturbances results in a quark falling into the junction. The exotic states were not considered here. These features of the classic behavior are important for describing hadron states on the RT's and for choosing and developing the most adequate QCD-based string hadron models.

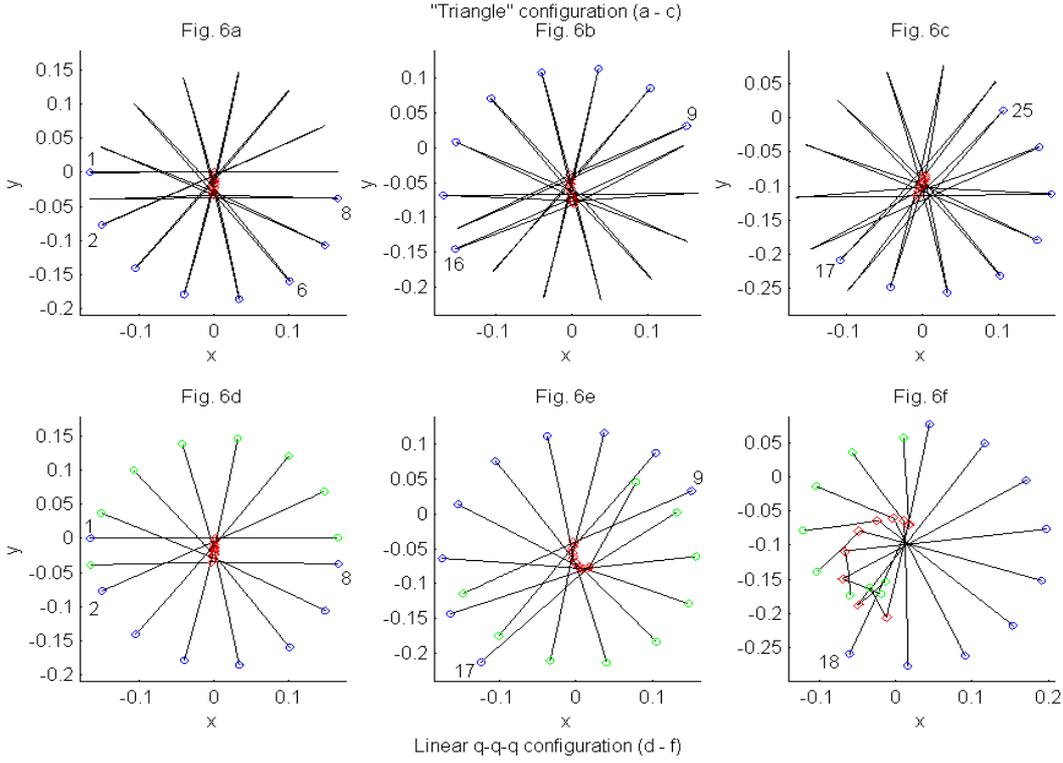

Suggested string model was also used to scrutinize RT's for mesons and baryons [33]. The standard string models predict linear trajectories at high angular momenta J with some form of nonlinearity at low *J*. We have to admit that all string models have problems with quantization, consistent inclusion of spin and parity. Only after this the predictions of quantum string models has to be confronted with data.

## II.4. Nonrelativistic Quark Models

### a) Inopin model

In our recent series of papers [34-39], based on Hamiltonian (1) and the method of hyperspherical functions [40], a description of *N, Δ, Ω* resonance spectra and partial widths was given

$$H = H_0 + H_{hyp},$$

$$H_0 = \sum_{i=1}^{3} m_i + \sum_{i=1}^{3} \frac{\overrightarrow{P_i^2}}{2m_i} - \frac{2}{3} \sum_{i<j} \left( \frac{\alpha_s}{r_{ij}} - b r_{ij} \right) + V_0$$

$$H_{hyp} = \frac{2}{3} \sum_{i<j} \frac{\pi C_\alpha}{m_i^2} \left( 1 + \frac{8}{3} \left( \vec{s}_i \vec{s}_j \right) \right) \delta \left( \overrightarrow{r_{ij}} \right) + \sum_{i<j} \frac{2C_t}{3m_i m_j r_{ij}^3} \left[ 3 \frac{\left( \vec{s}_i \vec{r}_{ij} \right) \left( \vec{s}_j \vec{r}_{ij} \right)}{r_{ij}^2} - \vec{s}_i \vec{s}_j \right] \qquad (1)$$

Following [41], we introduce the constants *αₛ, Cₐ* and *Cₜ*, which determine the strength of the Coulomb, contact and tensor potentials, respectively. The use of the Hamiltonian (1) allows us to obtain better agreement with experiment for resonances of positive and negative parity, and also to describe resonances with both large *J* and *M*, and with small *J* and *M*.

We showed [34-39] that it was appropriate for such a description to take advantage of the concept of yrast states and yrast lines from the theory of atomic nuclei rotational spectra, as well as to make use of the concept of RT. One of the main result was that both theoretical and experimental spectra are nonlinear trajectories in Chew-Frautschi plots throughout the experimental region. Our model predicts a whole series of high-lying $N, \Delta$ Resonances, represented in Table I.

**Table I. Masses of the predicted N, Δ resonances, in GeV**

| N 11/2⁺(2.39) | N 13/2⁻(2.90) | N 15/2⁺(3.00) |
|---|---|---|
| N 17/2⁻(3.45) | N 19/2⁺(3.45) | N 21/2⁻(3.95) |
| Δ 15/2⁻(3.48) | Δ 17/2⁺(3.49) | Δ 19/2⁻(4.00) |

The present paper generalized the model [34-39] to $u, d, s, c, b$ flavours (it is impossible to create top baryons and mesons) and a wider range of angular momenta L=0-20. We obtain the spectra of baryon resonances (BR) in the same way as in [34-39], by solving Schrödinger equation (SE) with the HF method.

When the hadron wave function (WF) is expanded in the HF basis and substituted in the SE, one generally finds an infinite system of the differential equations for the radial WF (RWF). However, as was shown in [40] for a system of identical u, d quarks, the coupling of channels is weak, therefore it is sufficient to include only several terms in K, grand orbital momentum (GOM), in the WF expansion. With increasing excitation energy, the contribution of $H_{hyp}$ vanishes, hence the coupling of channels in general can be neglected. On the other hand, as we will consider only symmetric baryons with increasing quark mass (u→s→c→b), the abovementioned arguments will apply even more strongly.

Now, let us introduce Jacobi and hyperspherical coordinates for a three-body problem. The Jacobi coordinates are defined as usual (please see [34-39] for all the details).

$$\vec{R} = \left(\vec{r_1} + \vec{r_2} + \vec{r_3}\right)/3 , \quad \vec{\eta} = \left(\vec{r_1} - \vec{r_2}\right)/\sqrt{2}$$
$$\vec{\xi} = (\sqrt{2/3})\left[\left(\vec{r_1} + \vec{r_2}\right)/2 - \vec{r_3}\right] \tag{2}$$

The hyperspherical angle θ and ρ are defined as follows:

$$\eta = \rho \cos\theta, \quad \xi = \rho \sin\theta , \quad 0<\theta<\pi/2$$
$$\rho = \left(\vec{\eta}^2 + \vec{\xi}^2\right)^{1/2} , \quad 0<\rho<\infty \tag{3}$$

So, our WF and q-q potentials will depend on new variables $\vec{\eta}$ and $\vec{\xi}$.

We will work in the hypercentral approximation (HCA), where only that part of the interaction which is invariant under rotation in six-dimensional space ($\rho, \Omega_5$) is taken into account. Because in HCA baryons are pure rotational states in six-dimensional space, and because a very soft potential is used in our model, no strong correlations generating a so – called quark-diquark state can occur. So, we will neglect the θ dependence of our q-q potential. The introduction of θ dependence in the potential mixed the states belonging to one value of L, and slightly shifts their positions in the baryonic mass spectrum. As was proved by Richard [40], the most commonly used potentials in hadron spectroscopy (including ours) are very close to hypercentral. Now let us discuss our results for mass spectra and RT of baryons.

The method of numerical solution of the SE was described in detail in [34-39], but we will briefly recapitulate it here. The system of differential equations (SDE) was reduced to a system of first-order differential equations, which we integrated by determining a full set of Cauchy solutions. We then constructed the required solution by imposing the following boundary conditions on the RWF: $F_i(\rho=0)=0$ and $F_i(\rho=R_\infty)=0$. The eigenvalues (EV) were determined by zeroing the determinant constructed from the fundamental system of solutions for SDE. For the parameter $R_\infty$ we always choose a much larger value than the radius of the corresponding excited state $<\rho>_i$, which grows with L and $N_R$ (the radial quantum number). We always choose $R_\infty$ such that if we take $R_\infty=2\,R_\infty$, then the corresponding EV of any excited state will not change more than 1%. We use in our computation the following set of parameters (Table II). With this input we have calculated the mass spectra, RT, mean radii $<\rho>_i$ and slopes for u, d, s, c, b flavours, $N_R$=0, 1, 2, and angular moment range L=0-20.

**Table II. Input parameters in the model [34-39]**

| $\alpha_s=C_\alpha=1$ | $C_t=0.2$ | b=350MeV/fm |
|---|---|---|
| $V_0$=-513MeV | $m_u$=330MeV | $m_d$=330MeV |
| $m_s$=607MeV | $m_c$=1500MeV | $m_b$=5170MeV |

We note that the solution of the SE in the single channel approximation with the centrifugal potential $V_C=(K+3/2)(K+5/2)/(2m\rho^2)$ is equivalent from the mathematical point of view to the Davydov-Chaban model [42], in which the rotational spectrum of a nonspherical nucleus with variable moment of inertia is calculated. Because the moment of inertia is not constant but varies linearly with the angular momentum J of the nucleus, the rotational spectrum, instead of the quadratic behavior $J(J+1)$, it would have for a constant moment of inertia, approaches one linear in $J$. Approximately the same thing happens with the hadron: for large excitation energies the spectrum becomes linear in $J$, and the hadron becomes a strongly extended system along its symmetry axis. This should lead to a linear nature of the baryonic mass spectra along yrast lines. Our predictions are in accord with findings by Hey [43] and Hendry [44].

Let us start a detailed, sector by sector comparison of our results for different flavours. Note that RT's $[M^2=M^2(J)]$ for baryons have different types of curvatures: some are convex; others are concave functions of $J$ (see Fig. 2). The u-d (N, Δ) trajectories are concave for $N_R$=0, 1, 2, whereas strange (sss) RT for $N_R$=0, 1, 2 are "stereotypical" – they are nearly straight lines with slowly varying slopes $\alpha'$ (Fig. 2). Charmed (ccc) and bottom (bbb) trajectories are convex for $N_r$=0, 1, 2.

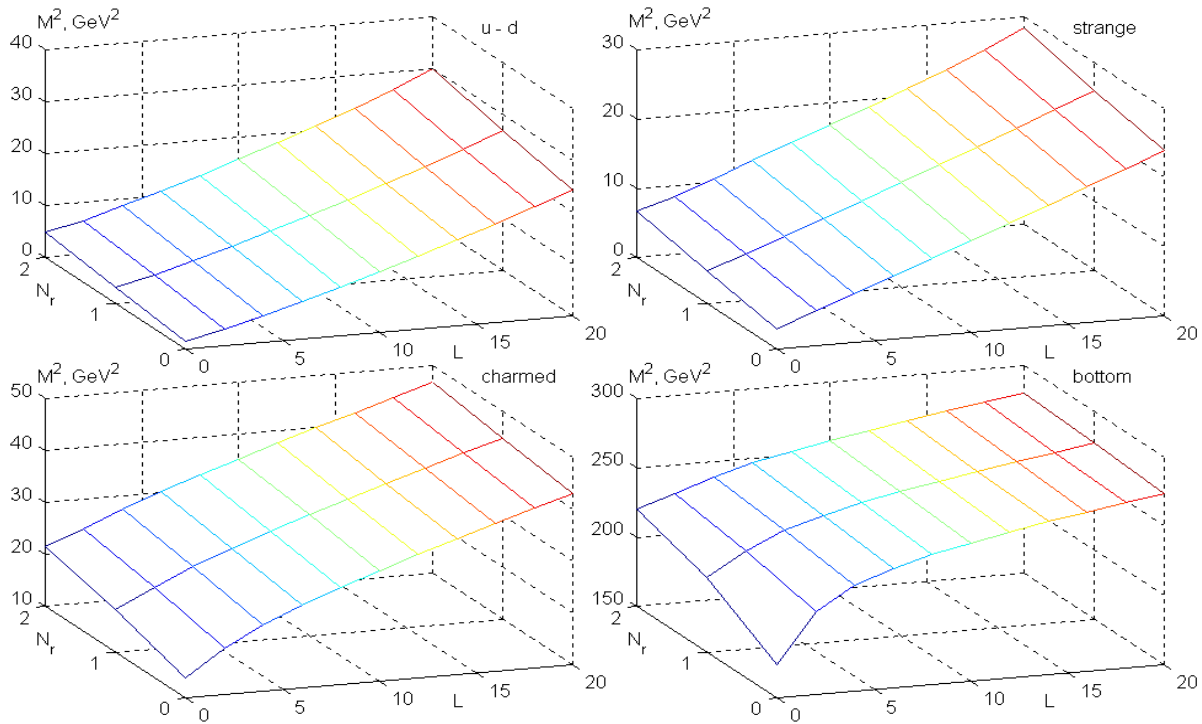

Fig.2 : Parent and two daughters RT in the potential quark model

When we analyze the mass spectra $M=M(J)$ for *u-d, s, c, b* baryons, they are all convex functions of $J$, with a rather complex dependence on $m_q$ (flavor) and $N_r$. If we fix $m_q$ and look at the Chew-Frautschi plot for $N_r=0,\ 1,\ 2$, we can see that these three curves generally are nonparallel (Fig.2), that strongly contradicts the conventional picture of the parallel RT. It is noteworthy that Olsson et al [19] recently examined the meson sector, using a relativistic flux-tube model, and noticed nonlinearity of RT at low angular momenta $J$. But it remains unclear whether model accounts adequately for physical observables, because it describes a very limited number of states in the mesonic sector.

Slopes for the *u-d* family start above 1 GeV$^{-2}$, with the largest value for $N_R=2$; the curves decrease almost monotonically with changing the rate of decreasing near $L=4$, for $N_r=0,\ 1,\ 2$ (see Fig. 3).

Slopes for the s family differ from all the other cases, because they are rather wear functions of $L$. Slopes for daughters $N_R=1,\ 2$ start just above unity, then slowly decrease to 0.87 and 0.78 GeV$^{-2}$, respectively, whereas the parent slope fluctuates slightly over the interval, spanning the range 0.97-0.91 GeV$^{-2}$ (see Fig. 3).

Slopes for the charm family are *highly nonlinear* functions of $L$. The $N_R=0,\ 1$, slopes increase monotonicaly, starting from 0.39 and 0.55 GeV$^{-2}$ and approaching 0.889 and 0.893 GeV$^{-2}$, respectively. The $N_R=2$ slope has a dip at $L=4$, and grow continuously, reaching 0.897 GeV$^{-2}$. These values are almost identical to Olsson's result for mesons for asymptotic $L$ [19]. Bottom baryon slopes differ sharply from

Fig.3 : Slopes for parent and two daughters RT in the potential quark model

Solid lines -- parent RT; dashed -- $N_r$=1 RT; dashed-dotted -- $N_r$=2 RT

the $u$-$d$-$s$-$c$ sector, first, by their small magnitude, and second, by the significant increase of the slopes along the trajectory (the $b$ slopes increase by about an order of magnitude, from 0.055 to 0.504 GeV$^{-2}$). Daughter slopes for $N_R$=2 fluctuate, but still increase with $L$. For convenience, we present the set of median values $<\alpha'>_i$ for the whole flavor multiplet (see Table III). It is interesting to note that the median values of $\alpha'_i$ are almost independent of $N_R$ for the $u$-$d$ and $s$ families.

**Table III. Median values $<\alpha'>_i$ for trajectories with $N_R$=0,1,2**

| $N_R$ | Up-down | Strange | Charm | Bottom |
|-------|---------|---------|-------|--------|
| 0 | 0.90 | 0.95 | 0.73 | 0.31 |
| 1 | 0.87 | 0.95 | 0.77 | 0.34 |
| 2 | 0.84 | 0.94 | 0.81 | 0.37 |

The expectation values of the hyperradius $<\rho>_i$ are basically smooth increasing functions of L and $N_R$, and decreasing functions of $m_q$.

We proved that the slope of the trajectories decreases with increasing quark mass in the mass region of the lowest excitations. This is due to the contribution of the color Coulomb interaction that increases with mass and results in a *curvature* of RT near the ground state. In the asymptotic regime the trajectories for all flavours are linear and have the same slope $\alpha'\approx0.9$ GeV$^{-2}$.

After careful numerical evaluation of baryon and meson spectra for all flavours we have shown that the point of establishment of linear RT depends on the exponent $\nu$ in the power law potential in Hamiltonian (1), and occurs at larger $L$ with larger $\nu$. The linear regime started only from $L>20$ for the oscillator confinement $(\nu=2)$ and it started from $L>18$ for the linear confinement $(\nu=1)$.

We proved that for mesons linear RT is established earlier than for baryons as a function of $L$. The reason lies in centrifugal energy term, which has $L(L+1)$ dependence for mesons and $(L+3/2)(L+5/2)$ dependence for baryons. As we see, three-dimensional corrections to $L(L+1)$-law are important for baryons.

One can see our predictions for RT's and comparison with data in Fig. 3-6, [39]. The particle data are taken from the PDG 2000 issue [45]. We see that our NRQM describe the data rather well, even the states with M>3000 MeV which need confirmation [44]. It describes *essentially nonlinear* RT's generated by $N(1535)1/2^-$ (Fig. 4, [39]) and $\Delta(1910)1/2^+$ (Fig. 5(b), [39]). The string model do not predict such nonlinear behavior (convexity) at small $J$.

The results of our potential model fits and predictions for baryon and meson spectra and RT reveals distinctive feature – RT in many cases are nonlinear functions of $J$. This fundamental feature is in accord with analysis of pure experimental RT from PDG 2000. Regge trajectories for mesons and baryons are not straight and parallel lines in general in the current resonance region both experimentally and theoretically, but very often have appreciable curvature, which is flavor-dependent. Our model predicts various forms of nonlinear behavior for thr RT with various flavours (Fig. 1, 2, [39]).

### b)Fabre de La Ripelle model.

Another type of potential model was designed for spectra and RT by Fabre de la Ripelle [46]. Author suggested to solve a Schrödinger equation (SE) with a power law potential with arbitrary exponent n.

The main purpose is to find which exponent n will produce hadron masses lying along straight RT. As a basic hypothesis author assume such confining potential:

$$V_{q\bar{q}}(r_{ij}) = V_0 + \alpha_{q\bar{q}} r_{ij}^n \qquad (1)$$

where $V_0$ and $\alpha_{q\bar{q}}$ are constants and where $r_{ij}$ is the distance between a $q\bar{q}$ pair. We search for the n leading to (hadron masses)$^2 \propto J$. In the nonrelativistic scheme, the spectrum of the $q\bar{q}$ meson masses,

$$M_{q\bar{q}} = V_0 + m_q + m_{\bar{q}} + E_{N,l} \ , \qquad (2)$$

is generated by the eigenenergies $E_{N,l}$ of the radial equation

$$\left\{ (\hbar^2/m)\left[ -(d^2/dr^2) + l(l+1)/r^2 \right] + \alpha_{q\bar{q}} r^n - E_{N,l} \right\} U_{N,l}(r) = 0 \ , \qquad (3)$$

where $m = 2m_q m_{\bar{q}}/(m_q + m_{\bar{q}})$ is the reduced mass, $N$ the number of nodes of $U_{N,l}(r)$ and $l$ the angular momentum. The change of scale $x=\beta r$, where $\beta = \left( \alpha_{q\bar{q}} m/\hbar^2 \right)^{n/(n+2)}$, provides the binding energy

$$E_{N,l} = \left(\hbar^2 / m\right)^{n/(n+2)} \left(\alpha_{q\bar{q}}\right)^{2/(n+2)} \varepsilon_{N,l} \qquad (4)$$

where $\varepsilon_{N,\,l}$ is an eigenvalue of the differential equation

$$y''_{N,l} = \left[l(l+1)/x^2 + x^n - \varepsilon_{N,l}\right] y_{N,l}(x), \qquad (5)$$

independent of both the quark masses and the strength $\alpha_{q\bar{q}}$. An approximate solution of (5),

$$\varepsilon_{N,\,l} \simeq (1 + n/2)\left[2l(l+1)/n\right]^{n/(n+2)}\left(1 + \frac{n(2N+1)}{\sqrt{l(l+1)(n+2)}}\right), \qquad (6)$$

is obtained for $l\neq 0$ by substitution of a parabola for the effective potential $l(l+1)/x^2 + x^n$ around its minimum. By using (6), (4) and (2) one finds that $M_{q\bar{q}}^2$ is asymptotically proportional to $J=l$ or $l\pm 1$ for $n=2/3$ when $l\rightarrow\infty$. For $n=2/3$, $l\geq 1$ and $N=0$, eq. (6) is the eigenvalue of (5) within a 1% accuracy. The slope of the RT obtained from meson masses

$$M_{q\bar{q}} = V_0 + m_q + m_{\bar{q}} + \left(\hbar^2 / m\right)^{1/4}\left(\alpha_{q\bar{q}}\right)^{3/4} \varepsilon_{N,l}, \qquad (7)$$

is a function of the reduced mass $m$ of the $q\bar{q}$ mesons.

For solving SE for the three quarks $q_i$ of mass $m_i$ baryon states author used the hyperspherical formalism (HF) in the hypercentral approximation (HCA) where only that part of the interaction which is invariant rotation in six-dimensional space is taken into accont. This method is very accurate for computing the baryon states, even with a complicated potential like the one of Ono and Schöberl [47]. The wave function (WF) $\Psi(r, \Omega)$ is written in terms of

(i) the hyperradius

$$r = \alpha_i\left[r_{jk}^2 + M\left(\vec{x}_i - \vec{x}\right)^2 / m_k\right]^{1/2},$$

where

$$\alpha_i^2 = 2m_j m_k / m(m_j + m_k), \qquad (8)$$

$m = \sum_C m_i m_j / M$ , (c cyclic), $M = m_1 + m_2 + m_3$, invariant in any exchange of two quarks;

(ii) the set $\Omega$ of five angular coordinates defining the direction of the polar vector $\vec{x}(r,\Omega)$ in six-dimensional space.

The hypercentral component of $\sum_c r_{ij}^{2/3}$ is $C\, r^{2/3}$, where

$$C = 4\pi^{-1/2}\Gamma\left(\frac{11}{6}\right) / \Gamma\left(\frac{10}{3}\right)\sum_c\left(\frac{m(m_j + m_k)}{2m_j m_k}\right)^{1/3}, \qquad (9)$$

is a function of the quark masses.

The three-quark WF is the product

$$\Psi = H_L\left(\vec{x}\right) U_{N,K}\left(r\right) / r^{K+1}$$

of a suitably symmetrized harmonic polynomial $H_L\left(\vec{x}\right)$ of degree $L$, and a radial wave solution of

$$\left\{\left(\hbar^2 / m\right)\left[-d^2 / dr^2 + K(K+1)/r^2\right] + C\alpha_{qq}r^{2/3} - E_{N,K}\right\} U_{N,K}(r) = 0, \tag{10}$$

where $\alpha_{qq}$ is the strength of the $qq$ potential $V_{qq} = V_B + \alpha_{qq}r_{ij}^{2/3}$ and $K = L + 3/2$ (so, author consider only yrast states).

Eq. (10) for baryons is similar to eq. (3) for mesons and leads to a similar mass formula:

$$M_{3q} = 3V_B + M + \left(\hbar^2 / m\right)^{1/4}\left(C\alpha_{qq}\right)^{3/4} \varepsilon_{NK} \tag{11}$$

Vector mesons and spin 3/2 baryons, for which the knowledge of only the triplet potential is needed, are investigated.

For total angular momentum $J$, the lower baryon states are obtained when $K$ in (10) is a minimum i.e for $L=l$ and $J=l+3/2=K$. These states are described by the symmetrical harmonic polynomials $H_{l,m}\left(\vec{x}\right) = C_l \sum_c r_{ij}^l Y_l^m\left(\omega_{ij}\right)$ combined to s=3/2 states to give J. They are yrast states. The interacting pair is in the $l$ orbital state while the spectator particle is in the $S$ state.

In this HCA baryons are pure rotational states in six-dimensional space and no strong correlations generating a so – called quark-diquark state can occur with the very soft potential used in the quark potential model.

Nevertheless, the interacting pair is strongly correlated for high orbitals by the spatial deformation induced by the spherical harmonic involved, while the spectator quark is in an $S$ state.

Eqs. (7) and (10) have the same structure:

$$M = A\left(m\right) + B\left(m\right)\varepsilon_{N,l}.$$

Where A and B are functions of the quark masses. For testing the potential (1) author plotted in Figs. 1, 2 [46]

$$\varepsilon = \frac{\left(\exp erimental \quad mass\right) - A\left(m\right)}{B\left(m\right)} \tag{12}$$

versus $\lambda=J-1$ for mesons and $\lambda=J$ for baryons together with the curves of the eigenvalues $\varepsilon_{N,\lambda}$ of (5) for $l\equiv\lambda$.

As the spin-orbit short-range interaction was not introduced in our potential only the $\varepsilon$ of the highest $J$ hadrons are plotted (yrast lines). The following masses: $m_u=m_d=270$ and $m_s=516$ for the $u$, $d$ and strangle quarks, and the potential parameters $V_0 = -1210$ and $\left(\hbar^2/m_u\right)^{1/4}\left(\alpha_{q\bar{q}}\right)^{3/4} = 738$, are used for the mesons.

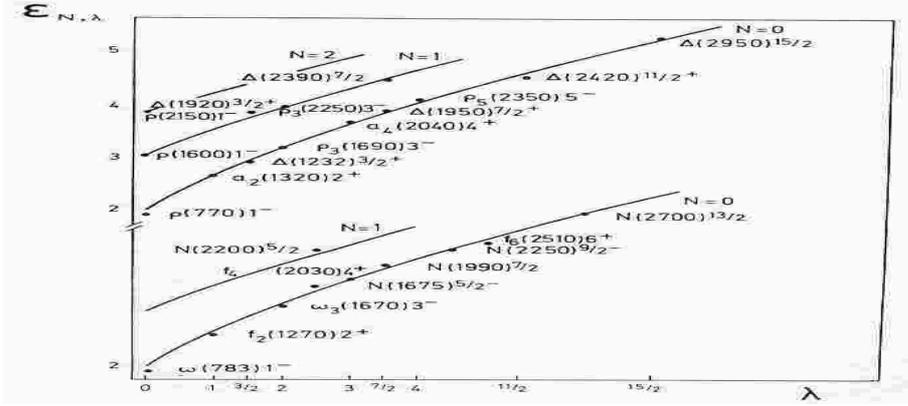

The Lipkin rule states that the $qq$ potential is half the $q\bar{q}$ potential, thus $V_B = \dfrac{1}{2}V_0$. But it applies only to the one-gluon-exchange potential, and for fitting the baryon masses author used $\alpha_{qq} = 0.45\alpha_{q\bar{q}}$, instead of $0.5\alpha_{q\bar{q}}$ for the strength of the confining potential.

One notices that the first monopole resonances ($l=0$, $N\neq0$) of the $\rho$, $K^*$ and $\varphi$ mesons are well reproduced. The $\Delta(1920)3/2^+$, which is a broad resonance, can be obtained either with $l=0$ and $s=3/2$, or as the member $(l=2 \otimes s=3/2)$ $J=3/2$ of the $l=2$, $s=3/2$, $J=1/2 \to 7/2$ states, both states are in the right position in fig. 1, [46]. On the average the $\varepsilon$ generated from masses of the various meson and baryon families are lying well along the trajectories of the eigenvalues $\varepsilon_{N, l}$ of (5).

This agreement results from the behavior of the coefficients $A(m)$ and $B(m)$ deduced from the mass transformations (3), (8), (9), in the SE.

Why such a nonrelativistic equation, which should not be valid for light quarks, works so well is a mystery. It might be related to the proportionality of the $\varepsilon_{N, l}$ and the eigenvalue $E_{N,l}$ of the relativistic equation proposed by Basdevant [48] for computing baryon spectra $(E_{0\lambda} \approx 1.125\varepsilon_{0,\lambda}$ for $3/2 \leq \lambda \leq 11/2)$.

For $N=\lambda=0$ all points are below the theoretical $\varepsilon_{00}$. The Coulomb-like attractive one-gluon-exchange component missing in our potential might be responsible for this discrepancy. By introducing this component author found that even with the modified potential

$$V_{q\bar{q}}(r) = V_0 + \alpha_{q\bar{q}}r^{2/3} - \beta_{qq}/r \qquad (13)$$

it is impossible to fully adjust simultaneously the $\rho$, $J/\Psi$ and $\Upsilon$ mass spectra.

The potential obtained by fitting the $\Upsilon$ spectrum

$$V_{q\bar{q}} = -1.1302 + 1.2514r^{2/3} - 0.07463/r \qquad (14)$$

is given in GeVfm units.

The quark masses used for the various meson families are (in MeV)

$m_u=m_d=257$,

$m_s=501.5$,     $m_c=1784$,     $m_b=5202.2$     (15)

The theoretical and experimental meson masses are exhibited in Table 1, [46].

The masses of the ρ family have been investigated for $l$ ranging from *0* to *4* and *N=0, 1, 2* for *S* states. All masses but the one of *ρ(770)* are predicted within error bars. For the *l=1* states author used the experimental average of the *J=1* and *J=2* states, because as $\overrightarrow{LS}\left(\left|J=1\right\rangle+\left|J=2\right\rangle\right)=0$ for *P* states it cancels the effect of the spin-orbit force. A good argument has been obtained for the *ϒ* family for the *1S*, *2S* and *3S* states and for the *1P* and *2P* states, but for the other *S* states above the $B\overline{B}$ threshold an irregular distribution of the spectrum prevents as accurate fit to the data. The masses of the $K^*$ and *φ* family are in good agreement with the experimental data.

As already mentioned the mass spectrum of the *J/Ψ* family exhibited in table 2, [46], does not fit the experimental data as well. Nevertheless, a good fit can still be obtained but with other potential parameters and at the expense of either the V or the ρ spectrum, which are impaired.

When $V_0+2m_u$ is substituted for $V_0$ in (13) with the parameters (14) and (15) one obtains a potential shape, which is in nearly perfect agreement with the Richardson potential [49] for $r\leq1.5$ fm.

We have to note that author [46] in fact obtained a linear RT's for *n=2/3* only for asymptotical regime. But the current resonance region has very little to do with asymptotically high *L*, and there low and moderate *L* prevail.

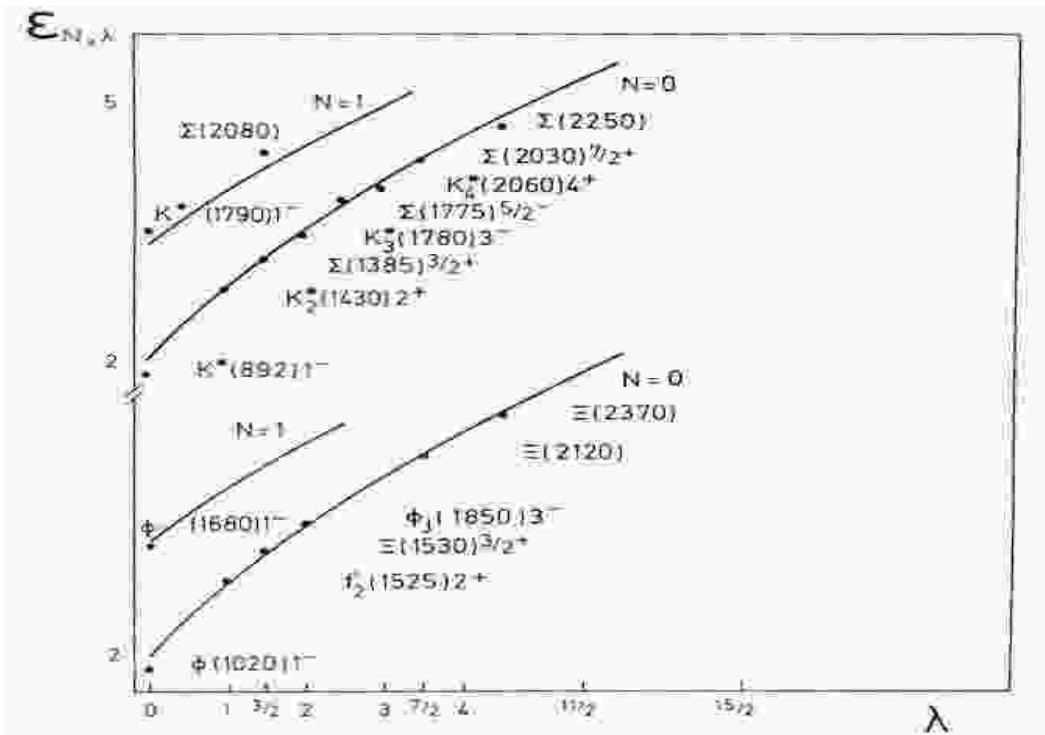

## c) Martin model

Andre Martin considered classical approach to mesons and baryons [50]. He proved that the large angular momentum behaviour of the leading RT of a meson or a baryon can be obtained by minimizing the classical energy of the system for given angular momentum. In particular, in nonrelativistic kinematics author derive the following formula for mesons with linearly rising potential

$$E_{clas} \sim const\ J^{2/3}, \tag{1}$$

which is clearly *nonlinear RT!*

Martin also explored baryonic RT. For the interaction between the three quarks, two variants were considered:

i)      quark-diquark potential

For this potential, in nonrelativistic kinematics one gets:

$$E_C(J) = \left(\frac{3}{4}\right)^{4/3} \frac{(\lambda J)^{2/3}}{m^{1/3}} \tag{2}$$

We see that RT (2) is *nonlinear and flavor dependent*.

ii)      The potential energy could be proportional to the minimum length of a Y-shaped string connecting the three quarks.

For the Y-potential plus nonrelativistic kinematics one gets

$$E = \left(\frac{3}{2}\right)^{5/3} \frac{(\lambda J)^{2/3}}{m^{1/3}} \tag{3}$$

Again author gets *nonlinear RT* with flavor dependence. From the comparison of (2) and (3) it's evident that quark-diquark configuration leads to minimal energy.

The drawback of this approach is that all results are classical and asymptotical in nature and valid only for leading RT.

## II.5. Spectrum generating algebra model(s)

In 1965, Dothan, Gell-Mann and Ne'eman [51] and, independently Barut and Bohm [52] suggested a different approach to hadron spectroscopy. In this approach, an algebra G [called spectrum-generating algebra, SGA] is chosen, and all operators relevant to hadron structure are expanded onto elements of G. In the special case in which the operators appearing in the expansion are invariant (Casimir) operators of the algebra G and its subalgebras G', G'', ..., one has a dynamic symmetry (DS).One can then solve the problem analytically in closed form. In the particular case of the energy operator, mass formulas, characteristic of the DS, arise. By acting with the transition operator $\widehat{T}$ on the states, one can also obtain, in closed form, transition matrix elements and thus decay widths. This method is quite general, and can be applied to both nonrelativistic and relativistic situations, though the expansion of the operators in terms of elements of the algebras is different in the two cases.

Recently, it has been suggested [53, 54] that SGA's for any combination of quarks, antiquarks and gluons can be constructed by taking products of appropriate space and internal algebras. In the simple case [54] of a quark and antiquark bound in a meson, the suggested space algebra is $U(3, 1)$, originating from the fact that one wants to include within the same representation of $G$ all states, corresponding to rotations and vibrations of a string with quarks at its ends. In view of the difficulty in dealing with noncompact algebras, we prefer to use, in this article, the compact form $U(4)$. There is a correspondence between the infinite–dimensional discrete representations of $U(3,1)$ and those of $U(4)$ when the dimension of the representations of $U(4)$ goes to infinity. Thus, $U(3,1)$ [or its compact form U(4)] describes the quantized geometric excitations of the string. We start a systematic investigation of hadronic properties in terms of the SGA

$$G = U(4) \otimes SU_S(2) \otimes SU_f(6) \otimes SU_C(3) \qquad (1)$$

for $q\bar{q}$ mesons, and its generalizations to multiquark and multigluon configurations.

Although the application of the SGA to the $q\bar{q}$ mesons appears, on the face of it, somewhat trivial, we do it for at least two reasons: (1) to set the stage for more complex calculations, such as the case of qqq baryons, and of strong, electromagnetic and weak decay widths [55] of hadrons, for which the use of an SGA is of crucial importance; 2) to emphasize the fact that the method simply becomes an expansion in terms of quantum numbers defining the representations of SGA and its subalgebras, thus producing simple formulas that can be easily compared with experiments. In other words, the presence of a symmetry gives relations between properties of hadronic states, which are, to a great extent, independent of the particular values of the model parameters. This formulation of the hadronic structure problem in terms of algebras allows one to test in a straightforward way features of QCD, without numerical solutions of equations of Schrödinger or Bethe-Salpeter type.

It has been suggested [53] that U(4) be taken as the SGA of geometric excitations and all states of the $q\bar{q}$ mesons belong to a single irreducible representation of U(4), characterized by the Young tableau.

$$[N] = [N, 0, 0, 0] = \qquad \ldots, \qquad (2)$$

where there are N boxes on the right. The representation (2) is totally symmetric, corresponding to the fact that the string excitations (rotations and vibrations) are bosonic in nature. The algebra of U(4) can be split in two ways that contain the angular momentum algebra SO(3):

$$
\begin{array}{ll}
\text{U(3)} \supset \text{SO(3)} \supset \text{SO(2)} & \text{(I)} \\
\text{U(4)} & \\
\text{SO(4)} \supset \text{SO(3)} \supset \text{SO(2)} & \text{(II)}
\end{array}
\qquad (3)
$$

If we consider the totally symmetric representations (2), states in the chain (I) are characterized by the quantum numbers

$$\left| \begin{array}{cccc} U(4) \supset & U(3) \supset & SO(3) \supset & SO(2) \\ \downarrow & \downarrow & \downarrow & \downarrow \\ N & n & L & M_L \end{array} \right\rangle \qquad (I) \qquad (4)$$

The allowed values of n, L, $M_L$ are given by the reduction of the representation *[N, 0, 0, 0]* into those of subalgebra of G. One obtains, *n=N, N-1, ..., 0; L=n-2, ..., 1* or *0* (n=odd or even) and *-L≤$M_L$≤+L*
For the chain (II) we have instead

$$\left| \begin{array}{cccc} U(4) \supset & SO(4) \supset & SO(3) \supset & SO(2) \\ \downarrow & \downarrow & \downarrow & \downarrow \\ N & \omega & L & M_L \end{array} \right\rangle \qquad (II) \qquad (5)$$

The allowed values of $\omega$, L, $M_L$ are given by *$\omega$=N, N-2, ..., 1* or *0* (N=odd or even); *L=$\omega$, $\omega$-1, .. , 0*, and *-L≤$M_L$≤L.*

We how briefly mention the connection between the two classification schemes (I) and (II) admitted by the SGA of *U(4)*, and the nonrelativistic quark model. If states of mesons were generated directly by the solution of SE, the chain (I) would be appropriate to problems involving harmonic oscillator (HO) potentials, since the degeneracy group of *3D* HO is *U(3)*. On the other hand, the chain (II) would be appropriate to Coulomb-like problems or those with a linear potential, since the exact degeneracy group of the *3D* Coulomb problems is *SO(4)* and numerical solutions of the SE with a linear potential have approximate degeneracy pattern of *SO(4).* Indeed, QCD suggests a linear plus Coulomb-like $q\overline{q}$ potential of the form

$$V(r) = -\frac{4}{3}\frac{\alpha_S}{r} + \sigma r \qquad (6)$$

Most importantly, the observation of rotational trajectories for mesons can be easily accommodated within a single representation of *SO(4),* but not of *U(3).* We therefore, use in this paper the *SO(4)* basis as one for developing the method of SGA for mesons. By assuming a dynamic *SO(4)* symmetry for the space part of the $q\overline{q}$ WF, we build in, from the outset, some constraints of the QCD nature of the quark interaction. In contrast, the same result is only obtained by delicate interplay of several dynamical effects in solutions of SE with suitable interquark potential.

If we assume a dynamic U(4)⊃SO(4) symmetry, the mass formula for geometric excitations must be constructed in terms of the Casimir operators of (5). The algebra of *SO(4)* has two invariants. Denoting the generators of *SO(4)* by $\vec{L}$ and $\vec{D}$, the two invariants are

$$C(SO(4)) = \vec{L}^2 + \vec{D}^2, \qquad C' = (SO(4)) = \vec{L} \cdot \vec{D} \qquad (7)$$

The eigenvalues of (7) in the representation (5) are

$$\langle C \rangle = \omega(\omega + 2), \qquad\qquad \langle C' \rangle = 0 \qquad\qquad (8)$$

We introduce the vibrational quantum number $\nu = \dfrac{N - \omega}{2}$, and subtract, for convenience, a constant term $N(N+2)$ from $C$. We then have

$$\langle C - N(N+2) \rangle = \omega(\omega + 2) - N(N+2) = -4(N+1)\left[\nu - \frac{1}{N+1}\nu^2\right] \qquad (9)$$

The algebra of SO(3) has only one invariant:

$$C(SO(3)) = \vec{L}^2, \qquad\qquad (10)$$

which eigenvalues $L(L+1)$. We do not break the $L$ degeneracy by introducing invariants of *SO(2)*. Thus the mass formula must be a functional only of (7) and (10).

In contrasting the mass formula, we consider the mass squared operator $M^2$, which is more appropriate for relativistic situations. Johnson and Thorn [56] and Bars and Hanson [56] have suggested that one expects a linear dependence of $M^2$ on $L$. This is very different from the usual nonrelativistic rigid string, for which the rotational energy grows [57] as $L(L+1)$, and is a crucial property of soft QCD strings. It implies that the string elongates, as it rotates. The elongation of the string is proportional to $\sqrt{L}$. Also, 't Hooft has shown by explicit calculation [58] in *1+1* dimensions, that one expects a linear dependence of $M^2$ on $\nu$. In order to reproduce these QCD result within the SGA approach, we write the $M^2$ operator as

$$M^2 = M_0^2 + A'\big[C(SO(4)) - N(N+2)\big] + B\left\{\left[C(SO(3)) + \frac{1}{4}\right]^{1/2} - \frac{1}{2}\right\} \qquad (11)$$

We emphasize here that dynamic – symmetric arguments do not fix uniquely the functional form $M^2$ in terms of the relevant Casimir invariants. The eigenvalues of (10) are

$$M^2(\nu, L) = M_0^2 - 4(N+1)\cdot A'\left[\nu - \frac{1}{N+1}\nu^2\right] + B\left\{\left[L(L+1) + \frac{1}{4}\right]^{1/2} - \frac{1}{2}\right\} \qquad (12)$$

Here the value of *(N/2)* represents the total number of vibrational states in the representation *[N]*. In view of confinement, the total number of bound states is infinite. We thus must take in our description $N \to \infty$. In practice it is sufficient to take $N$ large enough to include all known and unknown states up to a maximum value of $L$ and $\nu$. The maximum $\nu$ is, $\nu_{max} = N/2$ or *(N-1)/2*, while the maximum $L$ in each *SO(4)* representation is $N-2\nu$. The observed maximum number of $L$ is $\approx 5$, and the observed maximum value of $\nu$ is $\approx 4$. We take *N=100*. In the limit of large $N$, eq. (11) reduced to

$$M^2(\nu, L) = M_0^2 + A\nu + BL, \qquad\qquad (13)$$

with $A = -4A'(N+1)$

For internal excitations, we have the spin part and the flavor part. The Casimir operator of $SU_S(2)$ is

$$C(SU_S(2)) = \vec{s}^2 \tag{14}$$

with eigenvalues $S(S+1)$. The spin and orbital momentum must then be coupled to $\vec{J} = \vec{L} + \vec{s}$. Denoting by $SU_J(2)$ the combined algebra.

$$SU_S(2) \otimes SO_L(3) \supset SU_J(2) \tag{15}$$

we have $C(SU_J(2)) = \vec{J}^2 \tag{16}$

with eigenvalues $J(J+1)$. There is no simple QCD argument to tell whether the dependence on $s$ and $J$ is linear or quadratic. Perturbation arguments involving one gluon exchange suggest that the spin part appears in the potential approach, for a quark and antiquark of equal mass, with three terms: a spin-spin term of the type

$$V_S = V_S(r)\vec{S}_q \cdot \vec{S}_{\bar{q}}, \tag{17}$$

a spin-orbit term of the type

$$V_{SO} = V_{SO}(r)(\vec{S}_q + \vec{S}_{\bar{q}}) \cdot \vec{L}, \tag{18}$$

and a tensor interaction

$$V_T = V_T(r)S_{12}(q\bar{q}), \tag{19}$$

with exception of the tensor interaction, which is nondiagonal in the $L, S, J$ basis, the two terms can be written in terms of the Casimir operators of total $SU_S(2)$ and $SU_J(2)$ as

$$V_S = v_S C(SU_S(2)),$$
$$V_{SO} = v_{SO}[C(SU_J(2)) - C(SU_S(2)) - C(SO(3))], \tag{20}$$

with eigenvalues

$$\langle V_S \rangle = V_S S(S+1), \tag{21}$$

$$\langle V_{SO} \rangle = V_{SO}[J(J+1) - S(S+1) - L(L+1)], \tag{22}$$

Thus, these perturbations arguments suggest a quadratic dependence on $S$ and $J$. Unfortunately, experiments cannot tell whether the dependence on $S$ is linear or quadratic, since there are only two possible values for $S$ for mesons, $S=0$ and $S=1$. The test for the $J$ term in the meson mass squared is difficult, particularly for light mesons, since the error on the mass determinations for the relevant mesons is rather large. However, there are indications that the nonrelativistic $J(J+1)$ rule is not a good one for them. In the analysis of the following section, we take in analogy with (9) a linear dependence on the quantum number, i. e.,

$$M^2 = M_0^2 + A'\left[C(SO(4)) - N(N+2)\right] + B\left\{\left[C(SO(3)) + 1/4\right]^{1/2} - 1/2\right\} +$$
$$+ C\left\{\left[C(SU_S(2)) + 1/4\right]^{1/2} - 1/2\right\} + D\left\{\left[C(SU_J(2)) + 1/4\right]^{1/2} - 1/2\right\} \tag{23}$$

with eigenvalues

$$M^2(v, L, S, J) = M_0^2 + Av + BL + CS + DJ, \tag{24}$$

We do not consider here effects of the tensor interaction. The QCD-inspired arguments of Gürsey [59] suggest this term to be rather small, at least for light mesons.

The mass spectrum (24) is extremely simple, as shown in Fig. 3, [60]. Each *SO(4)* representation *v* provides a RT with *L=0, 1, 2, ...* There are an infinite number of such trajectories corresponding to *v=0, 1, ...* The slopes of the trajectories are directly related to the coefficients *A, B, C, D* in (24).

If one accounts for all flavours, the mass matrix must be diagonal and of the type

$$M^2(i, j; v, L, S, J) = (M_0^2)_{ij} + A_{ij}v + B_{ij}L + C_{ij}S + D_{ij}J \tag{25}$$

Following common usage we shall call a combination *i j* a family.

If we attempt a simultaneous fitting of masses for all families, we have to face two problems: quark masses $M_i$ are widely different; light mesons behave relativistically, while heavy mesons are nonrelativistic. In the spirit of a simple expansion of $M^2$ in terms of quantum numbers, we introduce the quantity $M_{ij} = M_i + M_j$, where $M_i$ and $M_j$ are the constituent masses of quark *i* and antiquark j and expand all coefficients in $M_{ij}$, keeping only the first-order term:

$$A_{ij} = a + a'M_{ij}, \qquad\qquad B_{ij} = b + b'M_{ij}$$

$$C_{ij} = C + C'M_{ij}, \qquad\qquad D_{ij} = d + d'M_{ij}, \tag{26}$$

$$(M_0^2)_{ij} = eM_{ij} + (M_{ij})^2.$$

This simple parametrization should allow us to go from relativistic to nonrelativistic situation in a simple manner. The rationale is that we should use the mass operator itself, rather than $M^2$, for nonrelativistic situations. If we then add to $M_0$ an interaction term U, we obtain

$$M = M_0 + U, \qquad\qquad M^2 = M_0^2 + 2M_0U + U^2 \tag{27}$$

If $M_0$ is small, we are in a relativistic situation; if $M_0$ is large, we are in a nonrelativistic domain. Both can be approximately covered by the parametrization of (25). Mass formula with all corrections has the form

$$\left\langle q_i \overline{q_i} \left| M^2 \right| q_{i'} \overline{q}_{j'} \right\rangle = \delta_{ii'}\delta_{jj'}\left[ (M_{ij})^2 + eM_{ij} + (a + a'M_{ij})\left[1 - \frac{1}{N+1}v\right] + (b + b'M_{ij})L + (c + c'M_{ij})S + (d + d'M_{ij})J \right] +$$
$$+ \left\langle M'^2 \right\rangle_{ij, i'j'} + \left\langle M''^2 \right\rangle_{ij, i'j'}$$

$$\tag{28}$$

There are 15 parameters in (28) and they were determined from a fit to 57 well – established states from the PDG 1990. The average deviation for the mass squared of these mesons is 5.7%.

The most striking situation happened for $\Psi$ and $\Upsilon$ families. We next plot, in Figs. 13 and 14, [60], the "vibrational" RT for the $\Psi$ and $\Upsilon$ families. We note that experimental masses no longer fall exactly on trajectories, but their trajectories are *slightly bent,* a feature also anticipated in the 't Hooft calculation [58]. The departure of the trajectories from linearity may indicate a breaking of the *SO(4)* symmetry. Another possible explanation is that the effective value of *N*, in our formula, appropriate for heavy mesons, is not $N \rightarrow \infty$ (*N=100* used in Figs. 13 and 14), but much smaller. This may be due to coupling with break-up channels that effectively terminate the rotational and vibrational bands. The *U(4)⊃0(4)* gives the following dependence on N [Eq. (11)]:

$$M^2(v) = M_0^2 - 4(N+1)A'\left[v - \frac{1}{N+1}v^2\right] \qquad (29)$$

with *N≈20*, we find the "vibrational" trajectories to be *significantly bent* [61], so as to actually describe the vibrational spectra of $\Psi$ and $\Upsilon$ observed so far. More experimental work is needed to extend these studies of the excited vibrational states, to clarify the picture.

SGA has also been applied to baryons [62]. The situation in baryons is much more complex since there are several passible configurations of the quarks. Corresponding orbital group is U(7). Authors fit the N, $\Delta$, $\Lambda$, $\Sigma$, $\Xi_i$, $\Omega$ spectra, but unfortunately they didn't notice the effects of nonlinearity in baryonic RT.

## II.6. Relativistic and semirelativistic models

### a) Basdevant model

Recently authors [63] studied the spectrum of a semirelativistic three-body Hamiltonian. The hyperspherical method proves to be very efficient. They show that the ground states of baryons can be calculated with good accuracy. However, when using the meson potential together with color assumption $V_{qq} = \frac{1}{2}V_{q\bar{q}}$ baryon Regge slopes came out noticeably too small. Authors analyze this problem and show that a quark-diquark structure for baryons cures this defect. Altogether the construction of a global unified potential model for baryons and mesons seems quite hopeful.

It is useful, owing to the empirical quality of the hyperradial approximation, to consider the following simple model for light quarks. Authors assume that the hyperradial approximation (HRA) describes the leading orbital excitations appropriately. Hence, one can use for three massless quarks the approximate Hamiltonian.

$$\widetilde{H} = \frac{32\sqrt{2}}{5\pi}\left(\frac{P}{\sqrt{3}} + \frac{1}{2}\mu^2 R\right) - \frac{3}{2}V_0, \qquad (1)$$

where $V_0$ is a constant and $\mu^2 r - V_0$ would represent a $q\bar{q}$ potential. If $J$ (or $K$) is the total three body orbital momentum, then a centrifugal barrier term of $(J+3/2)(J+5/2)/R^2$ is understood to be present in $P^2$. A good interpolating formula for the eigenvalues of (1) is

$$\left(\varepsilon_{nl}\right)^2 = \pi\left(2n + \frac{\lambda(n)l}{\pi} - \frac{1}{2} + \eta(n)\right),$$  (2)

with the values

| N | 1 | 2 | 3 | 4 |
|---|---|---|---|---|
| λ(n) | 3.9843 | 3.9485 | 3.9028 | 3.8596 |
| η(n) | 0.067 | -0.0338 | -0.0843 | -0.1146 |

The eigenvalues of $\widetilde{H} = \alpha P + \beta R - \lambda$ are $(\alpha\beta)^{1/2}\varepsilon_{nl} - \lambda$. Authors adjust $\mu^2$ and $V_0$ in (1) in order to reproduce $D_{33}$ and it's RT: from (2) that RT will be linear in agreement with experiment.

Adjusting $m_s$ in order to reproduce the *Σ\*(1385)*, they obtain

$\mu_0^2 = 0.167$          $V_0 = 0.570$          $m_s = 0.333,$  (3)

which is quite different from meson set:

$\mu_0^2 = 0.234$          $V_0 = 0.757$          $m_s = 0.308$  (4)

The string tension needed for mesons is noticeably larger than for baryons. An easy way to see this is to compare the Regge slopes predicted by (1) and its mesonic analog [64]

$$H_M = 2p + \mu^2 r - V_0$$  (5)

it is easy to obtain, through (2) that these models would predict for the radio of Regge slopes

$$\alpha'_B / \alpha'_M = \left(5\pi/16\right)^2\sqrt{3}/2 \sim 0.83$$  (6)

in contrast with the experimental value

$\alpha'_B/\alpha'_M \sim 1.1$  (7)

This is a serious problem in trying to build a unified potential-model approach for both mesons and baryons. Authors claim that considering baryons as a quark-diquark structures will improve the radio (7).

The center of the debate lies in the slopes of RT or, equivalently, in the masses of leading orbital excitations. Once more, authors insist on the fact that RT, which played a crucial role in classifying hadrons in the 70's, are a very fundamental feature of hadron spectroscopy. Analycity in the angular momentum is a direct consequence of confined potential models. The universal slope of the phenomenologically linear RT is a fundamental parameter of (light flavor) hadron dynamics, much more important that the mass of such or such a state.

Unfortunately authors [63, 64] totally missed the effects of nonlinearity of hadronic RT's.

## b) Semay model

C. Semay and R. Ceuleneer considered the spin-triplet spectra of light and heavy neutral mesons in the framework of a free two-body Dirac equation [65]. Two-body Dirac Hamiltonian is given, in the center–of–mass frame, by *(ℏ=c=1)*

$$H = (\vec{\alpha}_1 - \vec{\alpha}_2)\vec{p} + \beta_1 m_1 + \beta_2 m_2 + \frac{1}{2}(\beta_1 + \beta_2)\lambda r, \qquad (1)$$

The effective interaction is scalar in order to confine the particles, and the radial form factor is proportional to the interquark distance $r = |\vec{r}|$, as it is required to obtain linear RT.

As for as the energy spectrum is concerned, the DE and its Klein-Gordon counterpart lead, for the interaction considered in this work, to very similar spectra. Authors compare theoretical results for RT with the data and with the results of the nonrelativistic model [46]. Relativistic model [65] as well as NRQM [46] seems to fit well orbital RT's for *ρ, a, ω, f, φ, K\** families.

The predictions of their model concerning the radial excitations of light mesons are not in agreement with the experimental data. Some mesons, namely *ρ(1450), ω(1390),* and *K\*(1410),* seem to appear as extra states in Figs. 2 to 6 [65].

Authors also describe radial excitations in heavy mesons, and in particular, *J/Ψ* and *Υ* mesons. As one can see *J/Ψ* and *Υ* experimental radial spectra are in fact *nonlinear*, which is in accord with findings by Iachello [60].

The drawback of this model is that it describes very limited meson sector.

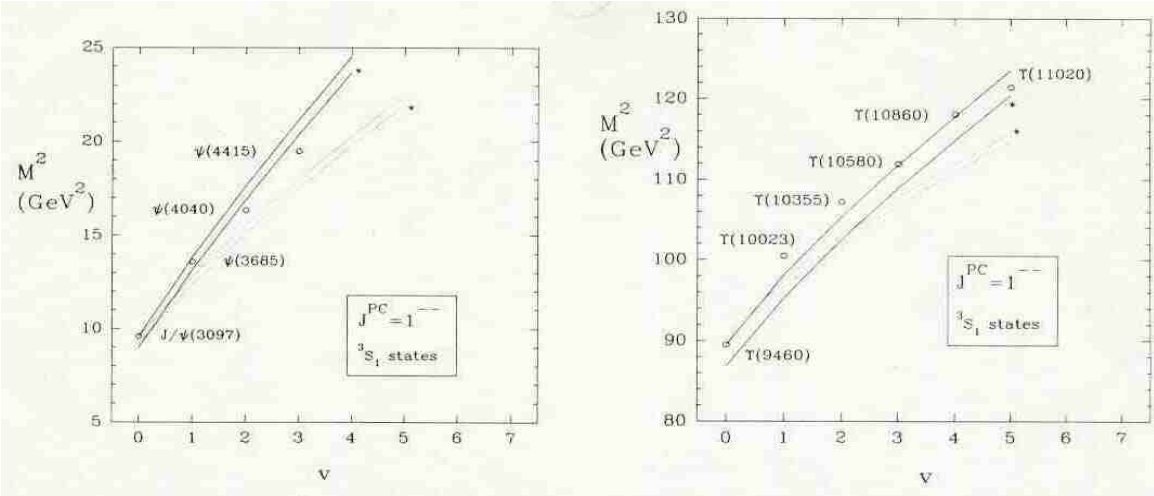

## c) Relativistic flux tube model (RFT)

The relativistic flux tube (RFT) model is a new effective model of QCD, which has been recently proposed as an alternative to the potential model for describing the hadrons [66]. Like the relativistic potential model, it produced linear RT in the ultrarelativistic limit, and it reduced to the usual SE with linear confinement potential in the heavy mass limit. Nevertheless, contrary to the potential approach, the relativistic corrections induced by the RFT model are in agreement with the predictions of the Wilson loop formulation of QCD. Moreover, an extension of this model for glueball and hybrid mesons is possible.

The RFT model relies on the assumption that the quark and the antiquark in a meson are connected by a rigid straight tubelike color flux field carrying both energy and momentum. In this model, the field becomes a real dynamical entity.

The equations of motion of an asymmetric flux tube model are given by set of three coupled nonlinear equations: two constraint equations for the total impulsion and the orbital angular momentum, respectively, and one equation defining the Hamiltonian of the system. These equations depend on the quark transverse velocity operator, which commutes neither with the position nor with the momentum operators. To solve this problem, authors use a matrix method relying on an iterative procedure.

In natural units $(\hbar = c = 1)$, the Lagrangian $L$ of the system is then given by

$$\mathcal{L} = \mathcal{L}_1 + \mathcal{L}_2 + V(r)$$

$$\mathcal{L}_i = -m_i \gamma_i^{-1} - a \int_0^r dr_i' \gamma_{i\perp}'^{-1} - \frac{C}{2} \gamma_{i\perp}^{-1}, \tag{1}$$

where $\gamma_i = \left(1 - v_i^2\right)^{-1/2}$ and $\gamma_{i\perp} = \left(1 - v_{i\perp}^2\right)^{-1/2}$. $C/2$ is an energy contribution from the extremity of the flux tube. The role of this constant is to take into account possible boundary effect due to the contact between the flux tube and the quarks. When $l=0$, we have $v_1=v_2\equiv0$, which simplifies greatly the equations of motion. In this case, the system is completely described by the equation

$$H = \sqrt{P_R^2 + m_1^2} + \sqrt{P_R^2 + m_2^2} + ar + C + V(r), \tag{2}$$

which is a spinless Salpeten equation, with an instantaneous interaction, for zero angular momentum ($p_r^2 = \vec{p}^2$). The contribution of the flux tube is then equivalent to the potential $ar+C$.

The short range interaction is parametrized by the usual Coulomb-like potential

$$V_C(r) = -\frac{k}{r} \quad , \tag{3}$$

where $k$ will be either a constant, or a function

$$k(r) = k_c \left[1 - \exp\left(-r / r_c\right)\right] \tag{4}$$

In the case of pseudoscalar mesons, the energy difference between $^1S_0$ and $^3S_1$ states is very large. Several relativistic models [67-69] have shown that the OGE mechanism cannot explain this large nondegeneracy. A possible solution to this problem is to use an interaction stemming from instanton effects. Unlike the OGE potential, the instanton interaction possesses an explicit spin dependence, since it contains a projector on spin $S=0$ states. Authors choose the following smeared Gaussian function with a range $r_0$:

$$h(r) = \frac{1}{\left(\pi r_0^2\right)^{3/2}} \exp\left(- r^2 / r_0^2\right) \qquad (5)$$

Despite the of a nonzero range instanton interaction (II), authors assume, that it acts only for $l=0$ mesons. In that case for the isovector $n\bar{n}$ mesons with quantum numbers $l=S=0$, the Hamiltonian takes the form

$$H = 2\sqrt{\vec{p} + m_n^2} + ar + C + V(r) - gh(r), \qquad (6)$$

where we assume that $m_u = m_d = m_n$.

To solve all the eigenvalue equations considered in this work, authors expand the eigenstates in a harmonic oscillator (HO) basis $\{\Phi_i\}$ with a limited number of quanta:

$$\Psi(\vec{r}) = \sum_{i=1}^{N} C_i \Phi_i(\vec{r}, b), \qquad (7)$$

where $b$ is the oscillator length. Authors have checked that a very good approximation (to about 1 MeV) can be obtained with $N=10$-$15$.

Fit to pseudoscalar sector shows that the interaction used here to simulate instanton effects is too simple. Also, a proper treatment of the spin could greatly improve the description of pseudoscalar states in the RFT model. The Regge and vibrational trajectories of light mesons are presented in Figs. 1 and 2, [66], respectively. The orbital excitations of $n\bar{s}$ mesons are well reproduced. The main problem for the strange mesons is the description of the $2^3S_1$ state for which the experimental situation is a little bit confused. Two candidates exist and these theoretical predictions lie between the two experimental values. It seems impossible to obtain within the same set of parameters a good description of both Regge and vibrational trajectories: when the predictions for the first radial excitations of the light mesons are good, the slope of the RT's is found to be too small. The cause of this problem could be that retardation effects are not taken into account in this model. One can expect that these effects play a more important role in vibrational dynamics than in rotational dynamics. So they could modify appreciably the connection existing between the slope of Regge and vibrational trajectories.

It is clear that a complete reliable RFT model must take into account the fermionic nature of quarks, and must include the retardation effects which could play an important role in the description of vibrational excitations. It is possible that such improvements of the original RFT allow a better determination of the mass of the lightest quarks in this model.

### d) <u>Point form of relativistic quantum mechanics (RQM)</u>

In this paper authors apply the point form of relativistic quantum mechanics (RQM) to the description of meson RT [70]. RQM is also known in the literature as relativistic Hamiltonian dynamics or Poincare-invariant quantum mechanics with direct interaction. The description in the point form implies that the operators $\hat{M}^{\mu\nu}$ are the same as for noninteracting particles, i.e. $\hat{M}^{\mu\nu} = M^{\mu\nu}$, and these interaction terms can be presented only in the form of the four-momentum operator $\hat{P}$.

Authors calculated mass spectra of mesons containing *u, d* and *s* quarks with the effective potential in the oscillator form

$$\bar{W}(r) = W_0 \delta(\vec{r}) + \beta^4 r^2 \qquad (1)$$

The spectra of mesons composed of quarks with equal masses, are given by

$$M^2 = 4(m^2 + W_0) + 8\beta^2(2n + l + 3/2) \qquad (2)$$

The parameters *(m²+W₀)* and *β* have been found from the fitting the ρ-*a* RT,

$(m^2+W_0)=0.261\ GeV^2,$ $\qquad\qquad$ *β=0.375 GeV*

They observe that all experimental data are in good agreement with model spectrum for *l≥1* and *S=2* (Fig. 2). As in the case of bound states with *S=0,* the agreement between theoretical predictions and the existing experimental data is not good (Figs. 3-4). Such deviations can be explained by absence of spin-dependent terms and short-distance term of the potential.

When the masses of the quark and antiquark are different they obtain

$$M^2 = 2W_0 + 2\beta^2(2n + l + 3/2) + m_1^2 + m_2^2 + 2\sqrt{(W_0 + 2\beta^2(2n + l + 3/2))} \times$$
$$\times \sqrt{(W_0 + 2\beta^2(2n + l + 3/2) + m_1^2 + m_2^2) + m_1^2 m_2^2} \qquad (3)$$

One can see that the dependence (3) is, generally, *nonlinear* and became linear only at asymptotically large *l* .

Authors apply (3) for the strange meson RT and agreement with data is good (see Fig. 5). But for mesons with S=0, the agreement between the model and the data is not good (see Fig. 9).

As wee see, generally, RQM leads to *nonlinear RT's* for mesons with mixed flavours and to linear RT's for the meson with identical quarks.

### e) Semiclassical relativistic model

Authors present a semiclassical relativistic model for the orbital spectra of mesons, based on the assumption of a universal, flavor-independent linear confining interaction. Flavor dependence of the spectra arises from the quark masses [71].

Most hadrons which consist of light quarks have been grouped in rotational families where

$$J = \alpha' M^2 + \alpha_0 \qquad (1)$$

is the relation between spin (*J*) and mass squared. In the case of mesons all integer *J* values are included with alternating parity for the mesons as *J* increases (exchange degeneracy). In same cases the assignment to a particular family is not certain, or the formula (1) fails. This failure is most acute for the rotational family to which the pseudoscalar mesons (*π* and *K*) belong. The mass spectrum (1) is obtained in any relativistic model where the orbital angular momentum is carried by a rotating linear field (constant rest energy per unit length) such as in the dual string model, or (approximately) in the bag model with massless colored quarks and gluons.

It has already been pointed out how (1) becomes modified in the same kind of model when equal-mass particles are attached at the ends of the string or are the sources of stretched color-electric flux lines. Here authors extend this trivially to allow unequal masses at the ends. Authors treat the system as a string, but the same results will be obtained when quarks of various masses are treated classically in the colored-quark bag model. The model is classical, but authors argue that a slightly modified version of it still could give an accurate representation of the mass spectrum in a full quantum theory. For example, the classical string model gives (1) without the intercept $\alpha_0$. The full quantum treatment provides only the "quantum defect" $\alpha_0$ in $J$. It has been suggested that in the absence of complete theory a useful approximation might be obtained by calculating the ground-state mass and using that result together with (1) to determine $\alpha_0$. One can test this suggestion in the limit opposite to the massless relativistic string. Consider two equal-mass quarks moving nonrelativistically in a linear potential with slope $1/2\pi\alpha'$. The classical (Bohr) model with a "quantum defect" in $l$ gives

$$l = l_0 + \frac{4\pi}{3^{3/2}} \alpha' \sqrt{m} \left(M - 2m\right)^{3/2} \tag{2}$$

Authors compare (2) with the predictions of the SE. They determine the "defect" $l_0$ by fitting (2) to the exact ground state ($l=0$). The SE was solved numerically for several low values of $l$ and the agreement is impressive. Johnson guess that such a quantum-defect formula can give a reasonable representation of the excited states of a system which corresponds to the balancing of an infinitely rising long-range attraction against a centrifugal repulsion, since in this case the quantum wave function is well localized about a classical orbit. When would we expect such a guess to fail? It should fail in those situations where in addition to a long-range attraction there are strong attractive short-range interactions. In states with $l\neq0$, these could be relatively unimportant, but in the $l=0$ state they would have an important effect. In such a case one would not expect that the energies of the excited states would be simply determined by the quantum defect approximation with $l_0$ obtained from the $l=0$ mass. The WF would first have to <u>climb out of the attractive hole</u> before finding itself in the long-range rising potential whose dominance for finite $l$ is the physics of the quantum-defect approximation. On the other hand, if the short-range interactions are repulsive then the particles are already a part in the ground state and orbital excitation would just move hem farther into the long-range part of the interaction. Hence one might anticipate that the classical formula with the quantum defect could work quite well even in going from $l=0$ to $l=1$.

In the case where the relationship fails for the $l=0$ state, one might still expect that the excited states are well represented by such a formula since short-range effects are greatly reduced when $l\neq0$. In this circumstance the defect would be better determined by fitting the mass of the $l=1$ state for example.

We would like to remark that the meson spectrum indicates that this kind of dynamics may be operating. The states $\pi$ and $K$ which do not lie on linear RT are ones where the short-range quark spin-spin interaction is strong and attractive.

A classical relativistic string with masses $m_1$ and $m_2$ attached at its extremities, and moving in its own rest system with the maximum angular momentum consistent with a given total mass $M$, is straight and rotates rigidly with mass and angular momentum

$$M = \frac{m_1}{\left(1-v_1^2\right)^{1/2}} + \frac{m_2}{\left(1-v_2^2\right)^{1/2}} + \frac{1}{2\pi\alpha'\omega} \int_{-v_2}^{v_1} \frac{dv}{\left(1-v^2\right)^{1/2}},$$

$$J = \frac{m_1 v_1^2}{\omega\left(1-v_1^2\right)^{1/2}} + \frac{m_2 v_2^2}{\omega\left(1-v_2^2\right)^{1/2}} + \frac{1}{2\pi\alpha'\omega^2} \int_{-v_2}^{v_1} dv \frac{v^2}{\left(1-v^2\right)^{1/2}}$$

(3)

The speeds $v_1$ and $v_2$ the ends are determined so that the masses $m_1$ and $m_2$ move under the tension at their respective ends of the string. This gives the relation between the ends speeds $v_1$ and $v_2$ and the rotational frequency $\omega$,

$$v_i = \left(1 + m_i^2\pi^2\alpha'^2\omega^2\right)^{1/2} - m_i\pi\alpha'\omega$$

(4)

with the speed determined by (4), (3) may be regarded as a parametric ($\omega$) representation of the dependence $J=J(M)$.

Authors suggest as a semiclassical version of (3) the same formulas with $J$ in (3) replaced by $J-J_0$ where $J_0$ is taken as a quantum defect. We will now discuss various limiting cases of (3) modified by the inclusion of the "defect". As $m_1$ and $m_2$ tend to zero, the ends move at the velocity of light, and (3) reduces to the linear RT

$$M = \frac{1}{2\alpha'\omega}, \qquad J - J_0 = \frac{1}{4\alpha'\omega^2}, \text{ or } J = \alpha'M^2 + J_0$$

(5)

If we apply (3) to the case of one massless quark and one heavy quark, it can be written in the simpler parametric form based upon the velocity of the heavy quark.

$$J = \pi m^2\alpha'\left[\frac{v^3}{\left(1-v^2\right)^{3/2}} + \frac{v^2}{\left(1-v^2\right)^2}\left(\pi/2 + \sin^{-1}v\right)\right],$$

$$M = m\left[\frac{1}{\left(1-v^2\right)^{1/2}} + \frac{v}{\left(1-v^2\right)}\left(\pi/2 + \sin^{-1}v\right)\right]$$

(6)

In the limit when $(M-m)/m \ll 1$ and the velocity $v \ll 1$, we obtain the approximate form

$$J = 2\alpha'\left(M-m\right)^2\left[1 - \left(\frac{2}{\pi}\right)^2 \frac{M-m}{m} + ...\right],$$

(7)

which is clearly *nonlinear RT*.

In the opposite limit $v \rightarrow 1$ ($M \gg m$) one of course recover the linear dependence, $J \rightarrow \alpha' M^2$. In the case of the D's, the relativistic dependence is sufficiently important that (6) is about as convenient to use as the approximate form (7). In the case of bottom quarks ($m \simeq 5\ GeV$), the simple form $J-J_0=2\alpha'(M-m)^2$ will be sufficiently accurate for the lowest angular excited states of mesons with one heavy quark and one light quark.

Authors have shown in the case of the quark mesons that the observed flavor variations in the Chew-Frautschi plots can be accounted for by quark mass differences in a semiclassical relativistic model based upon a universal, flavor independent linear confining interaction.

### f) t'Hooft 1/N$_c$ model

A recently proposed gauge theory for strong interactions, in which the set of planar diagrams play a dominant role, is considered in one space and one time dimension. In this case, the planar diagrams can be reduced to self-energy and ladder diagrams, and they can be summed. The gauge field interactions resemble those of the quantized dual string and the physical mass spectrum consist of a nearly straight Regge trajectory [58].

In the model there is only space and one time dimension. There is a local gauge group $U(N)$, of which the parameter $N$ is so large that the perturbation expansion with respect to $1/N$ is reasonable. Our Lagrangian is

$$\mathcal{L} = \frac{1}{4} G_{\mu\nu i}{}^{j} G_{\mu\nu j}{}^{i} - \bar{q}^{ai} \left( \gamma_\mu D_\mu + m_{(a)} \right) q_i^a ,$$ (1)

where

$$G_{\mu\nu i}{}^{j} = \partial_\mu A_{i\nu}^{i} - \partial_\nu A_{i\mu}^{i} + g \left[ A_\mu, A_\nu \right]_i^{j} ;$$ (2a)

$$D_\mu q_i^a = \partial_\mu q_i^a + g A_{i\mu}^{j} q_j^a ;$$ (2b)

The model becomes particularly simple if one impose the light-cone gauge condition:

$$A_- = A^+ = 0$$ (3)

In that gauge we have

$$G_{+-} = -\partial_- A_+$$ (4)

and

$$\mathcal{L} = -\frac{1}{2} Tr(\partial_- A_+)^2 - \bar{q}^a \left( \gamma \partial + m_{(a)} + g\gamma_- A_+ \right) q^a$$ (5)

Let us consider the limit $N \rightarrow \infty$; $g^2 N$ fixed, which corresponds to taking only the planar diagrams with no Fermion loops. They are of the type of Fig. 2, [58]. All gauge field lines must be between the fermion lines and may not cross each other. The gauge fields do not interact with themselves in such diagrams. We have nothing but ladder diagrams with self-energy insertions for the fermions.

The ladder diagrams satisfy a Bethe-Salpeter equation, depicted in Fig. 4, [58]. The equation for the WF $\varphi(x)$ can be written as

$$\mu^2 \varphi(x) = \left(\frac{\alpha_1}{x} + \frac{\alpha_2}{1-x}\right)\varphi(x) - P\int_0^1 \frac{\varphi(y)}{(y-x)^2}dy \qquad (6)$$

Author was unable to solve this equation analytically. But much can be said, in particular about the spectrum. First, one must settle the boundary condition. At the boundary $x=0$ the solutions $\varphi(x)$ may behave like $x^{\pm\beta_1}$, with

$$\pi\beta_1 \cot g\pi\beta_1 + \alpha_1 = 0 \qquad (7)$$

but only in the Hilbert space of functions that vanish at the boundary the Hamiltonian (the right-hand side of (6)) is Hermitian:

$$(\psi, H\varphi) = \int_0^1 \left(\frac{\alpha_1+1}{x_1} + \frac{\alpha_2+1}{1-x}\right)\varphi(x)\psi^*(x)dx + \frac{1}{2}\int_0^1 dx\int_0^1 dy \frac{(\varphi(x)-\varphi(y))(\psi^*(x)-\psi^*(y))}{(x-y)^2} \qquad (8)$$

In particular, the "eigenstate"

$$\varphi(x) = \left(\frac{x}{1-x}\right)^{\beta_1}$$

in the case $\alpha_1 = \alpha_2$, is not orthogonal to the ground state that does satisfy $\varphi(0) = \varphi(1) = 0$.

Also, from (8) it can be shown that the eigenstates $\phi^k(0) = \phi^k(1) = 0$ form a complete set. Author concludes that this is the correct boundary condition. A rough approximation for the eigenstates $\phi^k$ is the following. The integral in (6) gives its main contribution if $y$ is close to $x$. For a periodic function we have

$$P\int_0^1 \frac{e^{i\omega y}}{(y-x)^2}dy \simeq P\int_{-\infty}^{\infty} \frac{e^{i\omega y}}{(y-x)^2}dy = -\pi|\omega|e^{i\omega x}$$

The boundary condition is $\varphi(0) = \varphi(1) = 0$. So if $\alpha_1, \alpha_2 \approx \simeq 0$ then the eigenfunctions can be approximated by

$$\varphi^k(x) \simeq \sin k\pi x, \quad k=1,2,\ldots \qquad (9)$$

with eigenvalues

$$\mu^2_{(k)} \simeq \pi^2 k \qquad (10)$$

This is a straight RT, and there is no continuum in the spectrum! The approximation is valid for large k, so (10) will determine the asymptotic form of the trajectories whereas deviation from the straight line are expected near the origin as a consequence of the finiteness of the region of integration and the contribution of the mass terms.

The physical interpretation is clear. The Coulomb force in a one-dimensional world has the form

$$V \propto |x_1 - x_2|,$$

which gives rise to an insurmountable potential well. Single quarks have no finite dressed propagators because they cannot be produced. Only colorless states can escape the Coulomb potential and are therefore free of infra-red ambiguities. Authors result is completely different from the exact solution of two dimensional massless QED, which should correspond to $N=1$ in our case. The perturbation expansion with respect to $1/N$ is then evidently not a good approximation; in two dimensional massless QED the spectrum consist of only one massive particle with the quantum numbers of the photon.

In order to check our ideas on the solutions of eq. (6) we devised a computer program that generates accurately the first 40 or so eigenvalues $\mu^2$. We used a set of trial functions of the type

$$A x^{\beta_1} (1-x)^{2-\beta_1} + B (1-x)^{\beta_2} x^{2-\beta_2} + \sum_{k-1}^{N} C_k \sin k\pi x .$$ The accuracy is typically of the order of 6 decimal places for the lowest eigenvalues, decreasing to 4 for the 40th eigenvalue, and less beyond the 40th.

A certain WKB approximation that yields the form

$$\mu^2_{(n)} \to \pi^2 n + (\alpha_1 + \alpha_2) \log n + c^{st}(\alpha_1, \alpha_2), \quad \text{n=0, 1, ...} \tag{11}$$

was confirmed qualitatively (the constant in front of the logarithm could not be checked accurately). As we see (11) clearly leads to nonlinear RT.

In Fig.5 author shows the mass spectra for mesons built from equal mass quarks. In the case $m_q = m_{\bar{q}} = 1$ (or $\alpha_1 = \alpha_2 = 0$) the straight line is approached rapidly, and the constant in eq. (11) is likely to be exactly $3/4\pi^2$.

In Fig.6 author give same results for quarks with different masses. The mass difference for the nonets built from two triplets are shown in two cases:

(a)     $m_1=0$;          $m_2=0.200$;          $m_3=0.400$,

(b)     $m_1=0.80$;       $m_2=1.00$;          $m_3=1.20$.

in units of $g/\sqrt{\pi}$. The higher states seem to *spread logarithmically*, in accordance with eq. (11). But, contrary to eq. (11), it is rather the average mass, than the average squared mass of the quarks that determines the mass of the lower bound states.

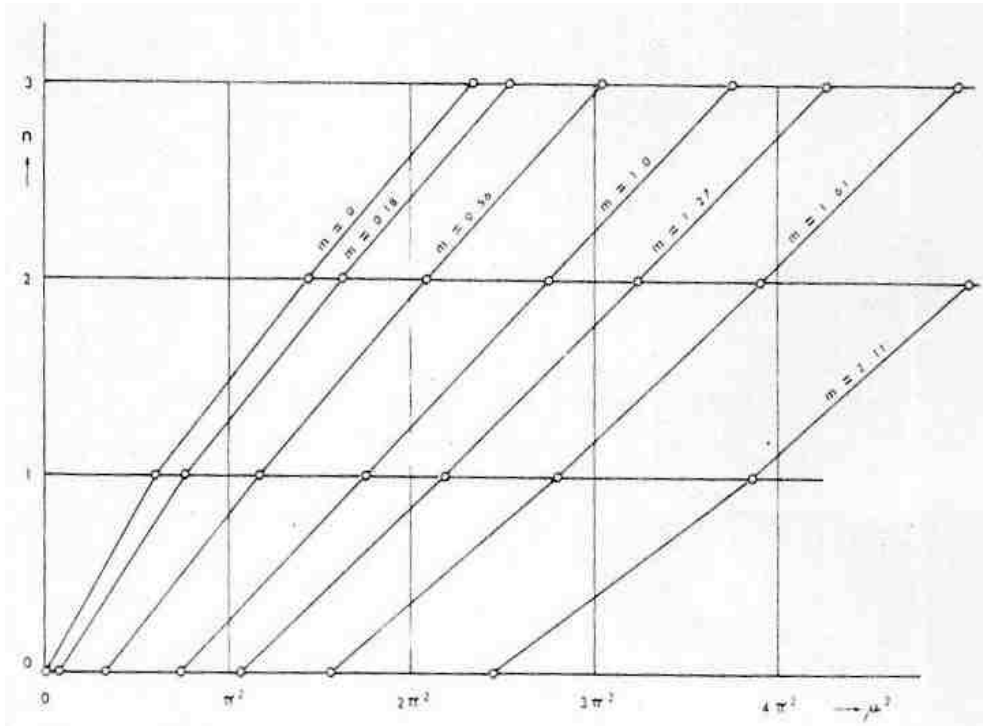

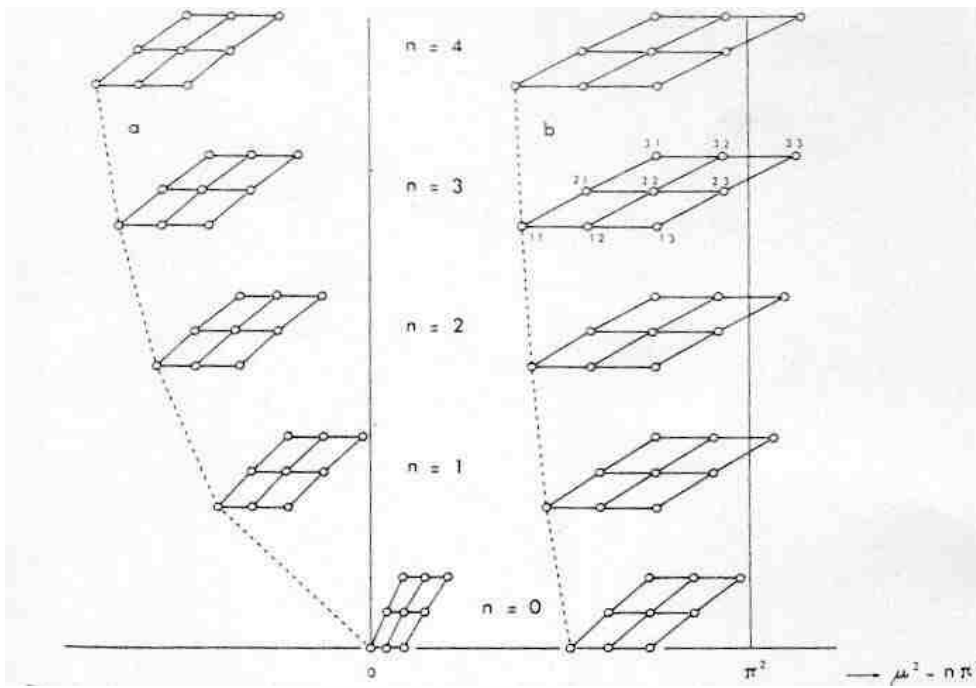

## g) **Dominantly orbital state description (DOS)**

The dominantly orbital state description is applied to the study of light mesons (DOS) [72]. (See also [73] for the DOS applied to heavy-light mesons). In this approach, the orbitally excited states are obtained as a classical result, while the radially excited states can be treated semiclassically. In this work [72] authors consider the case of a quark and an antiquark with finite but different masses.

We consider a system composed of two particles with masses $m_1$ and $m_2$ interacting via a scalar potential $S(r)$ and a vector potential $V(r)$ which depend only on the distance $r$ between the particles. In the center-of-mass laboratory, the classical mass $M$ of the system characterized by a total orbital angular momentum $J$ is given by

$$M\left(r, P, J\right) = \sqrt{P_r^2 + \frac{J^2}{r^r} + \left[m_1 + \alpha_1 S(r)\right]^2} + \sqrt{P_r^2 + \frac{J^2}{r^2} + \left[m_2 + \alpha_2 S(r)\right]^2} + V\left(r\right), \quad (1)$$

where $P_r$ is the radial internal momentum. The parameters $\alpha_1$ and $\alpha_2$ indicate how the scalar potential $S(r)$ is shared among the two masses. These quantities must satisfy the following conditions to ensure a good nonrelativistic limit:

$$\alpha_1 + \alpha_2 = 1 \text{ and } \lim_{m_i \to \infty} \alpha_i = 0 \qquad (2)$$

A natural choice is to take

$$\alpha_1 = \frac{m_2}{m_1 + m_2} \text{ and } \alpha_2 = \frac{m_1}{m_1 + m_2} \qquad (3)$$

The idea of the DOS description is to make a classical approximation by considering uniquely the classical circular orbits, that is to say the lowest energy states with a given $J$ (*yrast lines*). This state is defined by $r = r_0(J)$, and thus $dr/dt = 0$ and $P_r = 0$. Let us denote $M_0(J) = M(r = r_0, P_r = 0, J)$. In order to get the radial excitations, a harmonic approximation around a classical circular orbit is calculated. If the harmonic quantum energy is given by $\Omega(J)$, then the square mass of the system with orbital excitation $J$ and radial excitation $n$ *(0, 1, ...)* is given by

$$M^2\left(J, n\right) = M_0^2\left(J\right) + M_0\left(J\right)\Omega\left(J\right)\left(2n + 1\right) \qquad (4)$$

The long-range part of the interaction between a quark and an antiquark is dominated by the confinement, which is assumed to be linear function of $r$. As its Lorentz structure is not yet determined, we assume that the confinement is partly scalar and partly vector. The importance of each part is fixed by a mixing parameter $f$ whose value is 0 for a pure vector and $1$ for a pure scalar.

The short-range part of the interaction is assumed to be of vector type and given by the usual Coulomb-like potential. Thus we have

$$S\left(r\right) = far \qquad (5)$$

in which a is the usual string tension, whose value should be around *0.2 GeV²*, and

$$V\left(r\right) = \left(1 - f\right)ar - \frac{k}{2}. \qquad (6)$$

After heavy calculations with MATHEMATICA, authors arrived at the following general expression for the meson square mass

$$M^2 = \alpha A\left(f, \beta\right)J + B\left(f, \beta\right)m\sqrt{aJ} + C\left(f, \beta\right)m^2 + aD\left(f, \beta\right)k + aE\left(f, \beta\right)\left(2n + 1\right) + O\left(J^{-1/2}\right) \qquad (7)$$

where *0≤f≤1, 0≤β≤1*, and $m$ is the mass of the heaviest quark. Though coefficients of the formula (7) cannot be obtained analytically, they are calculated for all possible physical situations. The values of coefficients $A$ and $E$ as a function of the mixing parameter f and the asymmetry parameter $\beta$ are given in Figs. 1, [72].

A term proportional to $\sqrt{J}$ *deforms* the RT for a low value of $J$. This term vanishes if *m=0*, that is to say the two quarks are massless, or if $\beta=0$, that is to say the lightest quark is massless.

Finally, let us remark that there is no coupling between orbital and radial motion for large $J$ values (absence of terms n $J$). This is only a consequence of the Coulomb + linear nature of the quark-antiquark potential. This may not be true for other types of potentials.

It remains to be seen, how the predictions of the DOS model will be verified by data.

### h) Durand model

In this paper authors present the results of a rather extensive analysis of the spin-averaged heavy – and light-quark $q\bar{q}$ spectra using the reduced Salpeter equation [74]. Spin-dependent effects are treated in a separate paper. Although they obtain ostensibly reasonable fits to the $b\bar{b}$, $c\bar{c}$ and $s\bar{s}$ data using an interaction containing a short-range Lorentz-vector one-gluon-exchange term and a long-range Lorentz-scalar confining interaction, the results are misleading.

After few approximations authors obtained the standard reduced Salpeter equation:

$$\left(M - \omega_1 - \omega_2\right)\Phi\left(\vec{p}\right) = \frac{1}{\left(2\pi\right)^3}\int d^3p'\Lambda^+\left(\vec{p}\right)\gamma_0 V\left(\vec{P},\vec{P}'\right)\Phi\left(\vec{P}'\right)\gamma_0\Lambda^-\left(-\vec{P}\right) \tag{1}$$

The general features of the $q\bar{q}$ interaction in QCD are well known. The details are not, especially for light quarks. For large $q\bar{q}$ separations, both lattice QCD and hadronic string models predict an asymptotically linear interaction between heavy quarks:

$$V\left(r\right) \approx Br - \frac{\beta}{r} + ... \tag{2}$$

This confining part of the potential is expected to have a Lorentz-scalar structure, a result confirmed by the spin dependence of $V$ found in lattice calculations [75]. The *1/r* or Lüscher term [76] in (2) arises from the transverse zero-point oscillations of the string, and can be identified as Casimir energy. For a standard Nambu string, the most appropriate string to idently with a QCD flux tube, $12\beta/\pi = \frac{1}{2}\left(d-2\right) = 1$ in d=4 dimensions.

Authors use a fairly flexible parametrization of the potentials in fitting the data in $q\bar{q}$ systems, and take the interaction as a sum of scalar and vector terms with

$$V_V\left(r\right) = -\frac{4}{3}\frac{\alpha_S\left(r\right)}{r}e^{-\mu'r} + \delta\left(-\beta/r + Br\right)\left(1 - e^{-\mu r}\right) \tag{3}$$

and

$$V_S\left(r\right) = \left(1-\delta\right)\left(-\beta/r + Br\right)\left(1 - e^{-\mu r}\right) + V_0 + \left(C_0 + C_1 r + C_2 r^2\right)\left(1 - e^{-\mu r}\right)e^{-\mu r} \tag{4}$$

where $\beta=\pi/12$. The vector term incorporates the expected short-distance behavior from single-gluon exchange but with a damping factor $e^{-\mu'r}$ to eliminate this term at large $r$, where $r^{-1}$ dependence is associated with the Lüscher term in $V_s$. They have also included a multiple of the long-range interaction in $V_V$ to see if we can determine the vector-scalar nature of the confining interaction. It is expected, $\delta=0$. $V_S$ includes the expected long-range interaction and a purely phenomenological intermediate-range term. In the parametrization above, $\mu'^{-1}$ acts as a confinement radius, around which the nature of the $q\bar{q}$ interaction changes.

Authors conclude from the fitting the spectra, that Lorentz structure of the confining interaction cannot be determined using the spin-averaged data alone. In Fig. 2, [74] authors show the RT calculated for light-quark ($l\bar{l}$) systems using the full relativistic wave equation with $m_l=200\ MeV$ and scalar confinement ($\delta=0$), and compare the results with the observed RT for the spin-triplet and spin-singlet $l\bar{l}$ mesons. The slopes of the calculated trajectories are strikingly large compared to those observed for the $\rho$, $\omega$, and $\pi$ trajectories and these RT are clearly nonlinear. RT's are more nonlinear at small masses and $L$. The $L=0$ Regge intercept fort the leading trajectory corresponds to a mass of the lowest $1S$ $l\bar{l}$ state of $540\ MeV$, somewhat below the spin average of the $\rho$ and $\pi$ masses, $613\ MeV$. A change in the light-quark mass to $m_l\approx320\ MeV$ removes this discrepancy but still leaves the RT's much too steep, which slopes greater than twice the observed slopes. The large slopes correspond to overly close spacings of masses with increasing L. Thus the first two spacings on the leading trajectory are calculated as $440\ MeV$ and $290\ MeV$, to be compared with the observed spacings $M(a_2)-M(\rho_1)=550\ MeV$ and $M(\rho_3)-M(a_2)=370\ MeV$ on the $\rho$ trajectory. It is customary to determine the asymptotic slope of the scalar confining potential $V_s\approx Br$ for $r$ large by using the string theory result, $B=1/2\pi\alpha'$, where $\alpha'$ is the slope of the (approximately linear) RT. The value $B$ given by the fit, $B\approx0.177\ GeV^2$, agrees essentially exactly with the string model result for the observed slope of the $\rho$ trajectory, $\alpha'\approx0.9\ GeV^2$, but has no relation to the calculated slopes of the $l\bar{l}$ RT's in Fig. 2 (!) They have found, in fact, that there is no value of $B$, reasonable or unreasonable, which will lead to an $l\bar{l}$ Regge slope consistent with experiment. The light-quark systems clearly satisfy stringlike dynamics, but not the relativistic potential dynamics considered here.

The calculated separation of the $1S$ and $2S$ $l\bar{l}$ states of $440\ MeV$ given above continues the trend evident in the $\delta=0$ column in Table I of a decrease in the separation of the $1S$ and $2S$ states with decreasing quark mass, i.e., 591, 576, 552, and 440 $MeV$ for the $b\bar{b}$, $c\bar{c}$, $s\bar{s}$, and $l\bar{l}$ systems. This trend is opposite to that in the spin-averaged $b\bar{b}$, $c\bar{c}$ and $s\bar{s}$ data where the $1S$-$2S$ separation are 577, 595, and 624 $MeV$, hence increasing with decreasing quark mass.

The effect will of course be reduced somewhat when the angular averages which enter the partial-wave projections to states of definite $L$ are taken into account, but will not disappear entirely. The residual effect seems, in fact, to account for the systematic trends discussed in the preceding subsection. As may be seen from Fig. 1, light-quark systems are much more sensitive than heavy-quark systems to the behavior of the interaction at large distances. The progressive weakening (or *flattening)* of the effective long range confining interaction with decreasing quark mass accounts for both the compression of the energy spacings with decreasing quark mass found in these calculations, and the steepness of the $l\bar{l}$ RT. The corresponding strengthening of the vector gluon-exchange interaction at short distances is relatively less important as the lighter quarks are not especially sensitive to this region of the potential.

The result of this work are discouraging with respect to the utility of the reduced Salpeter equation for the description of the light- or strange-quark systems. Authors conclude in fact, that this approach is fundamentally flawed. The problems recited above – the incorrect systematic trends in mass differences, and the drastically incorrect slopes of RT's – are intrinsic to an approach based on the use of the reduced Salpeter equation with static scalar confinement. However, the failure of the model to reproduce the observed "stringy" behavior of the $l\bar{l}$ RT's suggest that the problems would be eliminated in a theory which included the dynamical energy of the color field between quark and antiquark as well as the static field energy represented by the static one-gluon exchange and confining interactions.

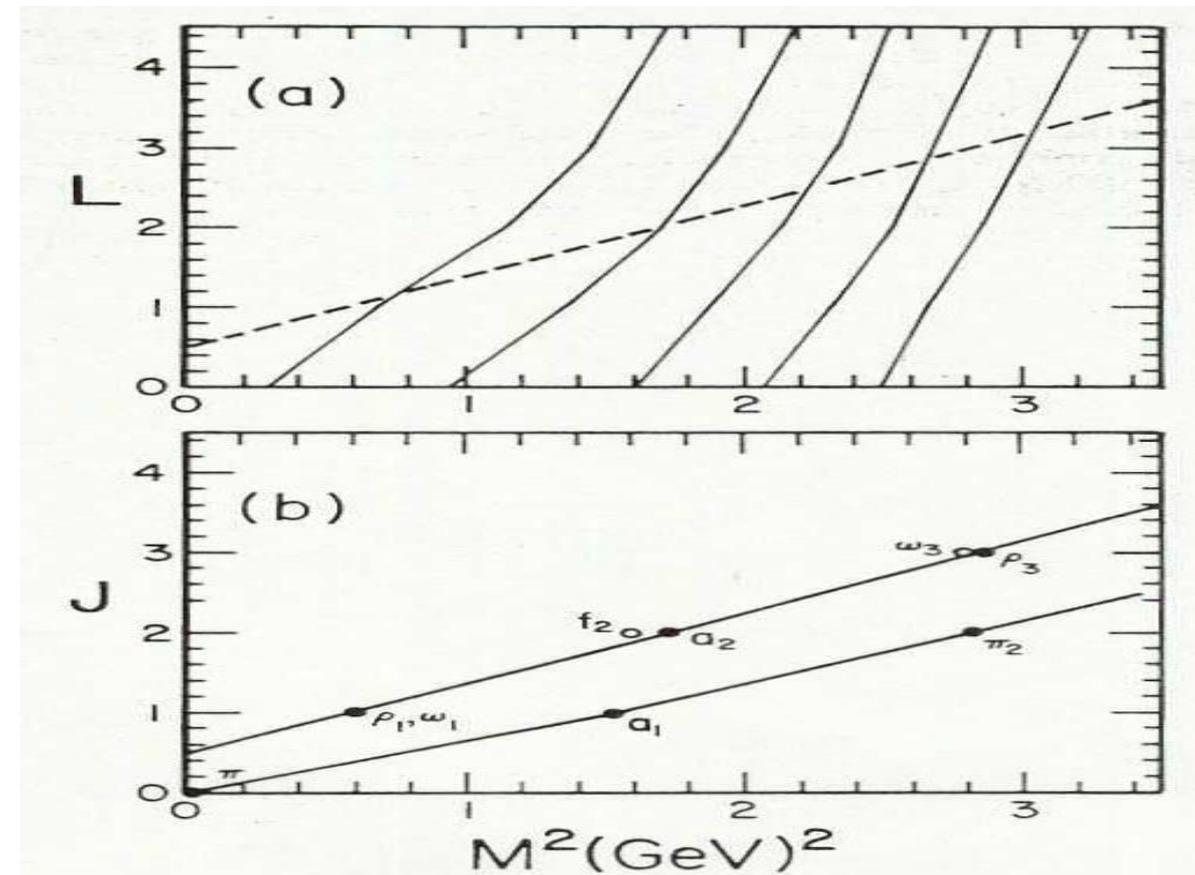

### i) Relativistic model of Martin

In his famous paper [50], Martin considered relativistic model for mesons and baryons. For $q\bar{q}$ author employs the following Hamiltonian ($c=1$)

$$H = |\vec{p}_1| + |\vec{p}_2| + V(r_{1r}),$$ (1)

or, in the c. m. system,

$$H = 2p + V(r)$$ (2)

Martin considers a purely linear potential,

$$V(r) = \lambda r$$ (3)

The Hamiltonian could be rewritten in different form:

$$H = 2\sqrt{\frac{-d^2}{dr^2} + \frac{J(J+1)}{r^2}} + \lambda r$$ (4)

Neglecting the lack of commutativity of the two operators under the square root, we write

$$\sqrt{\frac{J(J+1)}{r^2} - \frac{d^2}{dr^2}} \simeq \frac{\sqrt{J(J+1)}}{r} + \frac{1}{2}\frac{r_{min}}{\sqrt{J(J+1)}}\frac{d^2}{dr^2},$$ (5)

where $r_{min}$ minimizes $2\sqrt{[J(J+1)]}/r + \lambda r$, and making the harmonic oscillator approximation, one get for $\lambda=1$,

$$H \simeq 2[J(J+1)]^{1/4} + \frac{(r-r_{min})^2}{[J(J+1)]^{1/4}} - \frac{1}{2}\frac{1}{[J(J+1)]^{1/4}}\frac{d^2}{dr^2},$$ (6)

and

$$E(n,J) \simeq 2[J(J+1)]^{1/4} + \frac{(n+1/2)\sqrt{2}}{[J(J+1)]^{1/4}}$$ (7)

This ev1idently show that corresponding RT's for $q\bar{q}$ will be *nonlinear* (curved) in the region of low and moderate $J$, corresponding to current resonance energy region. If we take $n=0$, $J=1$, the nonlinear term (second) gives 25% correction to the energy(!)

### j) Nambu-Jona-Lasinio model of Shakin

In two recent works authors [77] used generalized Nambu-Jona-Lasinio (NJL) model to describe a very large number of light meson states [78-79]. The Lagrangian of NJL is

$$L = \overline{q}\left(i\partial - m^0\right)q + \frac{G_s}{2}\sum_{i=0}^{8}\left[\left(\overline{q}_i\lambda^i q\right)^2 + \left(\overline{q}_i\gamma_s\lambda^i q\right)^2\right] - \frac{G_V}{2}\sum_{i=0}^{8}\left[\left(\overline{q}\gamma^\mu\lambda^i q\right)^2 + \left(\overline{q}\gamma^\mu\gamma_s\lambda^i q\right)^2\right] +$$

$$+ \frac{G_D}{2}\{\det[\overline{q}(1+\gamma_s)q] + \det[\overline{q}(1-\gamma_s)q]\} + L_{tensor} + L_{conf} \tag{1}$$

Here, the fourth term is the 't Hooft interaction, $L_{tensor}$ denotes interactions added to study tensor mesons. In this work authors predicted $K_0^*(1730)$ meson. This is in accord with systematic features of light-meson spectra [80] and also with a recent analysis of $S$-wave, $K_\pi$ scattering [81].

A systematic phenomenological analysis of the spectra of light $q\overline{q}$ meson states has been reported in [80]. It was found that the states lay on "trajectories", which were straight lines with the equation $M^2 = M_0^2 + (n-1)\mu^2$, where n is the principal quantum number and $\mu^2$ is a phenomenological parameter, which was in the range $1.10 < \mu^2 < 1.40$ GeV$^2$. A particular meson can have several trajectories. For example in the case of the $\rho$ meson there are two trajectories corresponding to states that are either predominantly $^3S_1$ or $^3D_1$ states. For the $f_0$ mesons, there is a trajectory for the $n\overline{n} = \left(u\overline{u} + d\overline{d}\right)/\sqrt{2}$ states and one for the $s\overline{s}$ states, with the assumption that we have approximate ideal mixing for the $f_0$ states.

Authors [77] calculated RT for $n\overline{n}$ and $s\overline{s}$ states and get two linear RT with slope $\mu^2 = 0.82 GeV^2$, which is smaller than in [80] (see Fig. 3, [77]).

Authors [77] made comparison with pion spectrum Fig. 4. It is clearly seen that data suggests *nonlinear* RT, while authors [77], [80] tried to fit by linear RT with $\mu^2 = 1.39 GeV^2$. Cross deviations from the data for both models are evident.

On Fig.5 authors depicted radial RT for $\rho$ meson. It is clear that both models [77], [80] contradict each other and with data. In particular, results [77] leach to nonlinear RT for upper curve.

In Fig. 6 authors show the radial RT for $a_1(11^{++})$ mesons, NJL model leads to *nonlinear* RT, which deviate from data. Experimental RT for $a_1$ is nonlinear as well.

In Fig. 7 authors show the two trajectories for the $\eta(00^{-+})$ mesons. Here the lowest trajectory corresponds to $n\overline{n}$ states and the upper to $s\overline{s}$ states, where (approximate) ideal mixing is assumed. Clearly the results of [77] and [80] contradict with data and with themselves.

Next, authors [77] consider the $a_0(10^{++})$ and $f_0(00^{++})$ mesons. There is some uncertainly associated with the assignment of $q\overline{q}$ configurations for these mesons. In Fig. 8 authors show the states of the $a_0(10^{++})$ mesons. The solid line has a slope of $\mu^2 = 1.38 GeV^2$ and the dotted line has a slope of $1.25$ GeV$^2$. Both data and NJL clearly leads to *nonlinear* pattern, although there is discreapancy between theory and experiment.

In Fig. 9 authors show the RT for $f_0$ mesons. Upper curve corresponds to $s\bar{s}$ - and lower to $n\bar{n}$ - states. One can see that NJL contradicts to results of [80], and there is no clear correspondence with data. Data clearly indicate *nonlinear* character of this $f_0(00^{++})$ radial RT.

In NJL model the lowest scalar nonet is composed of the $a_0(980)$ and $K_0^*(1430)$, with the $f_0(980)$ and $f_0(1370)$ playing roles that are analogous to those of the $\omega(782)$ and $\varphi(1020)$ mesons of the vector nonet of states that includes the $\rho(770)$ and the $K^*(892)$. To emphasize this analogy authors show the $\omega$ and $\varphi$ RT's in Fig. 10. There is clear discreapance between the NJL and data, both clearly suggest *nonlinear* character of RT's.

From all the results and comparisons we conclude that NJL model for mesons support the *nonlinear* character of mesonic RT's. It's remained to be seen how the NJJL model is applicable to baryonic RT's.

### *k)* __Lagaë model__

Lagaë investigated several models of $q\bar{q}$ binding in light-quark mesons using the Bethe-Salpeter (BS) equation in the instantaneous approximation [82]. These were considered with regard to their ability to reproduce three essential features of the light-quark meson spectroscopy: the existence of light pseudoscalars, the linearity of RT, and the smallness of spin-orbit splittings. This study has led him to reject both the hypothesis of a vector and of a scalar confining potential. Moreover, it has been shown that the BS kernel has to contain a chiral-symmetry-breaking part in order to accommodate the observed spin-orbit splittings. This in turn implies that the BS kernel has to depend on the mass ($M$) of the bound state itself in order to maintain compatibility with the Goldstone theorem.

Under these circumstances he was able to produce a model based on a vector confining potential complemented by an M-dependent reversed spin-orbit coupling which satisfies the three requirements mentioned above. In all these considerations the spontaneous breaking of chiral symmetry plays an essential role not only because of its direct implications on the spectrum but also because it determines the kinetic energy and the quark form factors to be used in the BS equation.

When considering scalar potential, Lagaë clearly obtained nonlinear RT from the BS equation. The trajectories are *curved downward* (see Fig. 4, [82]) and becomes more and more curved as the quark mass is decreased. It is well known that scalar potential is considered now as the most adequate in hadron spectroscopy.

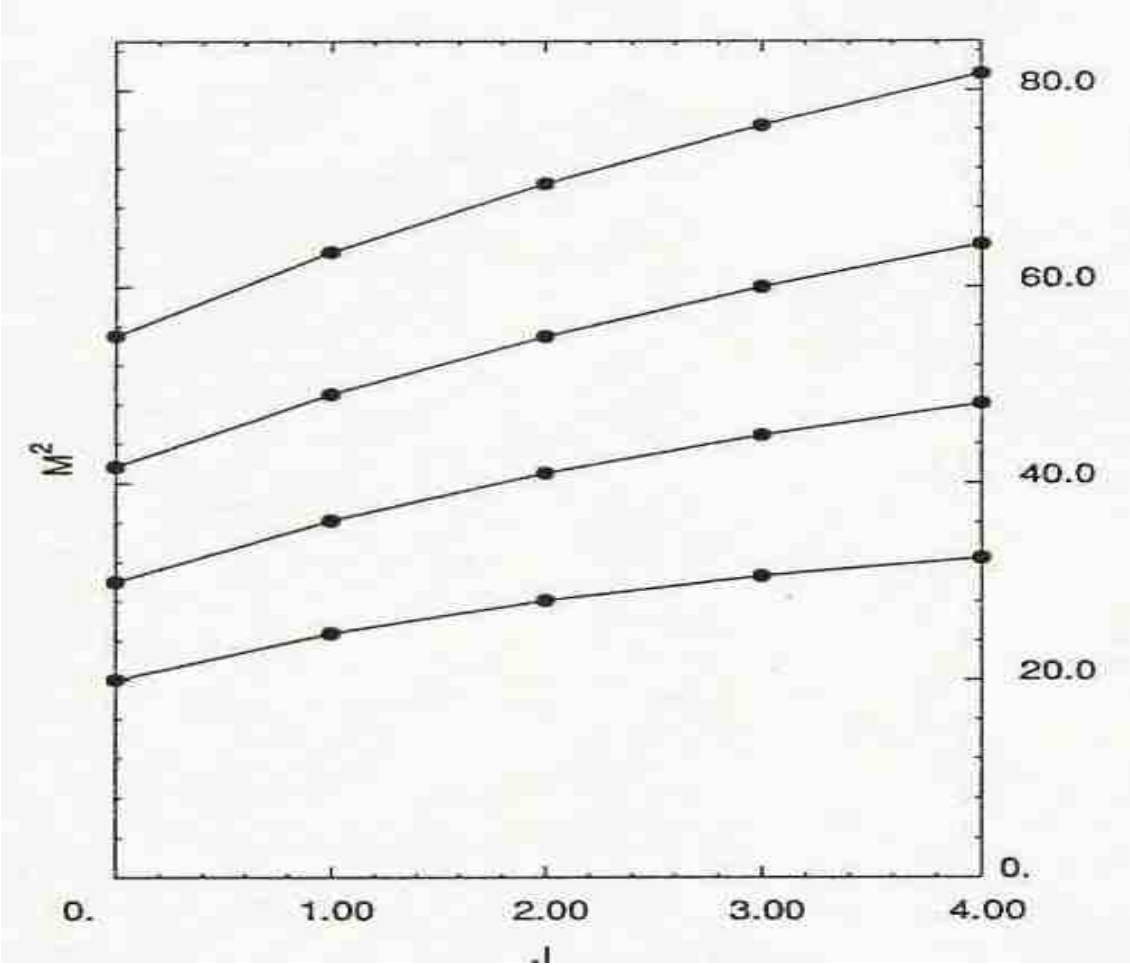

## II.7. WKB-type models.

### a) Bohr-Sommerfeld quantization and meson spectroscopy

It is the purpose of this work to show that the use of a Bohr-Sommerfeld quantization (BSQ) approach can provide unexpected insight into potential models [83]. Authors use this method to obtain some general characteristics of constituent quark models. In particular, he shows that it is possible to obtain analytical expressions for the spectra, root mean square radii, decay widths, electromagnetic mass splittings, or electric polarizabilities, which closely approximate, the numerically exact results obtained from full quantum Schrödinger or spinless Salpeter approaches.

The basic quantities in the BSQ approach are the action variables:

$$J_s = \oint P_s dq_s , \tag{1}$$

where $s$ labels the degrees of freedom of the system, and where $q_s$ and $p_s$ are the coordinates and conjugate momenta; the integration is performed over one cycle of the motion. The action variables are quantized according to the prescription

$$J_s = (n_s + c_s)h \tag{2}$$

where $n_s$ ($\geq 0$) is an integral quantum number and $c_s$ is some real constant, which according to Langer should be taken equal to $1/2$ [84].

The nonrelativistic Hamiltonian corresponding to the Cornell potential reads in natural units ($\hbar=c=1$),

$$H = \frac{1}{2\mu}\left(P_r^2 + \frac{P_\phi^2}{r^2}\right) - \frac{K}{r} + ar \qquad (3)$$

Quantization of $J_\phi$ trivially gives $L=l+C_\phi$; on the other hand, quantization of $J_r$ leads to the equation

$$2\mu r_+(Er_1 + 2k)K(\eta) + 2\mu Er_+ + (r_+ - r_1)E(\eta) - 3(l + c_\phi)^2\Pi(\pi/2, \gamma, \eta) - \frac{3}{2}\pi(n + c_r)r_+\sqrt{r_+ - r_1}\sqrt{2\mu a} = 0 \qquad (4)$$

where $\eta = \dfrac{r_+ - r_-}{r_+ - r_-}$ and $\gamma = 1 - r_- / r_+$, $\qquad (5)$

$K(X)$, $E(X)$, and $\Pi(\pi/2, x, y)$ are the complete elliptic integrals of the first, second, and third kinds, respectively ([85], p. 904), and n is the radial quantum number.

One of the most useful by–products of the BSQ method is to provide simple asymptotic expressions for the total energy for large values of $l$ and $n$. For large angular momenta ($l>>n$), the orbits become almost circular, and thus $r\approx r_+$ and $\eta\approx0$. In this limit, $\Theta=\pi$ and all the elliptic integrals are equal to $\pi/2$; keeping the leading terms in Eq. (4) we obtain

$$E \sim \frac{3}{2}\left(\frac{a^2}{\mu}\right)^{1/3} l^{2/3}, \qquad (l>>n) \qquad (6)$$

As we see, a linear potential lead to *nonlinear* RT.

For large radial quantum numbers ($n>>l$), the orbits have a large *eccentricity*, and thus $r_1\approx r\approx0$ and $\eta\approx1$. Setting $r_1=r_-=0$ into Eq. (1) leads to

$$E \sim \left(\frac{3\pi}{2}\right)^{2/3}\left(\frac{a^2}{2\mu}\right)^{1/3} n^{2/3} \qquad (n>>l) \qquad (7)$$

As we see, resulting RT's will be *nonparallel,* since $E^2 \approx n^{4/3}$ and $E^2\approx0$. These RT's will be dispersed in *n*.

This is a new result: the radial quantum number does not appear explicitly in the classical Hamiltonian, and thus naive semiclassical methods fail to reproduce correctly the spectrum in this sector. Equations (6) and (7) show that the spectrum has the same asymptotic behavior in $l$ and $n$. In fact, these properties remain valid for any power-law confining potential. Indeed for $V(r)\sim ar^\alpha$ $(\alpha>0)$, the large-$l$ behavior is found to be

$$E \sim \left(\frac{a^2}{(2\mu)^\alpha}\right)^{1/(\alpha+2)}\left(1 + \frac{\alpha}{2}\right)\left(\frac{2}{\alpha}\right)^{\alpha/(\alpha+2)} l^{2\alpha/(\alpha+2)}, \qquad (l>>n) \qquad (8)$$

As one can see Eq. (8) generally lead to *nonlinear* RT. For large *n*, the turning points behave as

$$r_+ \sim \left(\frac{E}{a}\right)^{1/\alpha}, \qquad\qquad r_- \sim 0,$$

and Eq. (1) leads to an elementary integral which gives

$$E \sim \left(\frac{a^2}{(2\mu)^\alpha}\right)^{1/(\alpha+2)} \left(\frac{\alpha\pi}{B(1/\alpha,3/2)}n\right)^{2\alpha/(\alpha+2)} \qquad\qquad (n>>l),$$

(9)

where $B(x, y)$ denotes the beta function ($p[85]$, p. 948). Since generally $E^2 \not\approx n$, appropriate RT's will be dispersed in n, except for $\alpha=2/3$.

The semirelativistic Hamiltonian corresponding to the nonrelativistic Hamiltonian (3) reads

$$H = 2\sqrt{P_r^2 + \frac{P_\phi^2}{r^2} + m^2} - \frac{K}{r} + ar \qquad\qquad (10)$$

After lengthy calculations for confining potential $V(r)=ar^\alpha$ $(\alpha>0)$ one can get

$$M \sim a^{1/(\alpha+1)}(\alpha+1)\left(\frac{2l}{\alpha}\right)^{\alpha/(\alpha+1)} \qquad\qquad (l>>n) \qquad\qquad (11)$$

It is easy to see that appropriate RT's from (11) are generally *nonlinear*, except for $\alpha=1$.

For large n the turning points have the forms

$$r_+ \sim \left(\frac{M}{a}\right)^{1/\alpha}, \qquad\qquad r_- \sim 0 \qquad\qquad (12)$$

and Eq. (1) leads to an elementary integral which gives

$$M \sim a^{1/(\alpha+1)}\left(\frac{2\pi(\alpha+1)}{\alpha}n\right)^{\alpha/(\alpha+1)} \qquad\qquad (n>>l) \qquad\qquad (13)$$

We see that expect for $\alpha=1$, RT's will be nonparallel and dispersed in *n*.

One-dimensional BSQ approach can be used to derive some simple formulas which can be applied to the three-dimensional *l=0* states. For a central potential, odd states of the one-dimensional SE remain solutions of the three-dimensional *l=0* SE (considering only the $x\geq0$ part of the *x* axis). In the nonrelativistic case, we have

$$P = \pm\sqrt{2\mu[E - V(x)]} \qquad\qquad (14)$$

With $V(x) = a|x|^\alpha$, the quantization of the action variable *J* leads to an equation which can be solved for the energy:

$$E = \left(\frac{a^2}{(2\mu)^\alpha}\right)^{1/(\alpha+2)} \left(\frac{\alpha\pi}{B(1/\alpha,3/2)}(n+3/4)\right)^{2\alpha/(\alpha+2)} \qquad\qquad , \qquad\qquad (15)$$

where authors has changed *n+1/2* into *(2n+1)+1/2* to only take into account odd states. This formula is different from Eq. (13) and again leads to generally nonlinear RT's, except for *α=2/3*.

Otherwise all RT's will be nonparallel and dispersed in n. Eq. (15) is very different from all previous formulas, because it should be valid for any *n*, and not just for large *n*. Eq. (15) shows that the energy behaves with the reduced mass as $E \propto \mu^{-\alpha/(\alpha+2)}$. It means that for all physically interesting cases (*α>0*) RT's will be decreasing functions of quark mass.

### b) **Fokker-Type confinement models of Duviryak**

It is well known that nonrelativistic potential model with the linear potential leads to the RT with the unsatisfactory asymptote $M \sim S^{2/3}$. Usually the RT's in the potential models are calculated in the oscillator approximation. Then the leading RT originates from the classical mechanics: it is close to the curve of circular motions on the ($M^2$, $S$) – plane [86].

Let us compare the classical RT of purely confinement time-asymmetric model to that which follows from the time-symmetric Fokker-type confinement model with the same parameters $m_1=m_2=m$ and *β*. In the nonrelativistic limit both models lead to the same relation:

$$M_c - 2m \approx 3\left(\frac{\beta S}{2\sqrt{m}}\right)^{2/3} \tag{1}$$

which is known from the nonrelativistic linear confinement model. Classical RT from the general time-asymmetric model as one from time-symmetric model are shown in Fig. 4 [86]. One can see a *big family of nonlinear RT's*.

These purely classical results give us the base to consideration of semiclassical quantization of the model. By analogy with WKB approximation method we put

$$S = \hbar\left(l + 1/2\right), \qquad l=0,\ 1,\ \dots \tag{2}$$

for the quantized internal momentum, and

$$\oint k_r dr = 2\pi\hbar\left(n_r + 1/2\right), \quad n_r=0,\ 1,\ \dots \tag{3}$$

for radial excitations; the integral runs over the classical phase trajectory.

In the case of purely confinement model we have

$$\int_{r_1}^{r_2} dr\sqrt{f\left(r, M, S\right)} = \pi\hbar\left(n_r + 1/2\right) \tag{4}$$

Using the oscillator approximation we expand the function *f(r, M, S)* about the circular orbit to first order in *ΔM=M-M_c* and *Δr=r-r_c*. The result is as follows:

$$f\left(r, M, S\right) \approx a^2\left(M, S\right) - b^2\left(S\right)\left(\Delta r\right)^2 \tag{5}$$

where $a^2(M,S) \equiv \dfrac{\partial f(r,M,S)}{\partial M}\Big|_c \Delta M = \dfrac{M_c\left(M_c^2 + 2m^2\right)}{\left(M_c^2 + 8m^2\right)} \Delta M$ $\qquad$ (6)

$$b^2(S) \equiv -\dfrac{\partial^2 f(r,M,S)}{2\partial r^2}\Big|_c = \dfrac{27\beta^2 M_c^4}{4\left(M_c^2 - 4m^2\right)\left(M_c^2 + 8m^2\right)} \qquad (7)$$

Then the integral in (4) is easily calculated:

$$\int_{-a/b}^{a/b} d(\Delta r)\sqrt{a^2 - b^2(\Delta r)^2} = \dfrac{\pi}{2}\dfrac{a^2}{b} \qquad (8)$$

Using (4)-(8) and assuming that $\Delta M$ is small compared to $M_c$ we obtain for $M^2$ the expression:

$$M^2 = M_c^2\left\{1 + \dfrac{6\sqrt{3}\beta\hbar}{M_c^2 + 2m^2}\dfrac{M_c^2 + 8m^2}{M_c^2 - 4m^2}\left(n_r + 1/2\right)\right\} \qquad (9)$$

which describes the leading and daughter RT's.

Similarly, in the case of general model we obtain the RT's determined in the implicit form by the equations

$$M^2 = M_c^2(\lambda)\left\{1 + \dfrac{\hbar}{\alpha}\Phi(\lambda)\left(n_r + 1/2\right)\right\} \qquad (10)$$

$$\Phi(\lambda) = \dfrac{\sqrt{\left[1 + 2\lambda + \nu(1+\nu)^2\right]\left[3 + (1+\nu)(1+\lambda)^2 + 3\nu(1+\lambda)^4\right]}}{(1+\lambda)^2\left[1 + \nu(1+\lambda)\right]\sqrt{\lambda(2+\lambda)}} \qquad (11)$$

and (2). Only at large l these RT's reduce to linear ones

$$M^2 \approx 6\sqrt{3}\beta\hbar\left(l + n_r + 1\right) + 6\left(m^2 - 3\alpha\beta\right) \qquad (12)$$

so that the daughters are parallel to leading trajectory. More over, states of unit internal momentum differences form into *degenerate towers* at a given mass. This tower structure is of interest for the meson spectroscopy. The number of relativistic potential models based on single-particle wave equation as well as two-particle models with oscillator interaction lead to degeneracy of $l + 2n_r$ type, but not $l + n_r$ type. The latter cannot be reproduced by single – particle relativistic models with the vector and scalar potentials. Figures 5 and 6, [86] present two examples of semiclassical RT's which are characteristic for heavy and light mesons respectively. Trajectories in Figs. 5(a) and 6(a) are calculated in the oscillator approximation which is good for $n_r \ll l$. These RT's are clearly *nonlinear*. It remains to be seen how considered model describe real resonance data.

## II.8. Cylindrically deformed quark bag model

Greiner et al consider the problem of constructing cylindrically deformed (CD) MIT quark bag solutions [87]. The motivations for considering quark bags of cylindrical geometry are several. Firstly, it is ideally suited to exhibit the string limit of the bag model. Secondly, one can hopefully recover the elegant result of Johnson and Thorn [88] (string limit of the bag model yields the asymptotic RT) – also in the region of low angular momenta. Thirdly, it is plausible that in high energy collisions hadrons are formed first as a "fire sausage" [89] which might then subsequently decay into more stable states. Therefore it is an interesting question to ask, whether the MIT bag model does have solutions corresponding to such geometries relevant at high excitations energies. Fourthly, the recent phenomenological model proposed by Mulders is an interpolation between the spherical bag model (SBM) and the cylindrical bag model (CBM) [88] and success implies that cylindrical bag structures (CBS) are possible not only for the usual hadrons, but also for the more exotic multiquark systems.

It is impossible to obtain physically interesting solutions for the cylindrical MIT bag, if one impose the linear confinement condition on all the surfaces of the cylinder. Authors use the condition $\Psi\overline{\Psi} = 0$ on the edges. This Lorentz-invariant quadratic condition says that the quark mass outside the bag is infinitely large.

Baryonic energy in the MIT bag model is given by

$$E_B = \sum_{\substack{i=occ \\ \text{mod } \epsilon}} \left[ m^2 c^4 + \frac{\hbar^2 c^2}{R^2} \left( (\mu R)_i^2 + \frac{(kL)_i^2}{x^2} \right) \right]^{1/2} + B \pi R^3 x - \sqrt{2} \frac{z_0}{R} \left( 1 + \frac{x^2}{4} \right)^{1/2} \qquad (1)$$

The size of the bag is specified by two parameters, its radius $R$ and its length $L$. In the limit of massless quarks it is more convenient to introduce, in place of $L$, the ratio of length to radius, $X=L/R$. The baryonic energy now depends on $R$ and $X$, and the size of the bag is determined by minimizing the energy with respect to $R$ and $X$. It turns out that the ground state of the cylindrical bag – determined without the zero-point energy term – corresponds to a value of $X_0\cong2$, and $R_0=0.9fm$. For a cylinder of this size it is possible to exactly inscribe a sphere of radius $R_0$ within it.

It is clear that for every eigenvalue of the radial momentum $\mu_{ni}^{(\alpha)}$, one has an *infinite* number of discrete allowed values of the *axial* momentum $K_{nj}^{(\alpha,i)}$.

In Fig. 1 Greiner shows how the lowest six eigenvalues of both parity types of solutions vary with $X$ for the lowest radial momentum excitation: $\mu R_{01}^{(+)} = 1.435$. It is seen that when $X\rightarrow\infty$, the first and second, third and fourth, fifth and sixth eigenvalues *coalesce* – for solutions of both parities. For even parity solutions the confluence of pairs of successive eigenvalues occurs at $kL\cot\dfrac{kL}{2} = 0$, i.e., when $kL=\pi,\ 3\pi,\ 5\pi,\ \dots$ For odd parity solutions on the other hand, the *confluence* occurs when $kL\cot\dfrac{kL}{2} = \infty$, i.e., when $kL=2\pi,\ 4\pi,\ 6\pi,\ \dots$ Figure 1 indeed confirms this behaviour.

The CBS are not eigenstates of the total $\vec{J}^2$, so that the comparison of these bag states with states of hadronic resonances is rendered difficult. However, to get a rough idea, Greiner calculates the mean value of the square of the orbital angular momentum for each of the baryonic states, and, after making the quantum correction $\left\langle \Psi \middle| \vec{L}^2 \middle| \Psi \right\rangle = \hbar^2 \left( l + 1/2 \right)^2$, obtain the mean angular momentum $l$ of the baryon. The energies of the even-parity baryonic states obtained by filling the lowest even-parity models of the bag are shown in Fig. 2, [87]. The stability of the energies is at once obvious. The difference between these and the corresponding energies of the MIT CB with rounded corners is not large, in particular for extremely string-like cylinders (see also Fig. 10).

Because these bag states are degenerate with respect to the sign of $J_z(=\hbar/2)$ as well as the $I$-spin these three quark states may be interpreted either as $N$ or as $\Delta$ states. Reasonably good agreement is obtained with the observed $N$ and $\Delta$ resonances, as seen from Fig. 3, [87]. The trajectories corresponding to different radial excitation are shown in Fig. 3. In the case of the RT's for the odd-parity $N$ states, the slope of the lowest band is bound to be close to the observed value. The energies of the set of even and odd parity states corresponding to radial excitations have profiles similar to that in Fig. 2. The leading RT in Fig. 3 shows an *upward bending* trend at higher excitations. It is reasonable to think that this has its origin in the energy minimas of Fig. 2 becoming less deep with the excitation energy.

One can see from Fig. 3 that RT's corresponding to different radial model have significantly different ($\alpha_R$=0.8, 0.77 and 0.4 GeV$^{-2}$).

The intercept of the present RT's do not agree with the experimental data. A possible reason for this could be that the present geometry is incapable of reaching the zero-angular momentum spherical limit. The baryonic states that lie on a given RT are single particle excitations built up on the ground state in which three quarks occupy the lowest bag mode. The states among this sequence which correspond to higher excitations correspond to cylinders that are longer and thinner. Thus, if one studied very high excitations, the bag energy would reach a minimum for exceedingly large lengths and short radii. This will be the string limit of the bag model. In Table 1 [87] authors show how the cylindrical bag solutions become stringlike at *very high excitations*.

At this stage of this model of non-interacting quarks, authors are unable to obtain the RT's for mesons. When the constant $Z_0$ is adjusted the slope for mesonic systems does decrease and approach 0.9, the experimental value, however simultaneously the good baryonic slope gets vitiated. Authors hope to overcome this deficiency by introducing quark-quark interactions.

## II.9. Phenomenological, and analytic interpolating models

## a) Sergeenko model

There exists a *conviction*, that the RT's $\alpha(t)$ of light flavor mesons are linear in a *whole region*, that is, not only in the *bound-state* region *(t>0)*, but in the *scattering* region *(t<0)* too. However in the experiment [92] far more complicated behavior of the $\rho$-meson trajectory, $\alpha_\rho(t)$, was discovered.

Presented in [92] experimental data on inclusive $\pi^0$ and $\eta$ production in GeV/c $\pi^\pm p$ collisions cover the kinematic region $0 \leq -t \leq 4$ *(GeV/c)*$^2$ and *x≥0.7* and have compared in detail with the predictions of triple Regge theory [93]. So far as there reactions are theoretically clean with $\rho$ ($\pi^0$ production) or $a_2$ ($\eta$ production) exchange there were extracted the RT in the *t* range of *0* to *–4 (GeV/c)*$^2$. A sample of high *–t*, *-t≤4 (GeV/c)*$^2$, has been fitted by the $\rho\rho P$ term, given by [93]

$$\frac{d^2\sigma}{dxdt} = G_\rho(t)(1-x)^{1-2\alpha_\rho(t)} \tag{1}$$

where the pomeron intercept, $\alpha_\rho(0)=1$, $G_\rho(t)$ is the residue function. There was shown that the $\rho$ trajectory *flattens off* at about *–0.6*. The uncertainty in this asymptote can be estimated by fitting in the region *0.81≤x≤0.98* which changes it up by *0.1* to $\alpha_\rho=-0.5$. This value $\alpha_\rho$, $\alpha_\rho=-0.5$, implies that the cross section behaves like *(1-x)*$^2$.

The constituent interchange model CIM [94] predicts a *leveling off* of $\alpha_\rho$ at *–1* or a *(1-x)*$^3$ cross section behaviour. However the exact value of $\alpha(t)$ in [92] is sensitive to the definition of *x*, where we have used *x* as the lab energy divided by the maximum possible energy at the given *t* value. Changing the definition so that the denominator is just the beam energy would decrease the fitted $\alpha$ by about *0.2* at *–t=4 (GeV/c)*$^2$. This means, that the $\rho$-trajectory *flattens off* at *–0.8* or lower, that is, the trajectory *level off* so that the Regge exchanges are the hard-scattering terms [92].

In this paper [91] Sergeenko derive an analytic expression for the RT's in the whole region,-$\infty<t<\infty$. Usually, the RT's of different hadrons are derived (in the framework of the potential models) for the bound state region, that is, at $t=E^2>0$. But for many purposes, for example, in the recombination and fragmentation models and other Regge model it is necessary to know the RT in the scattering region, at *t<0*, and in particular, the intercepts $\alpha(0)$ and slopes $\alpha'$ of RT.

In order to obtain the RT in a whole region it is necessary to know an analytic expression for the square of the total energy of quarkonium state, $E^2$, as a function of radial *n'*, and orbital *l*, quantum numbers. Then, if we invert $E^2(l)$ and express the angular momentum $l(E^2)$ as a function of $E^2$ we obtain RT. In this work author use only the fact, that the interquark potential has two asymptotics:

1)   $V(r) \propto -1/r$ at $r \rightarrow 0$ (Coulomb-like behavior motivated by the one-gluon exchange at small distances) and

2)      $V(r) \propto r$ at $r \to \infty$ (linear confining behavior which follows from lattice-gauge-theory computations).

The most reasonable possibility to construct an interquark potential, which satisfies both of the above constraints, is to simply add these two contributions. This leads to the so-called funnel-shaped (or Cornell) potential:

$$V(r) = -\frac{4}{3}\frac{\alpha_s}{r} + kr + V_0 \qquad (2)$$

It is known that heavy $Q\overline{Q}$ systems one can consider nonrelativictically. For low-lying states of heavy $Q\overline{Q}$ the main contribution to the energy comes from one-gluon exchange and in the first approximation one can neglect the confining term. In this case Schrödinger equation (SE) has exact analytical (hydrogen-like) solution and for energy eigenvalues we have approximately:

$$E'_n = -\frac{\widetilde{\alpha}_s^2 m}{4(n'+l+1)^2} + V_0 \qquad (3)$$

where $\widetilde{\alpha}_s = \frac{4}{3}\alpha_s$, m is the quark mass. The total energy of heavy quarkonium one can write in the form:

$E=2m+E'$. For the square of the energy in nonrelativistic approximation one have $E^2 \simeq 4m^2 + 4mE'$. Therefore, with the help of (3) we obtain for $E_n^2$:

$$E_n^2 \simeq -\frac{\widetilde{\alpha}_s^2 m^2}{(n'+l+1)^2} + 4m(m+V_0) \qquad (4)$$

The formula (4) is good to describe the lower states of heavy quarkonia, especially the bottomonium.

Now let us consider an extreme relativistic limit for higer excied states ($l, n \gg 1$) of light mesons. To describe bound states consisting of light quarks let us consider a static Klein-Gordon equation (KGE) of motion in which the potential has a Lorentz vector $V(r)$ and Lorentz scalar $S(r)$ parts:

$$\left[ \nabla^2 + \frac{1}{4}(E - V(r))^2 - (m + S(r))^2 \right]\Psi(\vec{r}) = 0 \qquad (5)$$

where the functions $V(r)$ and $S(r)$ we have chosen in the form:

$$V(r) = -\frac{\widetilde{\alpha}_s}{r}(1-c) + kr(1-d), \quad S(r) = -\frac{\widetilde{\alpha}_s}{r}c + krd \qquad (6)$$

Here in (6) c and d are the parameters.

Let us consider (4) for higher excited states. For large angular momenta, $l \gg 1$, one may expect that the bound states will feel only the confining part of the potential. We thus assume it is justified to ignore the Coulomb term. In this limit the WKB approximation can be used. For radial part of (4) we have:

$$\left[\frac{d^2}{dr^2} + \frac{1}{4}\left(E + \frac{\alpha_1}{r} - k_1 r\right)^2 - \left(m - \frac{\alpha_2}{r} + k_2 r\right)^2 - \frac{(l+1/2)^2}{r^2}\right]R(r) = 0 \qquad (7)$$

where $\alpha_1 = \tilde{\alpha}_s(1-c)$, $k_1 = k(1-d)$, $\alpha_2 = \tilde{\alpha}_s$ $k_2 = kd$, and the replacement $l(l+1) \rightarrow (l+1/2)^2$ has been made in accordance with the WKB method. The WKB quantization condition corresponding to (7) gives:

$$\int_{r_1}^{r_2} \sqrt{\frac{1}{4}\left(E + \frac{\alpha_1}{r} - k_1 r\right)^2 - \left(m - \frac{\alpha_2}{r} + k_2 r\right)^2 - \frac{(l+1/2)^2}{r^2}} = \pi(n' + 1/2), \qquad (8)$$

where $n'=0, 1, 2, ..., r_1$ and $r_2$ are the classical turning points. At $l >> 1$ the first turning point, $r_1$ is determined mainly by the term $\sim r^{-2}$ and the second one, $r_2$, by the quadratic term $\sim r^2$. Therefore, in this approximation we have for integral (8):

$$\frac{\pi}{2}\frac{E^2/4 - m^2 - (\alpha_1 k_1 - 4\alpha_2 k_2)/2}{\sqrt{4k_2^2 - k_1^2}} - \frac{\pi}{2}\sqrt{\alpha_2^2 - \frac{\alpha_1^2}{4} + (l+1/2)^2} = \pi(n' + 1/2) \qquad (9)$$

and for the quarkonium squared mass, $E_n^2$, at $l >> 1$ this gives

$$E_n^2 = 8k[a(2n' + l + 3/2) - b\tilde{\alpha}_s] + 4m^2 \qquad (10)$$

where $a = \sqrt{d^2 - (1-d)^2/4}$, $b = cd - (1-c)(1-d)/4$.

Note, that the value $a$ is real if $1/3 < d < -1$.

The RT's given by (10) is very similar to that of a harmonic oscillator-type Hamiltonian and good to describe the light-flavor mesons with $k \simeq 0.15$ GeV$^2$, $a \simeq 1$, $b \simeq 1$.

The energy spectrum in (4) lies in the region $E_n < 2m$ and the energy spectrum of the (10) lies in the region $E_n > 2m$.

It is known that an angular momentum in the Regge phenomenology is considered as an analytic function in the complex plane. We choose the leading RT's associated with $s=1$ $q\bar{q}$ states, with total angular $J=l+1$ and $n'=0$, and the square of the momentum as the function of continuous variable $z=l+1$:

$$P^2(z) = E_n^2 - 4m(m + V_0) \qquad (11)$$

Now let us suppore that (4) gives the asymptotic for the function $P^2(z)$ at small $z$, that is,

$$P^2(z) \simeq -\frac{\tilde{\alpha}_s^2 m^2}{z^2}, \quad z \rightarrow 0 \qquad (12)$$

Then (10) gives another asymptotic, but for large $z$:

$$P^2(z) \simeq 8kaz, \qquad z \rightarrow \infty \qquad (13)$$

The question arises: what is the possibility to construct an approximate formula for the $P^2(z)$ which satisfies both of these constraints?

For this let us consider the two-point Pade approximant

$$\left[K/N\right]_{f(z)} = \frac{\sum_{i=0}^{K} a_i z^i}{\sum_{j=0}^{N} b_j z^j} \tag{14}$$

To obtain the asymptotics (12), (13) let us choose $K=3$, $N=2$ in (14). Now it is easy to see that the Pade approximant (14) satisfies (12), (13), if: $a_0 = -\widetilde{\alpha}_s^2 m^2$, $a_1=0$, $a_2=0$, $a_3=8ka$, $b_0=0$, $b_1=0$, $b_2=1$.

Therefore, with the help of relation (11) we obtain the following interpolating formula for $E_n^2$:

$$E_n^2 = 8k\left[a(2n'+l+3/2) - b\widetilde{\alpha}_s\right] - \frac{\widetilde{\alpha}_s^2 m^2}{(n'+l+1)^2} + 4m^2 \tag{15}$$

Formula (15) reproduces the mass spectra of light and heavy quarkonia well if $c\simeq1$, $d\simeq1$ ($a\simeq1$, $b\simeq1$) (see Table 1). This means that the potential is a Lorentz scalar in an extreme relativistic limit for higher excited states. Consequently one can write (15) in the following simple form:

$$E_n^2 = 8k\left(2n'+l+3/2 - \widetilde{\alpha}_s\right) - \frac{\widetilde{\alpha}_s^2 m^2}{(n'+l+1)^2} + 4m^2 \tag{16}$$

Let us transform (16) into an equation for the $l$:

$$l^3 + c_z\left(E^2\right)l^2 + c_2\left(E^2\right)l + c_3\left(E^2\right) = 0 \tag{17}$$

where

$$c_1\left(E^2\right) = 4n' + 7/2 - \widetilde{\alpha}_s + \lambda\left(E^2\right),$$

$$c_2\left(E^2\right) = (n'+1)^2 + 2(n'+1)\left[2n'+3/2 - \widetilde{\alpha}_s + \lambda\left(E^2\right)\right],$$

$$c_3\left(E^2\right) = (n'+1)^2\left[2n'+3/2 - \widetilde{\alpha}_s + \lambda\left(E^2\right)\right] - \widetilde{\alpha}_s^2 m^2/(8k),$$

$$\lambda\left(E^2\right) = \left(4m^2 - E^2\right)/(8k).$$

Replacing $E^2$ by new variable $t$, let's consider the cubic equation (17) in the whole region $-\infty < t < \infty$. The investigation of solutions of this equation shows that at $Q(t)>0$ there is one real solution of the form:

$$\alpha(t) = \sqrt[3]{-q(t)/2 + \sqrt{Q(t)}} + \sqrt[3]{-q(t)/2 - \sqrt{Q(t)}} - c_1(t)/3, \quad Q(t)>0 \tag{18}$$

where $Q(t) = P^3(t)/27 + q^2(t)/4$,

$$P(t) = -c_1^2(t)/3 + c_2(t),$$

$$q(t) = 2c_1^3(t)/27 - c_1(t)c_2(t)/3 + c_3(t).$$

On the other hand, at $Q(t)\leq 0$ there are three real solutions. Only first of these solutions,

$$\alpha(t) = 2\sqrt{-p(t)/3}\,\cos\big[\beta(t)/3\big] - c_1(t)/3\,, \qquad Q(t)<0 \qquad (19)$$

where $\beta(t) = \arccos\big[-q(t)\big/\big(2\sqrt{-p^3(t)/27}\big)\big]$, smoothly go over at $Q(t)=0$ into solution (18). Therefore two functions (18), (19) produce the RT in the whole region $-\infty<t<\infty$.

The RT's (18), (19) are linear at $t\to\infty$ with the universal slope $\alpha'=(8k)^{-1}$ and *flatten off* at $t\simeq-1$ (see Figs. 1, 2). From figure *1* one can see that $\rho$ and $\varphi$ RT's intersect in scattering region ($t\approx -0.4$ *(GeV/c)²*). $\Psi$ and $\Upsilon$ RT's which are totally different for bound states, merge in the scattering region at $t\approx 15$ *(GeV/c)²* (see Fig. 2).

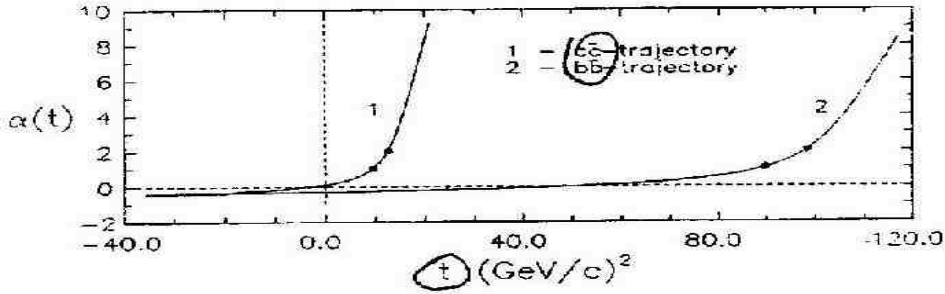

The first derivative, $\alpha'(t)$, is positive in the whole region $-\infty<t<\infty$. The intercepts $\alpha(0)$, and slopes $\alpha'$ of quarkonium RT's are the following:

$$\alpha_\rho(0)\simeq 0.52 \qquad \alpha'_\rho\simeq 0.79\ \text{GeV}^{-2}$$

$$\alpha_{ss}(0)\simeq 0.43 \qquad \alpha'_{ss}\simeq 0.43\ \text{GeV}^{-2} \qquad\qquad (20)$$

$$\alpha_{c\bar c}(0)\simeq 0.10 \qquad \alpha'_{c\bar c}\simeq 0.04\ \text{GeV}^{-2}$$

$$\alpha_{b\bar b}(0)\simeq -0.31 \qquad \alpha'_{b\bar b}\simeq 0.003\ \text{GeV}^{-2}$$

This trend of decreasing slopes with increasing quark mass resembled the results of our potential model, described earlier [34-39]. It remains to be seen how this model will be applicable to baryons.

### b) Filipponi model

Recently Filipponi et al [95] develop an approach to RT's to answer the generic question: How do mesonic masses depend upon the quark masses?

Authors made two crucial simplifying assumptions to obtain their formula:

i)     All trajectories were assumed linear in *(mass)²* of the hadronic state,

$$\alpha(s) = \alpha(0) + s\,\alpha' \qquad\qquad (1)$$

ii)     The functional dependence of $\alpha(0)$ and $\alpha'$ on quark masses is via *($m_1+m_2$)*.

After analysing the Review of Particles Properties and some theoretical models, Filipponi obtain the following (purely phenomenological) simple formula for slopes:

$$\alpha'(m_1 + m_2) = \frac{0.9 GeV^{-2}}{\left[1 + 0.22\left(\dfrac{m_1 + m_2}{GeV}\right)^{3/2}\right]} \qquad (2)$$

where $m_1$ and $m_2$ are the corresponding constituent quark masses for that trajectory. The comparison between Eq. (2) and input data can be seen in Fig. 1.

A similar analysis was performed for the intercept $\alpha_l(0)$, where the subscript refers to the leading trajectory. Authors consider only mesonic systems for which the lowest physical state is at $J=1$. A global description for these is given by

$$\alpha_I(m_1 + m_2; 0) = 0.57 - \frac{(m_1 + m_2)}{GeV} \qquad (3)$$

A comparison of Eq. (3) with input data is shown in Fig. 2. Two points from a theoretical analysis [96] for the $B_c$ system fall quite below the curve. Next authors consider secondary RT's and splitting between the energy levels for the same $J$ of a given system. Calling $\alpha_{I,II}(0)$ the leading and the secondary RT intercept, they estimated the "distance"

$$\Delta\alpha(0) = \alpha_I(0) - \alpha_{II}(0) \qquad (4)$$

from the data and phenomenological models. A rather loose bound was found

$$1.3 < \Delta\alpha(0) < 1.6 \qquad (5)$$

The result becomes more interesting when transposed in terms of the energy splitting

$$\Delta E_J^{II-I} = \left(E_J^{II} - E_J^{II}\right) \qquad (6)$$

between states of the same angular momentum of a given system. Quite strikingly, it is found to be a constant (between *0.5-0.8 GeV*) for all systems (composed of *u, d, s, c,* and *b* quarks). It is shown in Fig.3. Combining Eqs. (2) and (3) authors get an expression for the leading mesonic RT:

$$\alpha(m_1 + m_2; t) = 0.57 - \frac{(m_1 + m_2)}{GeV} + \frac{0.9 GeV^{-2}}{1 + 0.22\left(\dfrac{m_1 + m_2}{GeV}\right)^{3/2}} t \qquad (7)$$

Equation (7) allows us to make "predictions" (or consistency checks) about the energy spectra of excited mesons for the *D, D_S* and *B, B_S* systems, which were not used as input data. In Fig. 6 [97] the leading trajectories for all existing flavours are plotted to show the *intersection* region. Filliponi et al also considered RT's in scattering region [98].

This model [95-98] was seriously critisized by Goldman and Burakovsky [99]. Authors [99] noted that formula (7), [95] should satisfy the additivity of trajectory intercepts. This additivity is satisfied in two dimensional QCD-motivated models and therefore should be considered as a firmly established theoretical constraint on RT. Nevertheless Goldman proved [99] that main formula (7), [95] will necessarily result in violation of the intercept additivity constraint. Moreover, the numerical values of intercepts, given by (7), [95] in the light quark sector contradict data (see [99]).

Another constraint provided by the heavy quark limit is the additivity of inverse slopes [100]. Unfortunately the model of Filipponi does not satisfied this constraint as well.

## II.10. Results based solely on work with PDG.

### a) Analysis of Tang and Norbury

This paper is concerned with the properties of RT's which are graphs of the total quantum number $J$ versus mass squared $M^2$ over a set of particles of fixed principal quantum number $N$, isospin $I$, dimensionality of the symmetry group D, spin-parity and flavor. Variations in $J$ and $L$ *(J=L+S)* are equivalent when $S$ is fixed.

Scattering processes are usually analysed by the method of partial waves. The wave function (WF) in the far zone has the form

$$\Psi(\vec{r}) = e^{i\vec{k}\vec{r}} + f(k, \cos\theta)\frac{e^{i\vec{k}\vec{r}}}{r} \tag{1}$$

where $\theta$ is the angle between the wave vector $\vec{k}$ and the position vector $\vec{r}$. In the case of bound states, the plane wave term is absent. The form factor $f$ is written as a sum of partial waves as

$$f(k^2, \cos\theta) = \sum_{l=0}^{\infty} (2l+1)a_l(k^2)P_l(\cos\theta) \tag{2}$$

$$a_l(k^2) = \frac{1}{2}\int_{-1}^{1} f(k^2, \cos\theta)P_l(\cos\theta)d\cos\theta \tag{3}$$

In 1959, Regge generalized the solution of $f$ by complexifyng angular momenta. He interpreted the simple poles of $a_l(k^2)$ on the complex $l$-plane to be either resonances or bound states. Chew and Frautschi applied the Regge poles theory to investigate the analyticity of $a_l(k^2)$ in the case of strong interactions. They postulated that all strongly interacting particles are self-generating (the bootstrap hypothesis) and that they must lie on RT's (Chew-Frautschi) conjecture.

At first, linearity was just a convenient guide in constructing the Chew-Frautschi plots because data were scarce and there were few *a priori* rules to direct the mesons and baryons into the same trajectories. Once linearity was found to be a good working hypothesis, justification was given through certain assumptions in the Regge poles theory as follows: For *Rel≥-1/2*, the partial-wave components of the scattering amplitude $f$ have only simple poles and are functions of $k^2$,

$$a_l(k^2) \simeq \frac{\beta(k^2)}{l - \alpha(k^2)} \tag{4}$$

where $\beta$ is the residue and $\alpha$ the position (Regge trajectory) of the simple poles. One can use Watson transformation to rewrite Eq. (2) as the Sommerfeld-Watson formula to include the poles.

For the purpose of plotting, authors use $J = \alpha(M^2)$. $\alpha(t)$ represents a set of leading Regge poles on the complex $l$-plane and is called the Reggeon.

The linearity of RT's has been the object of investigation once again recently. On the theoretical front, Tang [102] used perturbative QCD to show that RT's are *nonlinear* by studying high energy elastic scattering with mesonic exchange in the case of both fixed and running coupling constants.

On the experimental side Brandt et al [28] affirmed the existence of nonlinear Pomeron trajectories from the data, analysis of the recent *UA8* and Intersecting Storage Rings (ISR) experiments at CERN. They published a parametrization of Pomeron trajectories containing a quadratic term

$$\alpha(t) = 1.10 + 0.25t + \alpha''t^2, \tag{5}$$

where $\alpha''$ is a constant.

In this paper, Tang check the claims of nonlinear RT by plotting the recent experimental data published in the 1998 PDG book. Their plots confirm the existence of *nonlinear* RT's. Early Chew-Frautschi plots also show that RT *fan out*. Tang refer to this nonintersecting property as "*divergence*". Authors also show that many trajectories *intersect*.

When trajectories of different principal quantum numbers $N$ but all other quantum number fixed are plotted together, they appear parallel. Authors call this property "*parallelism*".

The starting point for constructing a meson RT is the meson assignment table in PDG. Tang fix $I$ and flavor by selecting particles from a single column. From this column, they isolate different RT's by fixing $N$ and spin-parity, when they select particles with consecutive values of $J$. For example the $1\ ^1S_0$, $1\ ^1P_1$ and $1\ ^1D_2$ states constitute an $N=1$ singlet RT; $1\ ^3P_0$ and $1\ ^3D_1$ the $N=1$ first triplet; $1\ ^3P_1$ and $1\ ^3D_2$ the $N=1$ second triplet; $1\ ^3S_1$, $1\ ^3P_2$, $1\ ^3D_3$, and $1\ ^3F_4$ the $N=1$ third triplet; $2\ ^3S_1$ and $2\ ^3P_2$ the $N=2$ third triplet and so on. Tang use the experimental error instead of the width to measure the accuracy of the mass of a meson. The error of mass squared, $dM^2$, is calculated from the mass $M$ and its error $dM$ by the relation $dM^2 = 2M\ dM$. The end results are 13 RT's containing 2 particles each, 4 containing 3 particles each and 4 containing 4 particles each.

We have to admit here that authors [101] used only summary Tables from PDG and do not use Full Listings, which would lead to richer set of RT's.

In particular authors analyze only 3- and 4-star baryons from PDG. The baryon assignment table use a set of slightly different quantum numbers, such as $J^P$, $(D, L_N^P)$ and $s$. The new quantum number $D$ is the dimensionality of the symmetry group and has the value of either 56 or 70. $D$, $S$, flavor, strangeness and isospin are constant along a baryon RT. Only $L$ is allowed to vary $N$ changes with $L$ in the same integer steps so that a change in $N$, is the same as a change in $L$. Hence we can ignore the consideration of $N$.

Regge recurrences are separated by 2 units of $J$. In the case of mesons, we can plot two RT together in some cases because the cross channel forces between them vanish. It is known as the "exchange degeneracy" (EXD) [103], which arises out of the cross channel forces which split $a(l, k)$ into even (+) and odd (-) signatures as $a_{\pm}(l, k)$. The separation of the even and odd signatures correspond to the two different RT's. If the cross channel forces vanish (as in the case of mesons), the even and odd signatures coincide and the even and odd trajectories overlap. It means $\alpha_{+}=\alpha_{-}$ and $\beta_{+}=\beta_{-}$. These are called the EXD conditions. In the case of baryons, the cross channel forces persist.

Linearity means that all the particles of a RT must lie on the straight line $M^2=\alpha J+\beta$. In a graphical analysis, nonlinearity can be detected by simple inspection in only extreme cases. Linearity on the other hand is more different to judge. Therefore authors devise a method called "*zone test*" to facilitate this judgement. They test linearity by the "zone test" on RT's with 3 or more particles.

A test zone of an experimental RT is defined to be the area bounded by the error bars of the first and the last particles and the straight lines joining them. Figures 1-6 illustrate these test zones (regions enclosed by the dotted lines). A zone contains all the possible straight lines crossing the error bars of the first and the last particles. A RT can be a straight line if the error bars of all other particles intersect the zone. In most cases intersections are easily discernible by intersection. If ambiguity ever arises in borderline cases, an exact numerical version of the zone test is used.

Suppose we are given a sequence of $N$ mesons and their values of mass square, with errors $\left\{M_i^2 \pm dM_i^2\right\}$. We calculate the equation of the straight line connecting $M_i^2 \pm dM_i^2$ and $M_N^2 \pm dM_N^2$ and then the equation of the line connecting $M_1^2 \pm dM_1^2$ and $M_N^2 \pm dM_N^2$. These two lines define the boundaries of the zone. For each $J$, we can calculate the bounds to be intersected by the error bar to qualify as a linear RT. For a 3-particle trajectory in which the particles are labelled (1, 2, 3), the lower and upper bounds at $J=2$, are calculated as

$$lb(3,2)=\frac{\left(M_1^2-dM_1^2\right)+\left(M_3^2-dM_3^2\right)}{2}$$
$$ub(3,2)=\frac{\left(M_1^2+dM_1^2\right)+\left(M_3^2+dM_3^2\right)}{2}$$

(2)

where *lb (3, 2)* stands for the lower bound and *ub (3, 2)* the upper bound of particle 2 along a 3-particle trajectory. Similarly, one can calculate the bounds of particles 2 and 3 along a 4–particle trajectory as

$$lb(4,2)=\frac{2\left(M_1^2-dM_1^2\right)+\left(M_4^2-dM_4^2\right)}{3}$$
$$ub(4,2)=\frac{2\left(M_1^2+dM_1^2\right)+\left(M_4^2+dM_4^2\right)}{3}$$

(3)

$$lb(4,3) = \frac{\left(M_1^2 - dM_1^2\right) + 2\left(M_4^2 + dM_4^2\right)}{3}$$

$$ub(4,3) = \frac{\left(M_1^2 + dM_1^2\right) + 2\left(M_{41}^2 + dM_4^2\right)}{3}$$

(4)

One can generalize these results for particle *i* along an n-particle trajectory as

$$lb(N,i) = \frac{(N-i)\left(M_1^2 - dM_1^2\right) + (i-1)\left(M_N^2 - dM_N^2\right)}{N-1}$$

$$ub(N,i) = \frac{(N-i)\left(M_1^2 + dM_1^2\right) + (i-1)\left(M_M^2 + dM_N^2\right)}{N-1}$$

(5)

Tang use the zone test to check linearity by simple inspection in Figs.1-6. At least one the error bars of the intermediate particles fails to intersect the test zone in all of the figures except Fig. 3.

The zone test for baryon trajectories are illustrated in Figs. 7-9. The *Λ* RT in Fig. 8 is shown to be nonlinear by the numerical zone test. In summary, 6 of 8 RT's with 3 or more particles each are shown to be *nonlinear*.

Divergence seems to be a property of the RT's in the early Chew-Frautschi plots and is also a prediction of the numerical calculations by Kahana et al. [104]. Divergence is defined to be the conjunction of two properties: (1) nonintersection and (2) fanning out.

Authors check divergence by plotting families of meson RT's with the same isospin and spin-parity in Figs.10-15. It is observed that nonlinear trajectories of similar masses *intertwine*. In general, RT's are not evenly separated in a graph. Some trajectories can be *obscured* when many of them are plotted over a large mass range on the same graph. Tang adopt a numeration scheme which allows them to identify the obscured trajectories in separate plots. Divergence is clearly violated in Fig.14 when trajectories intersect.

Although individual meson trajectories do not fan out, it can be seen in Fig. 10, 11, 13 and 15 that groups of them diverge on a global level. Tang also notice that these groups can be labeled according to mass difference. In general, the mass of the intersecting trajectories does not differ significantly. On the other hand, divergent trajectories have large mass difference. In summary, trajectories of small mass difference do not diverge but those of large mass difference fan out in their plots.

Parallelism refers to the properties that RT's of different values of *N* (which are otherwise identical) are parallel. Two trajectories are parallel if the dynamics are similar. There is no *a priori* reason why parallelism must hold. There are only two *φ* RT's with *N=1* and *N=2* (see Fig. 18). However these RT's consist of only 2 or 3 mesons each. It is not clear how they will behave at *J>2*. The error of *f₂(2010)* is also quite large compared to the separation of the two RT. In conclusion, the status of parallelism as a candidate for a property of RT's is still uncertain.

Conclusion: The linearity of RT is clearly violated in Figs. 1, 2, 4, and 6 by simple inspection but is supported by the numerical zone test in Fig. 3. Divergence is not observed on an individual basis. On the other hand, divergence of groups of RT's of small mass difference is observed on a global level. Due to insufficient data, parallelism is inconclusive. Author's analysis disagree with those models which predict linear mesonic RT in the small $J$ limit. Tang also shows that meson RT's both with increasing and decreasing slopes exist experimentally. Our potential model [34-39] is in total accord with this result.

### b)Naked truth about hadronic RT

The aim of the present paper [105] is to dissect the naked truth about hadronic RT's. The whole issue is an eclectic mix of confusions, partly because of a huge number of quark models, which are in many ways contradictory to each other. This leads to ambiguous conclusions about the true nature of RT's. In this situation the hadronic data itself presents the purest imprint of the hadronic world. Therefore we will scrutinize the last issue of Review of Particle Physics [45], and reconstruct all possible RT for mesons and baryons. Then we will extract all the slopes, characterizing the given RT and examine how they deviate from the standard recipe $\alpha'=0.9 GeV^{-2}$. The best parameters which describe wild deviations of RT from linear and parallel lines are the dispersion $\sigma$ and the average value of slope for the given RT, $<\alpha'>$.

### Mesons

We will work with the full listings of PDG [45]. First on the list is the light unflavored mesonic sector, where massive experimental discoveries were made during the last decade. The most extensively investigated are scalar-isoscalar $f$-mesons, total 28 states. From these data [45, 80] we could construct four radial and two orbital $f$ trajectories. Radial RT for $f_2$ ($J^{PC}=2^{++}$) is a 13-plet and it is the most nonlinear RT in Nature, with $\sigma=7.91\ GeV^{-2}$ and $<\alpha'>=6.37\ GeV^{-2}$ (see Fig.1). This RT has two peak slopes of $17.54\ GeV^{-2}$ and $27.47\ GeV^{-2}$. The $f_0$ radial RT is nonet and it is also essentially nonlinear (EN), with $\sigma=1.69\ GeV^{-2}$, $<\alpha'>=2.30\ GeV^{-2}$ with peak slope value of $6.10\ GeV^{-2}$ (see Fig. 1). The $f_4$ radial RT has only three states, but it's EN with $\sigma=2.85\ GeV^{-2}$, $<\alpha'>=4.40\ GeV^{-2}$ and peak slope value of $6.41\ GeV^{-2}$. All the EN mesonic RT will be assembled in Table 1. With new PWA just coming from Crystal Barrel Data [106], it is possible to construct radial RT for the $f_1$ mesons ($J^{PC}=1^{++}$). This $f_1$ is a quartet including the newly discovered $f_1(1971)$. It is EN with $\sigma=1.55\ GeV^{-2}$, $<\alpha'>=2.29\ GeV^{-2}$ with peak slope value of $3.69\ GeV^{-2}$. Orbital RT for f-mesons include parent $f_0$ and daughter $f_0$. Parent $f_0$ is a quartet with $\sigma=0.65\ GeV^{-2}$, $<\alpha'>=1.29\ GeV^{-2}$ and with peak slope value of $2.03\ GeV^{-2}$ and it is EN. Daughter $f_0$ RT is a triplet with $\sigma=0.82\ GeV^{-2}$, $<\alpha'>=1.26\ GeV^{-2}$ and peak slope value of $1.84\ GeV^{-2}$. These two RT are EN and nonparallel.

We refer the reader to [105] for the full description of analysis of mesonic sector.

Table1: Slopes for EN meson RT ($\alpha'$, average $\langle\alpha'\rangle$, mean square deviation $\sigma$, in GeV$^{-2}$)

| RT for mesons | Slopes $\alpha'$ for neighbor pairs | $\langle\alpha'\rangle$ | $\sigma$ |
|---|---|---|---|
| $f_0(0^{++})$ parent | 3.00 0.78 0.94 | 1.58 | 1.01 |
| $f_0(0^{++})$ daughter | 1.84 0.68 | 1.26 | 0.82 |
| $f_0(0^{++})$ radial | 3.13 1.16 2.34 1.45 0.88 6.10 1.72 1.66 | 2.30 | 1.69 |
| $f_1(1^{++})$ radial | 2.56 3.69 0.63 | 2.29 | 1.55 |
| $f_2(2^{++})$ radial | 2.39 3.56 17.54 3.33 1.64 1.82 4.95 27.47 1.61 3.60 3.36 5.13 | 6.37 | 7.91 |
| $f_4(4^{++})$ radial | 2.38 6.41 | 4.40 | 2.85 |
| $a_2(2^{++})$ radial | 0.98 3.18 1.18 2.04 1.27 | 1.73 | 0.90 |
| $a_0(0^{++})$ parent | 2.61 0.86 1.03 | 1.50 | 0.96 |
| $a_0(0^{++})$ daughter | 3.06 1.58 | 2.32 | 1.05 |
| $a_0(0^{++})$ gr.daugh | 4.88 2.87 | 3.88 | 1.43 |
| $\eta$ radial | 0.72 2.65 0.96 1.06 1.10 3.70 | 1.70 | 1.20 |
| $h_1$ radial | 1.81 1.61 0.69 0.87 | 1.25 | 0.55 |
| $K(0^-)$ parent | 0.73 1.14 0.34 1.25 | 0.87 | 0.42 |
| $K(0^-)$ daughter | 1.69 2.38 | 2.04 | 0.49 |
| $K(0^-)$ radial | 0.53 1.92 1.44 | 1.30 | 0.71 |
| $K(1^+)$ radial | 2.90 1.32 | 2.11 | 1.12 |
| $K(2^-)$ radial | 1.54 6.49 0.57 | 2.87 | 3.18 |
| $J/\psi$ radial | 0.25 1.60 0.47 1.03 0.46 | 0.76 | 0.55 |
| $\chi_c(1P)$ parent | 1.51 3.11 | 2.31 | 1.13 |
| $\chi_b(1P)$ parent | 1.54 2.54 | 2.04 | 0.71 |
| $\chi_b(2P)$ parent | 2.11 3.66 | 2.89 | 1.10 |
| $\Upsilon$ radial | 0.09 0.15 0.21 0.16 0.30 | 0.18 | 0.08 |

It was found that out of 32 mesonic RT, 2 belong to the category of EN. Seven RT are fairly nonlinear, and only three RT are linear, which amounts to 9% share. (We did not account for doublets RT, which don't have a curvature).

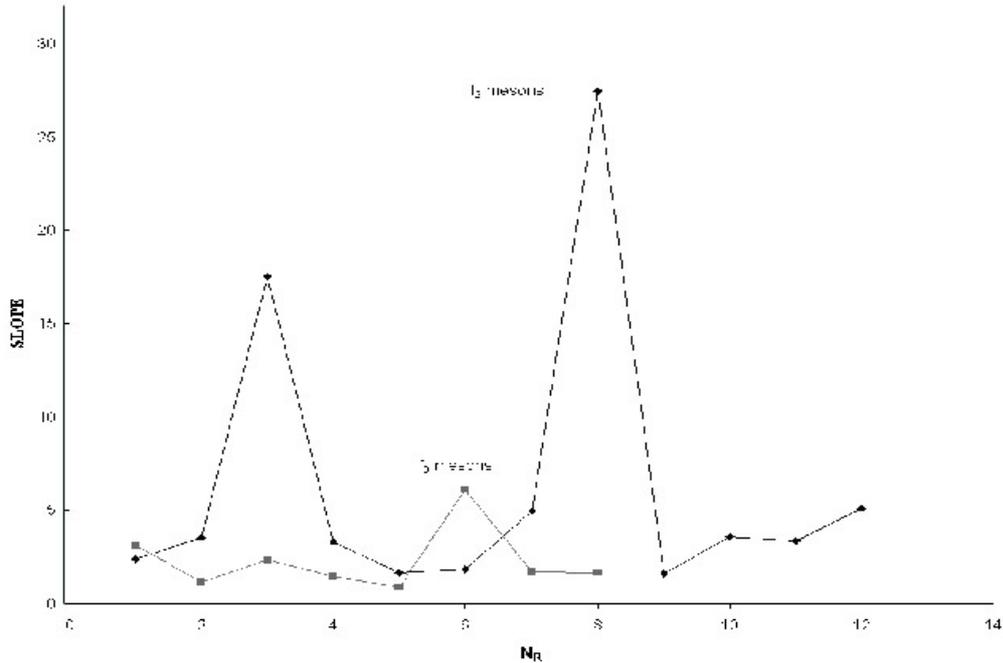

Fig.1. Slopes versus radial quantum number $N_r$ for $f_0$ and $f_2$ radial RT.

## Baryons

### N-Δ

In the baryonic sector we have many more trajectories than for mesons. Our strategy will be to discuss the most interesting cases, leaving the rest for the tables and figures.

The nonstrange sector is very rich, comprising *23N and 22Δ states* [45]. (We will neglect here isospin splitting effects). We will include in the analysis three new resonances just discovered at ELSA, SAPHIR [45]: $D_{13}^*(1895)$, $S_{11}^*(1897)$ and $P_{11}^*(1986)$. We will also include in analysis the so-called *N(3000* region) and *Δ(~3000* region), which are mostly the results of PWA by Hendry [44]. So, altogether we have *31N and 28Δ* resonances.

*N*, and *Δ* spectra exhibit a very interesting clustering structure. In nucleon sector we see the following four clusters: sextet $S_{11}(1650) - D_{15}(1675) - F_{15}(1680) - D_{13}(1700) - P_{11}(1710) - P_{13}(1720)$ is squeezed within *70 MeV* interval; triplet $D_{13}^*(1895) - S_{11}^*(1897) - P_{13}(1900)$ is squeezed within *5 MeV* interval; triplet $D_{13}(2080) - S_{11}(2090) - P_{11}(2100)$ is squeezed within *20 MeV* interval; and quartet $G_{17}(2190) - D_{15}(2200) - H_{19}(2220) - G_{19}(2250)$ is squeezed within *60 MeV* interval.

First cluster is split into three parity doublets (PD): $S_{11}(1650) - P_{11}(1710)$, $D_{13}(1700) - P_{13}(1720)$, $D_{15}(1675) - F_{15}(1680)$. Second cluster has one PD: $D_{13}^*(1895) - P_{13}(1900)$. Third cluster has one PD: $S_{11}(2090) - P_{11}(2100)$. Forth cluster has one PD: $H_{19}(2220) - G_{19}(2250)$.

In *Δ* sector we see the following two clusters: septet $S_{31}(1900) - F_{35}(1905) - P_{31}(1910) - P_{33}(1920) - D_{35}(1930) - D_{33}(1940) - F_{37}(1950)$ is squeezed within *50 MeV* interval and triplet $F_{37}(2390) - G_{39}(2400) - H_{311}(2420)$ is squeezed within *30 MeV* interval.

First cluster is split into three PD plus one extra state: $S_{31}(1900) - P_{31}(1910)$, $P_{33}(1920) - D_{33}(1940)$, $F_{35}(1905) - D_{35}(1930)$.

Second cluster has no PD. This clustering pattern and the precise mechanism of parity doubling in *N* and *Δ* spectra remained the challenges for the current quark models.

One promising approach introduced the so-called Rarita-Schwinger (RS) clusters [107]. As a result the author found three RS clusters both in *N* and *Δ* spectra, as opposed to the experimentally seen four clusters in *N,* and two in *Δ* sectors.

Now we turn to the analysis of *N*, and *Δ* RT. Major parent nucleon trajectory $P_{11}(938) - F_{15}(1680) - H_{19}(2220) - K_{113}(2700) - L_{117}(3500)$ is moderately nonlinear quintet, with $\sigma = 0.28\ GeV^{-2}$, $<\alpha'> = 0.81\ GeV^{-2}$. Major parent *Δ* RT $P_{33}(1232) - F_{37}(1950) - H_{311}(2300) - L_{315}(3700)$ is also just moderately nonlinear quartet, with $\sigma = 0.18\ GeV^{-2}$, $<\alpha'> = 0.78\ GeV^{-2}$.

Nevertheless there are plenty of nonlinear RT in the nonstrange sector (see Fig. 2). $P_{31}$ parent RT: $P_{31}(1750) - F_{35}(1905) - H_{39}(2300) - K_{313}(3200)$ is EN quartet with $\sigma = 1.63\ GeV^{-2}$, $<\alpha'> = 1.71\ GeV^{-2}$ with peak slope value of *3.53 GeV$^{-2}$*. $S_{31}$ parent RT: $S_{31}(1620) - D_{35}(1930) - G_{39}(2400) - I_{313}(2750) -$ $L_{317}(3300) - N_{321}(4100)$ stretches to the highest possible mass *4.1 GeV*. It is an EN sextet, with $\sigma=0.56$ $GeV^{-2}$, $<\alpha'>=0.97\ GeV^{-2}$, with peak slope value of *1.82 GeV⁻²*. $S_{11}$ parent RT: $S_{11}(1535) - D_{15}(1675) - G_{19}(2250)$ is extremely nonlinear triplet with $\sigma=2.51\ GeV^{-2}$, $<\alpha'>=2.67\ GeV^{-2}$, and peak slope value of *4.44 GeV⁻²*. $D_{13}$ parent RT: $D_{13}(1520) - G_{17}(2190) - I_{111}(2600) - L_{115}(3100) - N_{119}(3750)$ is essentially nonlinear quintet, with $\sigma=0.24\ GeV^{-2}$, $<\alpha'>=0.74\ GeV^{-2}$.

Among the radial RT in *N, Δ* sector there are few EN. $S_{11}$ radial RT is a quartet $S_{11}(1535) - S_{11}(1650) - S_{11}^{*}(1897) - S_{11}(2090)$. It is EN RT with $\sigma=0.87\ GeV^{-2}$, $<\alpha'>=1.72\ GeV^{-2}$ and peak slope value of *2.72 GeV⁻²*. $D_{13}$ radial RT is a quintet: $D_{13}(1520) - D_{13}(1700) - D_{13}^{*}(1895) - D_{13}(2600)$. It is EN RT with $\sigma=0.57\ GeV^{-2}$, $<\alpha'>=1.23\ GeV^{-2}$ and peak slope value of *1.72 GeV⁻²*. $P_{11}$ radial is a quintet: $P_{11}(939) - P_{11}(1440) - P_{11}(1710) - P_{11}^{*}(1986) - P_{11}(2100)$. It is EN <u>nonlinear</u> RT with $\sigma=0.59\ GeV^{-2}$, $<\alpha'>=1.29\ GeV^{-2}$ and peak slope value of *2.15 GeV⁻²*.

Some of the RT are too short (doublets) to judge on nonlinearity, but they have rather large slopes. $F_{35}$ radial RT is a doublet with $\alpha'=2.7\ GeV^{-2}$. $P_{31}$ daughter orbital RT is a doublet with $\alpha'=5.68\ GeV^{-2}$ and it is nonparallel to $P_{31}$ parent RT. $P_{31}$ parent orbital RT is a doublet with $\alpha'=2.0\ GeV^{-2}$. We conclude that in *N* and *Δ* sector we have five *N* and two *Δ* EN RT (see Table 2).

Table 2: Slopes for EN baryon RT ($\alpha'$, average $<\alpha'>$, mean square deviation $\sigma$, in $GeV^{-2}$)

| RT for baryons | Slopes $\alpha'$ for neighbor pairs | $<\alpha'>$ | $\sigma$ |
|---|---|---|---|
| N1/2⁻ parent | 4.44 0.89 | 2.67 | 2.51 |
| N3/2⁻ parent | 0.80 1.02 0.70 0.45 | 0.74 | 0.24 |
| N1/2⁺ radial | 0.84 1.18 0.98 2.15 | 1.29 | 0.59 |
| N1/2⁻ radial | 2.72 1.14 1.30 | 1.72 | 0.87 |
| N3/2⁻ radial | 1.72 1.43 1.36 0.41 | 1.23 | 0.57 |
| Δ1/2⁺ parent | 3.53 1.20 0.40 | 1.71 | 1.63 |
| Δ1/2⁻ parent | 1.82 0.98 1.11 0.60 0.34 | 0.97 | 0.56 |
| Λ1/2⁺ radial | 0.76 1.40 | 1.08 | 0.45 |
| Λ1/2⁻ radial | 1.23 2.22 1.32 | 1.59 | 0.55 |
| Λ3/2⁻ radial | 1.83 0.39 | 1.11 | 1.02 |
| Σ1/2⁺ radial | 0.83 7.58 2.65 2.49 | 3.39 | 2.91 |
| Σ1/2⁻ radial | 6.06 3.65 1.07 | 3.60 | 2.50 |
| Σ3/2⁻ radial | 3.41 1.03 | 2.22 | 1.68 |

### <u>Λ-Σ</u>

We turn now to the *Λ-Σ* sector. There are many interesting features in this *qqs* sector. One of them is exchange degeneracy (EXD) hypothesis, which happened to hold quite well in the *Λ-Σ* sector. As we will see later, EXD lead to trajectories with *negative* slopes, which never arise in *N-Δ* sector. Another feature is the *clustering* in the *Λ-Σ* sector, which is qualitatively different from clustering in the *N-Δ* sector. Third feature is the *existence of parity doubles* in the *Λ-Σ* sector.

Because *Λ-Σ* has only one strange quark their shape is still not so deformed. If we can imagine that we have three balls in a bag, and two of them are of almost the same weight, while the third a bit heavier than the two, the bag will get the form of a *pear*. It will be *reflectionally asymmetric*. Near the rest the deformation is perhaps still not so dramatic and the lowest excitations are similar to those of the nonstrange baryons. For that reason we observe also in Λ-spectrum the same sequence *1/2⁺, 1/2⁻, 3/2⁻* as in nonstrange baryons. If the heavier ball starts to gain rotational energy, the deformation will increase. The pear shape gets more pronounced and when the pear oscillated it goes to parity doublets, which we already see.

Full listings [45] give to us *18Λ* and *26Σ* resonances. Some of the states are lacking the $J^P$ assignments. Let's take a closer look at this. State *Λ(2000)* does not have $J^P$ but data from Cameron 78 (see [45]) allowed tentatively, the $J^P=1/2⁻$ assignment. Further evidence came from the recent paper by Iachello [108] and older one by Capsick-Isgur [109]. Therefore we assign $J^P=1/2⁻$ to the *Λ(2000)*. The *Λ*-states with highest masses, *Λ(2350)* and *Λ(2585)* were not described theoretically and there are no clear claims from the experiments. For this reason we will not include *Λ(2350), Λ(2585)* in our Regge analysis, and we have total of *16Λ* resonances to work with.

The situation with *Σ* hyperons is even more interesting. Two low-lying states, *Σ(1480)* and *Σ(1560)* do not have any $J^P$ assignments from the experiment, and theory can't predict them either. We will exclude *Σ(1480), Σ(1560)* from our analysis. The production experiments [45] give strong evidence for *Σ(1620),* tentatively claiming $J^P=1/2⁺$. This claim is in accord with calculations by Iachello [108]. So with newly defined *Σ\*1/2⁺(1620)*, we form an *exact PD Σ1/2⁻(1620) - Σ\*1/2⁺(1620).* The production experiments [45] give strong evidence for *Σ(1670)* bumps without $J^P$ assignments. Using predictions by Iachello [108] and Isgur [109], we clearly get $J^P=1/2⁻$ for *Σ(1670).* It's interesting that this way we have two resonances with the *same* mass and *different* $J^P$ (see [45]). Such a degeneracy waits proper theoretical explanation. The state *Σ(1690)* has most likely claim from the data [45] as $J^P=5/2⁺$. We will assign $J^P=5/2⁺$ to *Σ (1690)* in our analysis. Next *Σ* without $J^P$ assignment will be *Σ(2250).* Using the results [108, 109], we assigned $J^P=5/2⁻$ to *Σ(2250).* Last few *bumps*, *Σ(2455), Σ(2620), Σ(3000)* and *Σ(3170)* has no experimental claims for $J^P$, and there are no theoretical predictions so far for such a high masses. For this reason we will not include *Σ(2455), Σ(2620), Σ(3170)* in our analysis. Finally, we have total of *22 Σ* hyperons for our analysis.

Clustering pattern in *Σ* spectrum is very nontrivial. We clearly see three clusters there. Quartet *$P_{11}(1160) - D_{13}$ (1670) $- S_{11}^*(1670) - F_{15}^*(1690)$* is squeezed within *30 MeV* interval. There is one PD within this cluster: *$P_{11}(1660) - S_{11}^*(1670)$.* Triplet *$S_{11}(1750) - P_{11}(1770) - D_{15}(1775)$* is squeezed within *25 MeV* interval. There is one PD within this cluster: *$S_{11}(1750) - P_{11}(1770)$.* Triplet *$F_{15}(2070) - P_{13}(2080) - G_{17}(2100)$* is squeezed within *30 MeV* interval. There are no PD in this cluster. It's amazing that we have three *1/2⁺ - 1/2⁻* PD in the *Σ* sector: *$S_{11}(1620) - P_{11}^*(1620)$, $P_{11}(1660) - S_{11}^*(1670)$* and *$S_{11}(1750) - P_{11}(1770)$.*

In $\Lambda$ sector we witness only one cluster. This quartet $S_{01}(1800) - P_{01}(1810) - F_{05}(1820) - D_{05}(1830)$ is squeezed within *30 MeV* interval. The whole cluster is split into two PD: $S_{01}(1800) - P_{01}(1810)$ and $F_{05}(1820) - D_{05}(1830)$. Note that an author [107] suggested four clusters in $\Lambda$ sector. It is a big puzzle for current quark models to explain this difference in clustering and parity doubling between $N$-$\Delta$ and $\Lambda$-$\Sigma$ sectors.

Now we turn to Regge analysis of $\Lambda$-$\Sigma$ sector. We will mostly concentrate on EN RT. The $\Lambda 1/2^-$ radial RT is a quartet $S_{01}(1405) - S_{01}(1670) - S_{01}(1800) - S_{01}{}^*(2000)$. It is essentially nonlinear RT with $\sigma$=0.55 GeV$^{-2}$, $<\alpha'>$=1.59 GeV$^{-2}$ and peak slope value of *2.22 GeV$^{-2}$*. The $\Lambda 3/2^-$ radial RT is a triplet $D_{03}(1520) - D_{03}(1690) - D_{03}(2325)$. It is EN RT with $\sigma$=1.02 GeV$^{-2}$, $<\alpha'>$=1.11 GeV$^{-2}$ and peak slope value of *1.83 GeV$^{-2}$*. The $\Lambda 1/2^+$ radial RT is a triplet $P_{01}(1116) - P_{01}(1600) - P_{01}(1810)$. It is EN RT with $\sigma$=0.45 GeV$^{-2}$, $<\alpha'>$=1.08 GeV$^{-2}$. Many other $\Lambda$ RT's posses some degree of nonlinearity.

The $\Sigma 1/2^+$ radial RT is a quintet $P_{11}(1193) - P_{11}{}^*(1620) - P_{11}(1660) - P_{11}(1770) - P_{11}(1880)$. It is EN RT with $\sigma$=2.91 GeV$^{-2}$, $<\alpha'>$=3.39 GeV$^{-2}$ and peak slope value of *7.58 GeV$^{-2}$*. The $\Sigma 1/2^-$ radial RT is a quartet $S_{11}(1620) - S_{11}{}^*(1670) - S_{11}(1750) - S_{11}(2000)$.

It is EN RT with $\sigma$=2.91 GeV$^{-2}$, $<\alpha'>$=3.39 GeV$^{-2}$ and peak slope value of *6.06 GeV$^{-2}$*. The $\Sigma 3/2^-$ radial RT is a triplet $D_{13}(1580) - D_{13}(1670) - D_{13}(1940)$. It is EN RT with $\sigma$=1.68 GeV$^{-2}$, $<\alpha'>$=2.22 GeV$^{-2}$ and peak slope value of *3.41 GeV$^{-2}$*.

Amazingly, all EN radial RT in $\Lambda - \Sigma$ sectors are *mirroring* each other: $\Lambda 1/2^+$- $\Sigma 1/2^+$; $\Lambda 1/2^-$ - $\Sigma 1/2^-$; $\Lambda 3/2^-$ - $\Sigma 3/2^-$. There are no EN RT among the orbital $\Lambda - \Sigma$ trajectories.

## EXD IN $\Lambda$-$\Sigma$

It has been known for years that exchange degeneracy seems to exist experimentally at least for strange hyperons [5]. We will construct appropriate RT with tentatively assigned resonances. Major $\Sigma^-$ trajectory in this scheme will be a quartet: $P_{11}(1193) - P_{13}(1840) - F_{15}{}^*(1690) - F_{17}(2030)$. This EN RT has one *negative* slope $\alpha_2'$=-1.89 GeV$^{-2}$, $\sigma$=1.47 GeV$^{-2}$, $<\alpha'>$=-0.20 GeV$^{-2}$. Corresponding daughter RT is a triplet: $P_{11}{}^*(1620) - P_{13}(2080) - F_{15}(1915)$. This EN RT also has one *negative* slope $\alpha_2'$=1.52 GeV$^{-2}$, $\sigma$=1.49 GeV$^{-2}$, and trajectory is a quartet: $S_{11}(1620) - D_{13}(1580) - D_{15}(1775) - G_{17}(2100)$. This EN RT has one negative slope $\alpha'$=-7.81 GeV$^{-2}$, $\sigma$=5.32 GeV$^{-2}$, $<\alpha'>$=-1.67 GeV$^{-2}$. Corresponding $\Sigma 1/2^-$ daughter RT is a triplet: $S_{11}{}^*(1670) - D_{13}(1940) - D_{15}{}^*(2250)$. This is moderately nonlinear RT with $\sigma$=0.18 GeV$^{-2}$, $<\alpha'>$=0.90 GeV$^{-2}$ and it's nonparallel to the parent RT.

We start $\Lambda$ sector with major $\Lambda 1/2^+$ trajectory, which is a quintet: $P_{01}(1116) - P_{03}(1890) - F_{05}(1820) - F_{07}(2020) - H_{09}(2350)$. This EN RT has one *negative* slope $\alpha_2'$=-3.85 GeV$^{-2}$, $\sigma$=2.76 GeV$^{-2}$, $<\alpha'>$=-0.71 GeV$^{-2}$. Major negative parity $\Lambda 1/2^-$ trajectory is a quartet: $S_{01}(1405) - D_{03}(1520) - D_{05}(1830) - G_{07}(2100)$. This is ENRT with $\sigma$=1.20 GeV$^{-2}$, $<\alpha'>$=1.64 GeV$^{-2}$ and peak slope value of *3.03 GeV$^{-2}$*.

As we see, EXD in $\Lambda-\Sigma$ sector leads to a *new class* of trajectories, which are characterized by *negative average* slopes. Five out of six $\Lambda-\Sigma$ RT's are essentially nonlinear (see Table 3).

Table3: Slopes for baryonic essentially nonlinear EXD RT

| RT for baryons | Slopes $\alpha'$ for neighbor pairs | $<\alpha'>$ | $\sigma$ |
|---|---|---|---|
| $\Sigma 1/2^+$ parent | 0.51  -1.89  0.79 | -0.20 | 1.47 |
| $\Sigma 1/2^+$ daughter | 0.59  -1.52 | -0.47 | 1.49 |
| $\Sigma 1/2^-$ parent | -7.81  1.53  1.26 | -1.67 | 5.32 |
| $\Lambda 1/2^+$ parent | 0.43  -3.85  1.30  0.69 | -0.71 | 2.76 |
| $\Lambda 1/2^-$ parent | 3.03  0.96  0.94 | 1.64 | 1.20 |
| $\Xi 1/2^+$ parent | 0.48  3.14 | 1.81 | 1.88 |

## $\underline{\Xi, \Omega, \text{Charmed, Beauty baryons}}$

We still have to analyze double-strange hyperons, $\Xi$ *(qss)*. Full listings [45] give to us 11 $\Xi$'s. Some of the states are lacking the $J^P$ assignment. State $\Xi$ *(1620)* is a bump, which does not have $J^P$ assignment from the experiment, and it also could not be predicted by theory so far. For this reason, we will exclude $\Xi(1620)$ from our analysis. The state $\Xi(1690)$ does not have $J^P$ assignment from the experiment, but Iachello [108] predict this state with $J^P=1/2^+$. The state $\Xi(1950)$ does not have $J^P$ assignment from the experiment, but Iachello [108] predict this state with $J^P=3/2^-$. The state $\Xi(2030)$ has tentative assignment $J^P=5/2^?$ from the data [45], and we will consider it as a $\Xi^*5/2^-(2030)$. Resonance $\Xi(2120)$ does not have $J^P$ assignment from the experiment, but Iachello [108] predicts this state with $J^P=3/2^-$. Next bump, $\Xi(2250)$ does not have $J^P$ assignment from the experiment, but Iachello [108] predict $J^P=1/2^+$. The state $\Xi(2370)$ does not have $J^P$ assignment from the data, but Isgur [109] predict this state with $J^P=7/2^-$. The last resonance, $\Xi(2500)$, does not have $J^P$ assignment from the data, and it also could not be predicted by theory so far. For this reason, we will exclude $\Xi(2500)$ from our analysis. Finally, we have nine $\Xi$'s to work with: $P_{11}(1315)$, $P_{13}(1530)$, $P_{11}^*(1690)$, $D_{13}(1820)$, $P13^*(1950)$, $F_{15}^*(2030)$, $D_{13}^*(2120)$, $P_{11}^*(2250)$, $G_{17}^*(2370)$. There is no clustering in $\Xi$ sector. Major parent $\Xi$ RT is a doublet $P_{11}(1315) - F_{15}^*(2030)$, with $\alpha'=0.84$ $GeV^{-2}$, $\sigma=0$. Another RT is a parent $\Xi 3/2^-$. It is a doublet $D_{13}(1820) - G_{17}^*(2370)$, with $\alpha'=0.87$ $GeV^{-2}$, $\sigma=0$. $\Xi 1/2^+$ radial RT is a triplet $P_{11}(1315) - P_{11}^*(1690) - P_{11}^*(2250)$. It has $\sigma=0.31$ $GeV^{-2}$, $<\alpha'>=0.67$ $GeV^{-2}$ and it is fairly nonlinear RT. The $\Xi 3/2^+$ radial RT is a doublet $P_{13}(1530) - P_{13}^*(1950)$, with $\alpha'=0.68$ $GeV^{-2}$, $\sigma=0$. The $\Xi 3/2^-$ radial RT is a doublet $D_{13}(1820) - D_{13}^*(2120)$, with $\alpha'=0.85$ $GeV^{-2}$, $\sigma=0$. As we see, basically all RT's in $\Xi$ sector are too short to make a conclusion about their linear/nonlinear nature.

If we assume EXD for $\Xi$ hyperons, we can construct parent $\Xi 1/2^+$ triplet: $P_{11}(1315) - P_{13}^*(1950) - F_{15}^*(2030)$. This RT happened to be EN with $\sigma=1.88$ $GeV^{-2}$, $<\alpha'>=1.81$ $GeV^{-2}$ and peak slope value of $3.14$ $GeV^{-2}$ (see Table 3).

If we consider $\Omega$, charmed baryons and beauty baryons, there are not enough data to construct and analyze RT's. We conclude that out of total 21 baryons RT 13 are EN (62%). Five RT are fairly nonlinear (24%) and *only 3 RT (14%)* are in fact linear. (We don't account for doublet RT here, because they have no curvature).

## **Conclusions**

We have constructed and scrutinized here *all* possible RT for the full listings PDG2000 [45] and even accounted for the newest data on mesons [80, 106]. We want to stress that this approach leads to *minimal bias* in the interpretation of the results, unlike the results from any available quark models.

We have shown that in the mesonic sector out of total 32 RT, 22 trajectories are EN (or 69%), and 7 trajectories are fairly nonlinear (or 22%). Only three RT could be classified as linear, which amounts to 9% share. Among EN meson RT 10 are orbital and 12 are radial.

In baryonic sector out of total 21 RT, 13 RT's are EN (or 62%), and 5 RT's are fairly nonlinear (or 24%). Only 3 RT could be classified as linear, which amounts to 14% share. Among EN baryon RT 4 are orbital and 12 are radial.

Appropriate dispersion, $\sigma$, for nonlinear mesonic and baryonic RT span the range *1~27.5 GeV⁻²*.

We have four doublet RT in mesonic sector and 31 doublets RT's in baryonic sector, which don't have a *curvature*, and so this massive sector is used only for the evaluation of parallelism between different RT's.

So, our results strongly disagree with general opinion, that hadron RT are straight and parallel lines. As the data shows, in the currently available resonance energy region, meson and baryon RT are *grossly nonlinear*, and only small fraction (~12%) of all RT could be classified as linear, with $\sigma\sim 0$, $\alpha'\approx 0.9\ GeV^{-2}$. The existence of clusters in baryon spectra and their absence in mesonic sector is a big puzzle. In *N*-sector we have four clusters: sextet, quartet and two triplets. In $\Delta$-sector we have only two clusters: septet and triplet. In $\Sigma$-sector we have three clusters: quartet and two triplets. In $\Lambda$-sector we have one cluster, a quartet. The *N, $\Delta$* clusters have average spacing between the levels of *8.9 MeV*, and $\Lambda$, $\Sigma$ clusters have very similar average spacing of *8 MeV*. There are more interesting features between different types of clusters. First $\Sigma$ cluster *$\Sigma$(1660-1690)* overlaps and could be inserted inside first *N* cluster *N(1650-1720)*

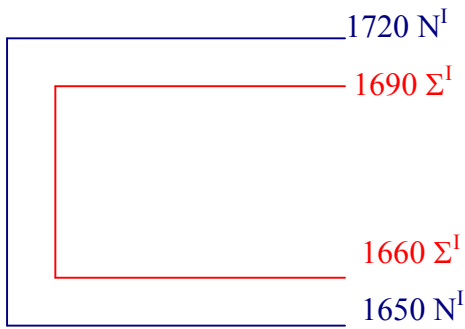

Third *N* cluster *N(2080-2100)* overlaps and could be inserted inside third $\Sigma$ cluster *$\Sigma$(2070-2100)*.

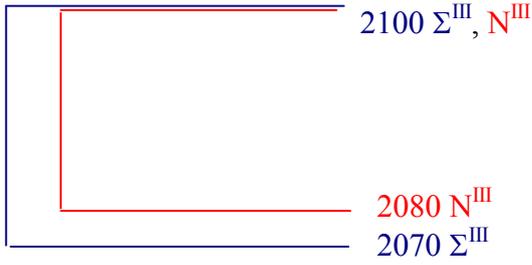

Total spread of *N, Δ, Λ, Σ* supercluster is *770 MeV*. Average level spacing is *19.3 MeV*.

First nucleon cluster is split into three PD: *1/2⁻ - 1/2⁺, 3/2⁻ - 3/2⁺, 5/2⁻ - 5/2⁺*.

First *Δ* cluster is also split into three PD: *1/2⁻ - 1/2⁺, 3/2⁻ - 3/2⁺, 5/2⁻ - 5/2⁺*. So they are exactly *mirroring* each other. Within *N* cluster, PD *1/2⁻ - 1/2⁺* and *3/2⁻ - 3/2⁺* occurred twice.

Second *N* cluster has one PD: *3/2⁻ - 3/2⁺*. Second *Δ* cluster has no PD at all.

Third *N* cluster has one PD, *1/2⁻ - 1/2⁺* and fourth *N* cluster has one PD, *9/2⁻ - 9/2⁺*.

Within *Σ* clusters we find two *1/2⁻ - 1/2⁺* PD which are exactly *mirroring* two *N1/2⁻ - 1/2⁺* PD.

In total, in *N, Δ, Λ* and *Σ* sectors we have six *1/2⁻ - 1/2⁺* PD, three *3/2⁻ - 3/2⁺* PD, three *5/2⁻ - 5/2⁺* and one *9/2⁻ - 9/2⁺* PD. The exact dynamical reasons for such clustering and PD patterns in baryons remains a big puzzle for theory.

## II.11. EXD and hadronic RT's

We have already touched briefly the EXD phenomenon in Tang, Norbury approach [101] and our analysis of worldwide data [105]. There exist group of authors, who devoted much attention to EXD recently [110-112].

### a) Kontros model

In the paper [110] two models for the Pomeron supplemented by exchange-degenerate sub-leading RT's are fitted to the forward scattering data for a number of reactions. By considering new Pomeron models, authors extend the recent results of the COMPAS group. Authors continue the program initiated before, by scrutinizing two more earlier unexplored models for the Pomeron. Authors show that they do not require the secondary RT's to be exchange-degenerate. As seen from Fig.1, [110] the 10 resonances belonging to the 4 different $I^G$ ($J^{PC}$) families ρ-ω-f₂-a₂ are compatible with a unique linear EXD RT with $\alpha(0)=0.48$. So, authors [110] have fitted two hitherto unexplored models for the Pomeron to the data and find that exchange degeneracy is a good approximation to reality.

### b) Desgrolard model

The exchange-degeneracy of the mesonic *f, ω, ρ* and *a₂* RT's dominant at moderate and high energies in hadron elastic scattering is analysed from two viewpoints [111]. The first concerns the masses of the resonances lying on these trajectories; the second deals with the total cross-sections of hadron and photon induced reactions. Neither set of data support exact EXD.

There are two kinds of EXD, qualified as *strong* and *weak*. In weak EXD, only the RT's with different quantum numbers coincide. In strong EXD, in addition, the residues of the corresponding hadronic amplitudes at the given pole in the *J*-plane. Conclusive and definite statements about weak EXD, however, are not possible without sufficiently precise experimental information about the hadrons lying on each RT. From this point of view, the most relevant RT's are the *f-, ω-, ρ-,* and *a₂-,* which can variously be exchanged in the *t*-channel of many elastic reactions.

To examine the agreement of weak EXD with data, authors first assume that the 4 RT's *f-, ω-, ρ-,* and *a₂-* are linear and coincide. Writing the relevant EXD linear EXD as

$$\alpha_{e-d}\left(m^2\right) = \alpha_{e-d}\left(0\right) + \alpha'_{e-d}m^2 \tag{1}$$

they determine the $\alpha_{e-d}\left(0\right)$ and $\alpha'_{e-d}$ by fitting 11 resonances lying on *f, ω, ρ,* and *a₂* RT's. Using MINUIT code they find

$$\alpha_{e-d} = 0.4494 \pm 0.0007, \; \alpha'_{e-d} = \left(0.9013 \pm 0.0011\right)GeV^{-2}, \; x^2/D_0F = 117.9 \tag{2}$$

The very high value of $x^2/D_0F$ is not surprising because (i) the data exhibit a known nonlinearity of the RT's and ii) the masses of the low lying resonances are measured with very high precision.

The conclusion is, thus, that in of an apparent agreement, with resonance data, *weak* EXD of the *f-ω-ρ-a₂-* RT's is not supported by the resonance data when a precise numerical analysis is performed. In order to verify the possibility of a limited validity of EXD authors have considered a weaker version where the RT's are grouped in pairs. For any grouping in pairs one can obtain the $\chi^2$ from Table 1, because each pair is considered independently of the other. An obvious general conclusion follows from this very simple analysis: under a careful numerical investigation, there are no experimental evidences from the resonances from the resonance region that the *f, ω, ρ, a₂-* RT's can be assumed to be EXD (perhaps with the exception of the single pair ρ-a₂).

The EXD hypothesis for the *f, ω, ρ, a₂* RT's can be checked also using elastic hadron scattering data. Authors restrict their analysis to the data on total cross-sections, $\sigma^{(t)}(s)$. In addition to the main goal authors test the hypothesis of two models of Pomeron, each one with two components. Thus they analyze the data using, the following expressions for the forward amplitudes $A_{ab}\left(s, t=0\right)$

$$A_{p\pm p}\left(s,0\right) = P_{NN}\left(s\right) + f_{NN}\left(s\right) + a_{NN}\left(s\right) \mp \omega_{NN}\left(s\right) \mp \rho_{NN}\left(s\right) \tag{2}$$

The conclusions are very brief. The fits with non degenerate RT's lead to somewhat better $\chi^2$ which can be taken as an indication in flavor of non degenerate RT's.

Thus, given that any model for scattering amplitudes should be in agreement with both types of data, from spectroscopy and from total cross-sections authors conclude that the hypothesis of exact exchange degeneracy, even in its weak formulation, is not supported by the present data.

### c) Fiore model

In present paper authors construct an explicit model for light unflavored mesonic RT. This model has correct threshold and asymptotic behaviour and fits the data on both the masses and widths of the observed mesonic resonances [112].

They start from a simple analytical model, where the imaginary part of the trajectory is chosen as a sum of the single threshold terms

$$\text{Im } \alpha(s) = \sum_n c_n (s - s_s)^{1/2} \left( \frac{s - s_n}{s} \right)^{\text{Re } \alpha(s_n)} \theta(s - s_n) \qquad (1)$$

with the correct asymptotic and threshold behaviour.

From the dispersion relation for the trajectory

$$\text{Re } \alpha(s) = \alpha(0) + \frac{s}{\pi} PV \int_0^\infty ds' \frac{\text{Im } \alpha(s')}{s'(s' - s)} \qquad (2)$$

where PV means the Cauchy principal value of the integral, the real part can be easily calculated. Authors got a lengthy formulas for $Re\alpha'(s)$. Then the width of a resonance is defined as

$$\Gamma(M^2) = \frac{\text{Im } \alpha(M^2)}{M \text{ Re } \alpha'(M^2)} \qquad (3)$$

Let us first consider the $\rho$ and $a_2$ RT. In fact only the EXD $\rho$-$a_2$ RT provides for a large set of input data: ten plus two, which need confirmation. As a first attempt, they keep all the resonsnces on an EXD $\rho$-$a_2$ RT and fit with five correction cycles. In agreement with (3) the widths of the resonances are affected to a larger extent by a breaking of EXD.

The study of the RT becomes, in this approach, important under many aspects. EXD must be badly broken since attempts to describe together $f$ and $\omega$ in this model failed. The EXD violation can be understood by noticing that the width of the $\omega(782)$ is only 4% of the width of the $f(1270)$ and it will be difficult to find an analytic function satisfying these constraints. Results of fit show that EXD breaking affects more the widths than the real part of the RT's (see Fig. 2 [112]).

Conclusion: In this paper authors attempt to systematize the light unflavored mesonic RT's by simultaneous fits to the masses and widths of the resonances. A deeper understanding of EXD plays an important role in this program. The imaginary part of the RT depends on the presence of one or more branch points, related to corresponding thresholds, for positive values of the argument. The first hypothesis is that these thresholds are additive. Analyticity places an upper bound on the asymptotic behaviour of RT. The second assumption fixes a particular asymptotic form compatible with this constraint. These two assumptions, regarding only the imaginary part of the RT, determine the analytic structure of the model.

The parameters of the model, the position of the thresholds and their weight together with the intercept of the RT, have been for the EXD $\rho$-$a_2$ RT's and for the $\rho$ and $f$ *alone*, and the EXD violation has been discussed and explained in the context of the model.

Due to the asymptotic behaviour of the real part only a *finite* number of resonances may lie on a given RT this model. However, for the examples considered here the bound on the resonance masses is too high for putting any constraint on a future search of new states. For the $f$ RT in fact, a higher effective threshold will be probably present.

## III. C O N C L U S I O N S

Regge trajectories continue to attract attention of many authors 40 years after their introduction. Large amount of papers devoted to hadron structure didn't notice or ignore the effects of nonlinearity in RT. We omit all these papers in present Review, because our focus is nonlinear RT's and their manifestation both in different experimental reactions and in theoretical models. We did not consider glueballs and exotica here because there are just a few papers devoted to RT's in that sector.

It was shown that well over 100 theoretical and experimental papers admit the nonlinear nature of hadronic RT. Our NRQM invent the quantitative method to estimate the degree of nonlinearity of given RT which is calculation of mean square deviation of slopes $\sigma$, for given RT.

The results of the potential model fits and predictions for baryon and meson spectra and RT reveals distinctive feature – RT in many cases are nonlinear functions of J. This fundamental feature is in accord with analysis of pure experimental RT from PDG2000, and with predictions of different quark models, reviewed here. RT's for mesons and baryons are not straight and parallel lines in general in the current resonance region both experimentally and theoretically, but very often have appreciable curvature, which is flavor-dependent.

It was established long ago, that the experimental RT's for N, $\Delta$ baryons are not strictly straight lines [43]. The authors of this seminal review and Hendry in [44] considered the facts of nonlinear behavior of the RT in mass squared. Rather, as Hendry concludes, baryon resonances seem more linear as a function of c.m. momentum.

As the analysis of PDG2000 shows [105], in the currently available resonance energy region, mesonic and baryonic RT are grossly nonlinear, and only small fraction (~12%) of all RT could be classified as a linear, with $\sigma \approx 0$, $\alpha' \approx 0.9$ GeV$^{-2}$. Appropriate dispersion $\sigma$, for nonlinear mesonic and baryonic RT span the range 1~7.9 GeV$^{-2}$, and slopes span the range 1~27.5 GeV$^{-2}$ .

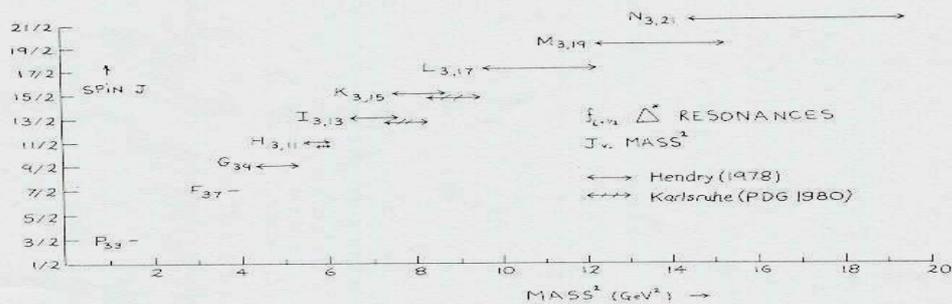

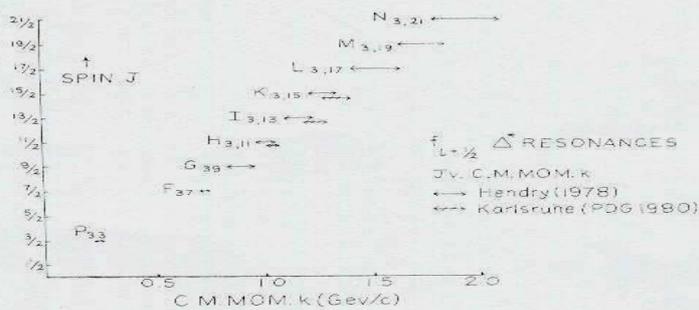

## Acknowledgements

Author is very grateful to G.S. Sharov for the help with manuscript and to the Internet café Omut, Kharkov, Ukraine, www.clubomut.kharkov.ua for the help with computing facilities.